\documentclass[12pt]{article}
\usepackage{epsfig,amsfonts,amssymb}
\usepackage{hyperref}
\usepackage{comment}
\usepackage{cite}
\input epsf.sty
\usepackage{cite}

\def\ZZZ{{\hbox{ Z\kern-1.6mm Z}}}
\def\RRR{{\hbox{ R\kern-2.4mm R}}}
\def\CCC{{\hbox{ C\kern-2.0mm C}}}
\def\zzz{{\hbox{z\kern-1mm z}}}

\newcommand{\qeq}{{\hbox{=\kern-2.3mm ? \kern.5mm }}}
\renewcommand{\qeq}{=}

\newcommand{\eps}{\epsilon}

\newcommand{\vp}{\varphi}
\newcommand{\ve}{\varepsilon}

\newcommand{\bJ}{{\bf J}}

\newcommand{\JJ}{{\cal J}}

\newcommand{\MM}{{\cal M}}

\newcommand{\OO}{{\cal O}}

\newcommand{\wt}{\widetilde}

\newcommand{\NN}{{\cal N}}

\newcommand{\be}{\begin{equation}}
\newcommand{\ee}{\end{equation}}
\newcommand{\ben}{\begin{eqnarray}\displaystyle}
\newcommand{\een}{\end{eqnarray}}

\newcommand{\refb}[1]{(\ref{#1})}
\newcommand{\p}{\partial}
\newcommand{\sectiono}[1]{\section{#1}\setcounter{equation}{0}}

\newcommand{\gsim}{\stackrel{>}{\sim}}

\newcommand{\ia}{i}
\newcommand{\ja}{j}

\def\one{{\hbox{ 1\kern-.8mm l}}}
\def\zero{{\hbox{ 0\kern-1.5mm 0}}}

\newcommand{\bea}[1]{\begin{eqnarray}\label{#1} }
\newcommand{\eea}{\end{eqnarray}}

\newcommand{\eqref}{\refb}

\usepackage{bm}
\usepackage[table]{xcolor}

\def\asnote#1{{\color{red} #1}}

\newcommand{\bM}{{\bf M}}

\def\bj{{\bf j}}

\def\figone{

\def\JPicScale{0.8}
\ifx\JPicScale\undefined\def\JPicScale{1}\fi
\unitlength \JPicScale mm
\begin{picture}(65,75)(0,0)
\linethickness{0.3mm}
\multiput(20,70)(0.12,-0.12){167}{\line(1,0){0.12}}
\linethickness{0.3mm}
\multiput(40,50)(0.12,0.12){167}{\line(1,0){0.12}}
\linethickness{0.3mm}
\put(40,30){\line(0,1){20}}
\linethickness{0.3mm}
\multiput(20,10)(0.12,0.12){167}{\line(1,0){0.12}}
\linethickness{0.3mm}
\multiput(40,30)(0.12,-0.12){167}{\line(1,0){0.12}}
\put(25,20){\makebox(0,0)[cc]{1}}

\put(55,20){\makebox(0,0)[cc]{2}}

\put(25,60){\makebox(0,0)[cc]{3}}

\put(55,60){\makebox(0,0)[cc]{4}}

\put(43,40){\makebox(0,0)[cc]{5}}

\put(15,10){\makebox(0,0)[cc]{A}}

\put(65,10){\makebox(0,0)[cc]{B}}

\put(17,70){\makebox(0,0)[cc]{C}}

\put(63,70){\makebox(0,0)[cc]{D}}

\put(43,48){\makebox(0,0)[cc]{E}}

\put(43,32){\makebox(0,0)[cc]{F}}

\end{picture}

}

\def\figtwo{

\def\JPicScale{0.8}
\ifx\JPicScale\undefined\def\JPicScale{1}\fi
\unitlength \JPicScale mm
\begin{picture}(70,80)(0,0)
\linethickness{0.3mm}
\multiput(40,10)(0.12,0.24){83}{\line(0,1){0.24}}
\linethickness{0.3mm}
\multiput(50,30)(0.12,-0.24){83}{\line(0,-1){0.24}}
\linethickness{0.3mm}
\multiput(50,30)(0.12,0.18){83}{\line(0,1){0.18}}
\linethickness{0.3mm}
\multiput(60,45)(0.12,-0.3){83}{\line(0,-1){0.3}}
\linethickness{0.3mm}
\put(60,45){\line(0,1){15}}
\linethickness{0.3mm}
\multiput(50,80)(0.12,-0.24){83}{\line(0,-1){0.24}}
\linethickness{0.3mm}
\multiput(60,60)(0.12,0.24){83}{\line(0,1){0.24}}
\linethickness{0.3mm}
\multiput(40,50)(0.12,-0.24){83}{\line(0,-1){0.24}}
\linethickness{0.3mm}
\multiput(50,30)(0.12,0.36){83}{\line(0,1){0.36}}
\linethickness{0.3mm}
\multiput(40,50)(0.12,0.71){42}{\line(0,1){0.71}}
\linethickness{0.3mm}
\multiput(10,80)(0.12,-0.12){250}{\line(1,0){0.12}}
\end{picture}

}

\def\asnote#1{{\color{magenta}#1}}

\def\asnotea#1{{\color{orange}#1}}

\def\asnote#1{{\color{black}#1}}
\def\asnotea#1{{\color{black}#1}}

\newcommand{\mmu}{\mu}

\begin{document}

\baselineskip 24pt

\begin{center}

{\Large \bf Gravity Waves  from Soft Theorem in General Dimensions}


\end{center}

\vskip .6cm
\medskip

\vspace*{4.0ex}

\baselineskip=18pt

\centerline{\large \rm Alok Laddha$^{a}$ and Ashoke Sen$^{b}$}

\vspace*{4.0ex}

\centerline{\large \it ~$^a$Chennai Mathematical Institute, Siruseri, Chennai, India}

\centerline{\large \it ~$^b$Harish-Chandra Research Institute, HBNI}
\centerline{\large \it  Chhatnag Road, Jhusi,
Allahabad 211019, India}


\vspace*{1.0ex}
\centerline{\small E-mail:  aladdha@cmi.ac.in, sen@mri.ernet.in}

\vspace*{5.0ex}

\centerline{\bf Abstract} \bigskip

Classical limit of multiple 
soft graviton theorem can be used to compute the angular power spectrum of
long wavelength gravitational 
radiation in classical scattering provided the
total energy carried away by the radiation is small compared to the energies of
the scatterers. We could ensure this either 
by taking the limit in which the impact parameter is
large compared to the 
Schwarzschild radii of the scatterers, or by taking 
the probe limit where one
object (the probe) has mass much smaller than the other object (the scatterer).
We compute the results to subsubleading order in soft momentum and test them
using explicit examples involving classical scattering. 
Our analysis also generalizes to the case
where there are multiple objects involved in the scattering \asnotea{and  }the objects 
exchange mass, fragment or fuse into each other during the scattering.
A similar analysis can be carried out for soft
photons to subleading order, reproducing standard textbook results. 
We also discuss the modification of soft expansion  in four
dimensions beyond the leading order due to infrared divergences.



\vfill \eject

\baselineskip 18pt

\tableofcontents

\section{Introduction and summary}

Soft theorems have a 
long history\cite{Gell-Mann,low,saito,burnett,bell,duca,
weinberg1,weinberg2,jackiw1,jackiw2,ademollo,shapiro}. 
They express amplitudes with one or more low 
momentum photon or graviton in terms of amplitudes without such photons or
gravitons. Recent surge of interest in soft theorem began with the
realization that they are related to asymptotic 
symmetries\cite{1312.2229,1401.7026,1408.2228,
1411.5745,1502.02318,1505.05346,
1506.05789,1509.01406,1605.09094,1605.09677,1608.00685,1612.08294,1701.00496,
1612.05886,1703.05448,1709.03850,1711.04371,progress}. 
During this investigation
it has also become clear that soft theorems extend beyond leading order in the
soft momentum\cite{1103.2981,1404.4091,1404.5551,1404.7749,1405.1015,
1405.1410,1405.2346,
1405.3413,1405.3533,1406.6574,1406.6987,1406.7184,1407.5936,
1407.5982,1408.4179,1410.6406,1412.3699,
1503.04816,1504.01364,1507.08882,
1509.07840,1604.00650,1604.02834,
1604.03893,1607.02700,1611.02172,1611.07534,1611.03137,
1702.02350,
1406.4172,1406.5155,1411.6661,1502.05258,1504.05558,1504.05559,
1505.05854,1507.00938,1507.08829,1511.04921,
1512.00803,1601.03457,1604.03355,1610.03481,1702.03934,1705.06175}.

There now exists general diagrammatic
proof of soft graviton theorem that is valid for generic
finite energy external states in generic theories in generic number
of dimensions\cite{1703.00024,1706.00759,1707.06803 }. 
In particular these results also hold when
the external states represent composite objects. Up to subleading order in the soft
momenta the result is universal, i.e. the \asnotea{relation }between the amplitude with soft
gravitons and the amplitude without soft gravitons \asnotea{is }independent of the theory or
the type of particles we have in the external state, and is sensitive only to the momenta
and angular momenta carried by these particles. At the subsubleading order there still
exists a relation between the two amplitudes, but it is not universal, i.e. it depends on
the theory and also the nature of the external particles 
involved\cite{1604.03355,1611.07534,1706.00759}. \asnotea{The }analysis of 
\cite{1703.00024,1706.00759,1707.06803} can be
easily extended to the electromagnetic case following \cite{1702.03934}, 
but in this case the soft theorem
exists only to subleading order, and only the leading term is 
universal.

Since soft theorems hold for all external states, they also hold for classical objects 
carrying large mass or charge. Therefore by taking appropriate limit of the soft theorem
we should be able to generate universal formula for electromagnetic and / or
gravitational radiation during a classical scattering process that depends only
on the initial and final momenta and angular momenta 
of the objects being scattered, but not on the details
of the scattering. For leading soft theorem this
has already been studied and shown to agree with classical 
results\cite{weinberg2}. Our goal
will be to use the subleading and subsubleading soft graviton theorem on the
scattering of a pair of massive objects, with classical limit corresponding to their
masses being much larger than the Planck mass. We shall also see that a 
similar analysis for soft photons
reproduce the standard textbook results for radiation from a moving 
charge\cite{jackson}  to
subleading order.

Our results will be valid for graviton wave-lengths large compared to 
the impact parameter and
the internal sizes of the  scatterers. 
The validity of our analysis also requires that
the total radiated energy is small
compared to the energy of the scatterers themselves. We can achieve this by taking
the limit in which the impact parameter \asnote{or the typical size of the scattarers }is 
large compared to the Schwarzschild radii
of the scatterers. 
On the other hand if we take the impact parameter \asnote{and the sizes }to 
be of the order of the Schwarzschild
radii, then we need to
work in the approximation in which one of the objects (called the probe) has
mass small compared to the other (called the scatterer). 

For the probe-scatterer approximation, our results in $D$ space-time dimensions 
can be summarized as follows. 
We shall work in the frame in which the scatterer is at rest initially.
Up to subleading order in $\omega$, the total energy carried by gravity
waves of polarization $\ve$, 
with frequency lying between $\omega$ and $\omega(1+\delta)$ and
within a solid angle $\Delta\Omega$ around the unit vector $\hat n$ pointing
out of the scatterer is given by
\be \label{ei1}
{1\over 2^D \pi^{D-1}} |S_{\rm gr}^{(0)}(\ve,k) +S_{\rm gr}^{(1)}(\ve,k) |^2  \omega^{D-1}\, \Delta\Omega\, \delta
\ee
where we have used $\hbar=c=8\pi G_N=1$ units, 
$k=-(\omega, \omega\hat n)$ denotes the momentum carried by the individual gravitons, 
and\footnote{The factor 
given in \refb{es00} has previously appeared in
\cite{1308.6285} in a different context.} 
\be \label{es00}
S_{\rm gr}^{(0)}= \sum_{a=1}^2 \left[{\ve_{\mu\nu} p_{(a)}^\mu p_{(a)}^\nu\over p_{(a)}.k}
+ \ve_{00}{p_{(a)}.k\over (k^0)^2} + 2 \, \ve_{0\nu} {p_{(a)}^\nu\over k^0}\right]\, ,
\ee
\be \label{esubleading0}
S_{\rm gr}^{(1)} = i\, \sum_{a=1}^2 \left[\left\{
{\ve_{\mu\nu} p_{(a)}^\mu k_\rho \over p_{(a)}. k} 
+ {\ve_{\nu 0} k_{\rho}\over k^0}
\right\}\bj_{(a)}^{\rho\nu}
+\, {\bJ^{ji}\over M_0}\, \left\{ {\ve_{i0} \, k_j \, p_{(a)}. k \over (k^0)^2}
+{\ve_{i\nu} \, p_{(a)}^\nu \, k_j\over k^0} \right\}\right]\, .
\ee
$p_{(1)}$ and $p_{(2)}$ are the  momenta of the probe before and
after the scattering, with the sign convention that momentum is always measured
in the ingoing direction. The indices are raised and lowered by flat metric $\eta^{\mu\nu}$ and
$\eta_{\mu\nu}$  with mostly plus signature. The scatterer
is initially taken to be at rest. 
$\bJ$ is the initial angular momentum of the scatterer and $M_0$ is
the mass of the scatterer. 
The indices $i,j,\ell$ run over spatial coordinates and the indices $\mu,\nu,\rho$ run over
all space-time coordinates.
$\bj_{(1)}$ and $\bj_{(2)}$ are
the  angular momenta of the probe before and after the scattering,
measured with respect to the point where the scatterer is situated
initially,\footnote{More precisely,
this is the \asnote{center of momentum of the scatterer }around which $\bJ^{i0}$ components vanish.}
again with the 
convention that \asnote{they are }counted with positive sign for ingoing and negative sign for
outgoing particles. 
Under the assumption that the internal size of the probe is small compared to the impact parameter, 
$\bj_{(a)}$'s receive dominant
contribution only from the orbital angular momenta.
The formula
for $\bj_{(1)}$ and $\bj_{(2)}$ in terms of the initial and final particle trajectories 
can be found in \refb{etraj}, 
\refb{ebjab}. \refb{ei1} can be shown to be invariant under the gauge 
transformation that shifts $\ve_{\mu\nu}$ by $k_\mu\xi_\nu+\xi_\mu k_\nu$ for
some vector $\xi$. It is also invariant under  
a shift in the origin of the time coordinate under which
$\bj_{(a)}^{0i}=-\bj_{(a)}^{i0}$ gets shifted by $c^0 p_{(a)}^i$ for some constant $c^0$. 
The expansion is in powers of max($\omega a, \omega b)$, 
where $b$ is the impact parameter and 
$a$ is the maximum of the sizes of the probe and the scatterer.
$S^{(0)}_{\rm gr}$ represents the leading term in the expansion and
$S^{(1)}_{\rm gr}$ is the subleading term. Since for real $\ve$ $S^{(1)}_{\rm gr}$ is purely
imaginary, its effect on \refb{ei1} vanishes to subleading order. However for circular
polarization for which $\ve$ is complex, $S^{(1)}_{\rm gr}$ contributes to subleading order. 

Our analysis is also valid when the probe fuses to the scatterer. In this case we have
to set $p_{(2)}$ and $\bj_{(2)}$ to 0 in \refb{es00}, \refb{esubleading0}. It is also valid
when there is more that one probe and / or when 
the probe splits up into fragments producing more than one object in the
final state
and / or exchanges mass with the scatterer i.e.\ \asnote{we }do not need to
demand that the initial and final masses of the objects coincide.
In all these case we have to extend the sum over $a$ in \refb{es00}, 
\refb{esubleading0} to all initial and final state objects of finite energy.

If we do not use the probe-scatterer approximation, but consider a general scattering
with arbitrary number of incoming and outgoing objects carrying energies of  
the same order, then we can still use \refb{ei1}-\refb{esubleading0} for
large impact parameter \asnote{or large size, but we need to }allow
the sum over $a$ to run over all external particles. To the experts this may appear somewhat unfamiliar
since the usual soft theorem does not have the non-covariant terms in \refb{es00}, \refb{esubleading0}
involving $\ve_{00}$ and $\ve_{0\nu}$. It is easy to see however that when all external states are
included in the sum over $a$,
the contribution from these terms
vanish using conservation of total momentum \asnote{and total angular momentum}. 
The same argument shows that
the term involving the $\bJ^{ij}/M_0$ vanishes. This is just  as well since now there is no distinguished
scatterer whose angular momentum represents $\bJ^{ij}$.

\refb{es00}, 
\refb{esubleading0} are universal independent of the type of probe or scatterer we have. 
For example the scatterer could be  Kerr-Newmann black hole in the 
Einstein-Maxwell theory,
or more general rotating black hole in a theory of gravity coupled of other massless 
gauge fields and scalars,
or some  other compact massive (rotating) object. The probe could be any object with
mass large compared to the Planck mass but much smaller than that of the scatterer.
We expect the universality to break down at the next order in the expansion in powers
of $\omega a$ where the results will be
sensitive to the internal structure of the scatterer. However we can still make some
meaningful prediction if we take the impact parameter $b$ to be much larger
than \asnotea{the size }$a$. In this case the non-universal terms give corrections of order $\omega^2 a^2$,
whereas the universal terms have larger contribution of order $\omega^2 b^2$ and 
$\omega^2 ab$. These universal contributions are captured by adding to $S_{\rm gr}^{(0)}
+S_{\rm gr}^{(1)}$ in \refb{ei1} a third term
\be \label{esubsubleading0}
S_{\rm gr}^{(2)} = -{1\over 2} \sum_{a=1}^2 (p_{(a)}. k)^{-1}   
\ve_{\mu\rho} k_\nu k_\sigma \bj_{(a)}^{\mu\nu} 
\bj_{(a)}^{\rho\sigma} - \sum_{a=1}^2 {1\over M_0 k^0} 
 \ve_{i\rho} k_j k_\sigma \bJ^{ij} 
\bj_{(a)}^{\rho\sigma} \, .
\ee
\asnote{If the macroscopic objects have energies that are comparable to each other, then 
eq.\refb{esubsubleading0} still holds, but the sum over $a$ runs over all the external states.
In this case the contribution from the second term in \refb{esubsubleading0} vanishes by
conservation of angular momentum. Also if we express the  $\bj_{(a)}^{\mu\nu}$'s as
sum of spin and orbital angular momenta, then the terms quadratic in the orbital
angular momenta and the cross terms between the orbital and the spin angular momenta
are of order $\omega^2 b^2$ and $\omega^2 ab$ and are large compared to the
non-universal terms. However the term proportional to the square of the spin angular
mometum is of order $\omega^2 a^2$ and are of the same order as the non-universal 
terms. Therefore we need to ignore this term in \refb{esubsubleading0}. }

Finally, by combining the result on the angular power spectrum described above with the result
of explicit computation given in section \ref{sfusion}, we can write down the expression for the 
radiative part of the metric field $h_{\alpha\beta}\equiv (g_{\alpha\beta}-\eta_{\alpha\beta})/2$
for long wavelengths. Denoting by $\tilde h_{\alpha\beta}(\omega,\vec x)$ the time Fourier
transform of $h_{\alpha,\beta}(t,\vec x)$, we have, for large $|\vec x|$ and up to subsubleading 
order in $\omega$,
\ben \label{ehabexp}
&& \tilde h_{\alpha\beta}(\omega,\vec x) = \tilde e_{\alpha\beta}(\omega,\vec x)
- {1\over D-2} \, \eta_{\alpha\beta}\, \tilde e_\gamma^{~\gamma}(\omega, \vec x)\, ,
\nonumber \\
&& \ve^{\alpha\beta}\, \tilde e_{\alpha\beta}(\omega, \vec x) = \NN\, e^{i\omega R}\, 
\left[S^{(0)}_{\rm gr}(\ve, k) + S^{(1)}_{\rm gr}(\ve, k) + S^{(2)}_{\rm gr}(\ve, k)\right]\, ,
\nonumber \\
&& R \equiv |\vec x|, \qquad \NN \equiv \left({\omega\over 2\pi i R}\right)^{(D-2)/2} 
{1\over 2\omega}  \, ,
\een
up to gauge transformation.\footnote{The additional power of $\omega$ coming from 
$\NN$ for $D>4$
is responsible for the absence of conventional memory effect in 
$D>4$\cite{1612.03290,1702.00095,1712.00873}. In \cite{1712.01204} a new definition of memory 
was proposed based on the ``Coloumbic modes" of $\tilde{h}_{\alpha\beta}$ that fall off as 
$1/R^{D-3}$.  It would be interesting to extend our analysis to this order in $1/R$ and check 
that these Coloumbic modes indeed have a non-trivial zero frequency limit.}

A similar formula for electrodynamics exists up to subleading order in powers of
the soft momentum, and is obtained by replacing 
$S^{(i)}_{\rm gr}$ by $S^{(i)}_{\rm
em}$
given in \refb{esoftphoton} and 
\refb{eph1}. 
Both for electromagnetism and gravity the relation between classical radiation
and soft theorem at the leading order has been noted before, see {\it e.g.} \cite{jackson}
for electromagnetism and \cite{weinbergbook} for gravity. However this relation at the  
subleading and subsubleading order does not seem to have been explored in detail 
before.\footnote{The relation between subleading terms in the
soft graviton theorem and a new
kind of memory effect in four space-time dimensions 
was discussed in \cite{1502.06120}. However, as we shall discuss in section \ref{sir}, in four
dimensions the usual soft expansion seems to get modified beyond leading order due to logarithmic
singularities in the expansion.}

\asnote{It may be possible to give a general proof of soft theorem in classical scattering using the 
effective field theory approach\cite{0409156,0605238,0511061} (see \cite{1601.04914} 
for a review). However in this paper we shall restrict our analysis to verifying the soft theorem
explicitly in various classical scattering processes. }

In
four space-time dimensions soft theorems are expected to be modified due to infrared
divergence effects\cite{1406.6987}. We shall avoid these divergences by working in 
general dimensions. In section \ref{sir} we shall discuss the special case of four dimensions and
explore how the classical limit of soft theorem is affected by the classical counterpart of 
infrared divergences -- long range force acting on objects taking part in the scattering.

It was argued in \cite{1411.5745,1502.06120,1502.07644,
1707.08016,1712.01204} that soft graviton
theorem is
related to memory effect\cite{mem1,mem2,mem3,mem4}. 
This correspondence was established in two steps -- first
relating soft graviton theorem to the radiative field produced in the scattering and then relating the
radiative field to the memory effect. The latter connection has so far been understood only in 
even space-time dimensions. In contrast the connection between soft theorem and power spectrum of
gravitational radiation described in \refb{ei1} makes use of the relation between radiative field and the
power spectrum and holds in all space-time dimensions. Our analysis further shows that the first
step in this relation -- the relation
between soft theorem and the classical radiative field --
is based not on single soft graviton theorem but multiple soft graviton
theorem, and relies on the vanishing of the contact term in the multiple soft
theorem in the classical limit.

The rest of the paper is organized as follows. In section \ref{eelectro} we review the
classical limit of the leading soft theorem in electrodynamics. In
section \ref{sgrav} we apply this to soft theorem in gravity to subleading order.  
In section \ref{spol} we compute the angular power spectrum of soft gravitons
by summing over polarizations and show that in the probe-scatterer approximation 
the effect of the scatterer does not fully
decouple even in the infinite mass limit. In section \ref{splunge} we consider a 
slightly different situation where the probe  fuses   with the scatterer to form a single
object, and analyze \asnote{the }angular power spectrum of soft gravitons emitted during this
process with the help of soft theorem. This will include in particular an object falling
into the black hole.
In section 
\ref{shigh} we extend the analysis of classical limit of soft theorem
to subleading order in electrodynamics and 
subsubleading order in gravity. Even though at this order the soft theorem receives
non-universal corrections, we argue that in the limit of large impact parameter the
universal part dominates over the non-universal part and we can make definite
predictions. In section \ref{stest} we test the general formulae\ derived from soft
theorem by analyzing the emission of electromagnetic and gravitational
radiation during various classical scattering processes   
to \asnote{respectively subleading and subsubleading order }in soft momentum. 
In section \ref{sir} we explore the origin of the corrections to the soft theorem in four 
dimensions due to infrared effects.
We conclude in section \ref{sdis} with a discussion of our analysis.

\sectiono{Electromagnetic radiation from classical scattering} \label{eelectro}

In this section we shall review the computation of long wavelength 
classical electromagnetic radiation
for the scattering of charged particles using soft theorem.

Let us consider the effect of emission of $\asnote{M }$ soft photons, carrying polarization and
momenta $(\ve_1, k_1),\cdots , \asnote{(\ve_M, k_M) }$ during the scattering of $n$ finite energy
particles carrying charge $q_{(a)}$ and momenta $p_{(a)}$ for $1\le a\le n$. 
All momenta are regarded as ingoing so that an outgoing particle carries negative 
energy.
According to the soft theorem, for
small $\{k_r\}$, this amplitude is given by
\be\label{esso1}
\left\{ \prod_{r=1}^{\asnote{M }} S^{(0)}_{\rm em}(\ve_r, k_r) \right\} \, \bM
\ee
where
\be \label{esoftphoton}
S^{(0)}_{\rm em}(\ve, k) = \sum_{a=1}^n q_{(a)}\, {\ve . p_{(a)}\over k. p_{(a)}}
\, ,
\ee
and $\bM$ is the amplitude of the process without soft gravitons.
We shall now use this to derive a formula for the classical electromagnetic 
radiation from such a scattering process.
Let us consider the differential cross section for emitting $N$ soft photons, each carrying polarization
$\ve$ and lying within a solid angle 
$\Delta\Omega$, and with frequency lying between $\omega$ and $\omega(1+\delta)$. 
This will be
given by
\be \label{eexpem}
|\bM|^2\,  A^N / N!\, ,
\ee
where \asnote{$A$ includes the contribution from the differential cross-section as well as
the phase space factor: }
\be\label{numberspecem}
A\equiv {1\over (2\pi)^{D-1}} \, {1\over 2\omega} \, 
|S^{(0)}_{\rm em}(\ve,k)|^2  \omega^{D-2} \,  \omega  \, \delta \, 
\Delta\Omega
= {1\over 2^D\pi^{D-1}} |S^{(0)}_{\rm em}(\ve,k)|^2  \omega^{D-2} \,
\Delta\Omega\, \delta \, .
\ee
For $q_{(a)}\sim q$, the leading term
in $S^{(0)}_{\rm em}(\ve,k)$ goes as $q/\omega$. This gives $A\propto q^2$ for fixed $\omega$, 
$\Delta\Omega$ and
$\delta$. Therefore
$A$ is a large number
for $q>>1$, which corresponds to the classical limit. 
\refb{eexpem} is maximized at
\be 
{\p\over \p N} (N\ln A - N\ln N + N) = 0 \, ,
\ee
where, in anticipation of the fact that $N$ will turn out to be large at the maximum, we have used 
Stirling's formula for $N!$.
This gives 
\be \label{enaem}
N=A= {1\over 2^D\pi^{D-1}} |S^{(0)}_{\rm em}(\ve,k)|^2  \omega^{D-2} \,
\Delta\Omega\, \delta \, .
\ee 
Before we proceed, let us make a few comments:
\begin{enumerate}
\item 
\refb{eexpem} is sharply peaked around the value of $N$ given in 
\refb{enaem} with a width of order $\sqrt N$.
\item We could also allow emission of other soft
photons in
different momentum bins. 
This would multiply the $|\bM|^2$ factor in \refb{eexpem} by other factors but
will not affect the $N$ dependence of this expression. As a result the value of $N$
given in \refb{enaem} remains unchanged.
\end{enumerate}
Therefore \refb{enaem} gives the 
number of soft photons emitted in this momentum bin.  

We can also compute the total power carried by the soft photons
in this bin by multiplying \refb{enaem} by the energy of each soft photon.
This is given by
\be\label{epower}
A \, \omega  = {1\over 2^D\pi^{D-1}} |S^{(0)}_{\rm em}(\ve,k)|^2  \omega^{D-1} \,
\Delta\Omega\, \delta\, .
\ee

As a special case of the configuration discussed above, we can consider the probe-scatterer
approximation in which we have a light particle (the probe) scattering off a heavy center
(the scatterer). 
We begin with the scattering of two  particles in $D$ space-time dimensions 
-- one with mass $m$ and charge $q$
and the other with mass $M_0$ and charge $Q$. We label the momenta of the incoming
and the outgoing particles as
\be\label{ep1}
p_{(1)} = (p_{(1)}^0, \vec p_{(1)}), \quad  p_{(2)} = (p_{(2)}^0, \vec p_{(2)}), \quad
p_{(3)}= (M_0,\vec 0), \quad p_{(4)}= (-\sqrt{M_0^2+\vec p^2}, -\vec p), 
\ee
where
\be \label{eonshell}
\vec p\equiv \vec p_{(1)}+\vec p_{(2)}, \qquad p_{(1)}^0 =\sqrt{\vec p_{(1)}^2 + m^2}, 
\quad p_{(2)}^0 =-\sqrt{\vec p_{(2)}^2 + m^2}\, .
\ee
Note that we have implemented the
conservation of spatial momenta  in the parametrization
used in \refb{ep1}. 
We now take the limit $M_0\to \infty$ keeping $p_{(1)}$, $p_{(2)}$ finite. In this limit
energy conservation equation reduces to
\be \label{econsem}
p_{(1)}^0+p_{(2)}^0 =0\, .
\ee
We denote by $\bM$ the amplitude of this process. 
In this limit 
the $a=3$ and $a=4$
terms in the sum in \refb{esoftphoton}  cancel
and we are left with
\begin{equation} \label{ephlead}
S^{(0)}_{\rm em}(\ve, k) = q \, \sum_{a=1}^2 (-1)^{a-1}\, {\ve . p_{(a)}\over k. p_{(a)}} \, .
\end{equation}
Substituting eq.(\ref{ephlead}) in (\ref{enaem}) and taking $D=4$, we see that this agrees with eq.(15.3)
of \cite{jackson} after appropriate change in the units.\footnote{In our normalization of
the charge the fine structure constant is given by $e^2/(4\pi)$ where $e$ is the electric
charge of the electron. 
In the normalization of \cite{jackson} the fine structure constant is given by
$e^2$. This leads to an extra factor of $4\pi$ in the expression given in \cite{jackson}.
\label{fo1}}

Note that in the analysis given above we have taken $M_0\to\infty$ limit keeping $Q$ fixed.
If instead we take the $M_0\to\infty$ limit keeping $Q/M_0$ fixed, as will be the case {\it e.g.}
for scattering from a near extremal black hole, the soft factor $S^{(0)}_{\rm em}$ acquires
an additional term
\be \label{edd1}
\Delta S^{(0)}_{\rm em} = {Q\over M_0} \, {1\over k^0}\, \left(\ve. p - \ve^0 
{k. p\over k^0}\right)\, .
\ee
One can easily check the invariance of this term under the gauge transformation that
shifts $\ve^\mu$ by a vector proportional to $k^\mu$.

\sectiono{Gravitational radiation from \asnote{classical }scattering} \label{sgrav}

We shall now repeat the analysis for emission of soft gravitons.
Since there are several limits involved, we shall begin by describing the order of limits
that we would like to take.
\begin{enumerate}
\item First we take limit in which 
the momenta / masses
of all the incoming and outgoing objects become large,
keeping their ratios fixed. We shall refer to them as macroscopic objects.
This is the classical limit  in which the number of emitted gravitons in any fixed
frequency range becomes
large. 
\item The classical limit ensures that the number of soft gravitons emitted during the
scattering is large. But we also need to ensure that the
total energy carried by the 
radiation emitted during the scattering is small compared to the energies
of individual macroscopic objects taking part in the scattering.
As will be explained 
below \refb{e321}, this is needed to ensure that in computing the soft factors
we can only include the contribution due to the macroscopic objects and ignore the
contribution due to the radiation that is produced during the scattering.
One way to ensure this is to keep the \asnote{distance between the centers of the macroscopic
objects large compared to their Schwarzschild radii during the scattering. }The 
other way is to take the probe limit. 
For $2\to 2$ scattering this means that 
the mass of one of the incoming 
macroscopic objects (the scatterer) is taken to be large
compared to the mass / energy of the 
other (the probe) in the rest frame of the scatterer. 
We shall see that in either of these cases, the ratio of the
total energy radiated
to the energies of the macroscopic objects becomes small. 
\item Next we make the soft expansion -- an expansion in powers of the momenta of
soft gravitons.  This will be an expansion in the larger of the two quantities
$a \omega$ and $b \omega$ where $a$ denotes the typical size of the systems
{\it e.g.} the Schwarzschild radius for a black hole, $b$ is the
impact parameter and $\omega$ is the frequency of the soft graviton. Therefore we take
$\omega a$ and $\omega b$ to be small. As we will see, at the leading and subleading order in the expansion, we can
work with $b\sim a$. However for the subsubleading terms -- to be discussed in
section \ref{shigh}, the non-universal 
contributions are of order $(a\omega)^2$, and therefore to get a useful result without
knowing the details of the internal structures of the probe and the scatterer, we
shall take $b>>a$. In this case terms of order $b^2 \omega^2$ and $ab\omega^2$
are larger than the unknown terms of order $a^2\omega^2$ and we can make
meaningful predictions.
\end{enumerate}
In actual computation usage of soft theorem forces us to carry out the soft expansion first
and then take the
other limits. Due to this reverse order of limits we shall encounter some subtleties involving 
the difference between eqs.\refb{efinsoft-} and \refb{efinsoft} below. 
We shall try to resolve this to the best of
our ability.

In the soft limit the amplitude of a scattering process with soft gravitons is given by a soft factor
acting on the amplitude without the soft graviton. 
For emission of \asnote{$M$ }soft gravitons of polarizations $(\ve_1,k_1), \cdots , (\ve_M, k_M)$, 
the leading term, replacing \refb{esso1}, 
\refb{esoftphoton}, is given by\cite{weinberg1,weinberg2}
\be
\left\{\prod_{r=1}^M S_{\rm gr}^{(0)}(\ve_r,k_r)\right\}\, \bM\, ,
\ee
where ${\bf M}$ is the amplitude without the soft gravitons,
and
\be \label{esoft0}
S_{\rm gr}^{(0)}(\ve, k)\equiv \sum_{a=1}^n {\ve_{\mu\nu} p_{(a)}^\mu p_{(a)}^\nu\over p_{(a)}.k}\, ,
\ee
where $p_{(a)}^\mu$ are the momenta of the macroscopic  objects involved in the scattering.
As before all momenta are counted as positive if ingoing.

Let us now discuss subleading corrections. 
The  amplitude with $M$ soft gravitons takes the following form to subleading
order\cite{1707.06803}
\ben \label{efullgenintro}
&& \hskip -.5in \Bigg[
\Bigg\{\prod_{r=1}^M \, S_{{\rm gr}}^{(0)}(\ve_r, k_r) \Bigg\} \ 
+ \sum_{s=1}^M \Bigg\{\prod_{r=1\atop r\ne s}^M \, S_{{\rm gr}}^{(0)}(\ve_r, k_r)  \Bigg\}\ 
S_{{\rm gr}}^{(1)}(\ve_s, k_s)  \nonumber \\
&& \hskip -.5in + \sum_{r,u=1\atop r<u}^M 
\Bigg\{ \prod_{s=1\atop s\ne r,u}^M \, S_{{\rm gr}}^{(0)}(\ve_s, k_s)  \Bigg\} 
\ \Bigg\{\sum_{a=1}^n \ \{p_{(a)}. (k_r+k_u)\}^{-1}\ 
\ \MM(p_{(a)}; \ve_r, k_r, \ve_u, k_u) \Bigg\}  \
 \Bigg]\,  {\bf M} \, , 
\een
\ben \label{edefmiintro}
&& \MM(p; \ve_1, k_1, \ve_2, k_2) \nonumber \\
&=&
(p. k_1)^{-1}  (p. k_2)^{-1} 
 \ \Bigg\{-k_1. k_2 \ p. \ve_1. p \ 
 p. \ve_2. p + \ 2 \ p. k_2 \ p. \ve_1. p \ p. \ve_2. k_1 
 \nonumber \\ && 
  + 2 \ p. k_1 \ p. \ve_2. p \ p. \ve_1. k_2
  - 2  \ p. k_1 \ p. k_2 \ p. \ve_1. \ve_2. p\Bigg\}
 \nonumber \\ &&  
 +\ (k_1. k_2)^{-1}
 \Bigg\{-(k_2.\ve_{1}.\ve_{2}. p)(k_2. p) 
 -(k_1.\ve_{2}.\ve_{1}. p)(k_1. p)  \nonumber\\
&&+\ (k_2.\ve_{1}.\ve_{2}. p)(k_1. p)
+(k_1.\ve_{2}.\ve_{1}. p)(k_2. p)  - \ve_1^{\gamma\delta}\ve_{2\gamma\delta}(k_{1} . p)( k_{2}. p)  \nonumber\\
&& -\ 2(p.\ve_{1}. k_2)(p.\ve_{2}. k_1) 
+ (p.\ve_{2}. p)(k_2.\ve_{1}. k_2)  
+ (p.\ve_{1}. p)(k_1.\ve_{2}. k_1)\Bigg\}\, ,
\een
\be \label{esublead}
S_{\rm gr}^{(1)}(\ve, k) 
= \sum_{a=1}^n {\ve_{\mu\nu} p_{(a)}^\mu k_\rho \over p_{(a)}. k} \JJ_{(a)}^{\rho\nu},
\qquad \JJ^{\rho\nu} \equiv p^\nu {\p\over \p p_\rho}-  p^\rho {\p\over \p p_\nu}
+ \Sigma^{\rho\nu}\, ,
\ee
where $\Sigma^{\rho\nu}$ is the generator of rotation acting on the
intrinsic spin of the state. 
When we take the convolution of this with the external wave-function $e^{-ip. x}$,
the $\p/\p p_\rho$, after integration by parts, produces a factor of $i x^\rho$.
Therefore in the classical limit the first two terms in $\JJ^{\rho\nu}$ reduce to 
$i (x^\rho p^\nu - x^\nu p^\rho)$. Since $(x^\rho p^\nu - x^\nu p^\rho)$ is the $(\rho\nu)$
component of the orbital angular momentum, the classical limit of $\JJ^{\rho\nu}$ can be
identified to
$i \, \bJ^{\rho\nu}$,
where $\bJ^{\rho\nu}$
denote components of the total angular momentum. \asnote{In this limit we can }express 
\refb{esublead} as
\be \label{esublead0}
S_{\rm gr}^{(1)}(\ve, k) 
= i\sum_{a=1}^n {\ve_{\mu\nu} p_{(a)}^\mu k_\rho \over p_{(a)}. k} \bJ_{(a)}^{\rho\nu}\, .
\ee

We shall now try to estimate the relative importance of the two 
subleading terms in \refb{efullgenintro}
in the classical limit. In this limit the momenta $p_{(a)}$ and the angular momenta $\bJ_{(a)}$
become large. Let us take $p_{(a)}\sim \mmu$ and 
$\bJ_{(a)}\sim \mmu\, \lambda$ for some large $\mmu$,  $\lambda$. Here $\mmu$ is the typical
mass / energy of the macroscopic objects and $\lambda$ is the typical length scale of
the problem, measured
in Planck units. $\lambda^{-1}$ will provide an effective upper cut-off on the frequency of soft radiation,
and hence, by definition, soft gravitons will carry energies small compared to $\lambda^{-1}$. So 
we take $k\sim \lambda^{-1}\, \tau$ where $\tau$ is the small number which controls the soft expansion.
Then we have 
\ben \label{escaling1}
&& S^{(0)}_{\rm gr}(\ve, k) \sim \mmu\, \lambda\, \tau^{-1}, \quad 
S^{(1)}_{\rm gr}(\ve, k) \sim \mmu\, \lambda, \quad
\MM(p; \ve_1, k_1, \ve_2, k_2)\sim \mmu^2, \nonumber \\ &&
\{p_{(a)}. (k_r+k_u)\}^{-1}\sim \mmu^{-1}\, \lambda\, \tau^{-1}\, .
\een
It follows from this that for finite number of soft gravitons, the three terms in \refb{efullgenintro} 
are of order $\mu^M\lambda^M\tau^{-M}$, $\mu^M \lambda^M \tau^{-M+1}$ and 
$\mu^{M-1}\lambda^{M-1}\tau^{-M+1}$ respectively. Therefore in the classical limit
$\mu,\lambda\to\infty$, we can drop the third term compared to the other two. However since
we shall also take $M$ to be large in this limit, we need to be more careful in our estimate.  
We shall proceed for now by dropping the third term, but
later confirm that it is indeed small compared to the second term for the value of $M$ where the
probability distribution peaks.

With this understanding we  replace \refb{efullgenintro} 
by
\be\label{efinsoft-}
\Bigg\{\prod_{r=1}^M \,  S_{{\rm gr}}^{(0)}(\ve_r, k_r) \Bigg\} \ 
+ \sum_{s=1}^M \Bigg\{\prod_{r=1\atop r\ne s}^M \, S^{(0)}_{{\rm gr}}(\ve_r, k_r)  
\Bigg\}\ 
S^{(1)}_{{\rm gr}}(\ve_s, k_s) \, .
\ee
Since $S_{\rm gr}^{(0)}$ and $S_{\rm gr}^{(1)}$ are both multiplicative $c$-numbers, to this order
\refb{efinsoft-} can also be written as\footnote{Note that \refb{efinsoft} contains 
subsubleading contributions that may be written
schematically, by dropping the arguments, as $(S_{\rm gr}^{(0)})^{M-2} (S_{\rm gr}^{(1)})^2$. 
One could ask in what sense this is more 
important than the subsubleading contributions of the form $(S_{\rm gr}^{(0)})^{M-1} 
S_{\rm gr}^{(2)}$ where $S_{\rm gr}^{(2)}$ is the subsubleading contribution to the single
soft graviton theorem. To this end note that there are ${M\choose 2}$ terms of the form
$(S_{\rm gr}^{(0)})^{M-2} (S_{\rm gr}^{(1)})^2$ and $M$ terms of the form 
$(S_{\rm gr}^{(0)})^{M-1} 
S_{\rm gr}^{(2)}$. Therefore while for finite $M$ they are of the same order, for large $M$
the $(S_{\rm gr}^{(0)})^{M-2} (S_{\rm gr}^{(1)})^2$ contributions dominate.}
\be\label{efinsoft}
\prod_{r=1}^M \,  S_{\rm gr}(\ve_r,k_r), \qquad S_{\rm gr}(\ve_r, k_r)\equiv 
\Bigg\{S_{\rm gr}^{(0)}(\ve_r,k_r) + S_{\rm gr}^{(1)}(\ve_r, k_r)\Bigg\}\, .
\ee
Going from \refb{efinsoft-} to \refb{efinsoft} requires some discussion. For finite
$M$, these two expressions are clearly equal up to the subleading order. However in
our analysis the number of soft gravitons will be large and for this the two expressions
are not equal. Therefore we need to decide on which one is the correct expression
for large $M$. This can be done by going back to the analysis of  \cite{1707.06803} and
repeating the analysis dropping all terms that vanish in the classical limit, i.e.\ are
suppressed by powers of $\mmu$ in the denominator. This includes the `contact terms'
leading to $\MM$ and also terms
that involve 
commutators of $\p/\p p_\mu$ and $p_\nu$ since this reduces the number of powers
of $\mmu$. In this case it becomes clear that the soft factor is given by the product of 
$r$ independent factors, one for each soft graviton, and is given by \refb{efinsoft}.
Note that \refb{efinsoft} is also the form  that we shall get if we take the soft limit
consecutively using single soft theorem multiple times. Before taking the classical
limit the result of consecutive soft limit is not symmetric under the exchange of the soft
gravitons, i.e. it preserves the memory of which one was taken to be soft first, but
in the classical limit the symmetry is restored since $S_{\rm gr}^{(0)}$ and 
$S_{\rm gr}^{(1)}$ commute.
This also provides further support to the fact that \refb{efinsoft} is the correct form
of the subleading multiple soft theorem for large number of soft gravitons.

It now follows that to subleading order in the expansion parameter $\omega a$, the 
number of soft gravitons in given range of energy and solid angle, and the energy
carried by them, are given respectively by \refb{enaem} and \refb{epower} 
with $S_{\rm em}^{(0)}(\ve, k)$ replaced by
$S_{\rm gr}(\ve, k) \equiv 
S_{\rm gr}^{(0)}(\ve, k)+S_{\rm gr}^{(1)}(\ve, k)$:
\be \label{enaemgr}
N= {1\over 2^D\pi^{D-1}} |S_{\rm gr}(\ve,k)|^2  \omega^{D-2} \,
\Delta\Omega\, \delta \, ,
\ee 
\be \label{epowergr}
E=N\, \omega= {1\over 2^D\pi^{D-1}} |S_{\rm gr}(\ve,k)|^2  \omega^{D-1} \,
\Delta\Omega\, \delta \, .
\ee 
However for the validity of this equation we need to satisfy several conditions: 
\begin{itemize}
\item The number of
soft gravitons emitted in a given bin must be large.
\item  The third term in \refb{efullgenintro},
encoding the effect of the contact term $\MM$,
must be small compared to the second term.
\item
The total energy / angular momentum carried away by the radiation must be small compared
to the energy and the angular momentum carried by the macroscopic objects.
\end{itemize}
We shall now describe under what condition this holds. In particular we shall see that the last
two conditions lead to identical constraints.

\asnote{Using the scaling described in \refb{escaling1} and the paragraph above it,
the number of soft gravitons emitted in a given bin  given by \refb{enaemgr} can be
estimated as:
\be \label{esti1}
N
\sim \mmu^2 \, \lambda^{-(D-4)}\, \tau^{D-4}\, \Delta\Omega\, \delta\, .
\ee
 }For fixed $\tau$, $\delta$ and $\Omega$, whether this is large or not depends on the relative
magnitude of $\mmu$ and $\lambda$. Now in $D$ space-time dimensions, the mass $m$ of a black
hole scales with its Schwarzschild radius $a$ as
\be\label{est11}
a \sim m^{1/(D-3)}\, .
\ee
Since $\mmu$ and $\lambda$ denote the typical mass and length scales of the classical object, 
it is natural to introduce a parameter $\sigma$ via the relation
\be \label{ebetlam}
\lambda \sim \mmu^{1/(D-3)} \, \sigma\, ,
\ee
where $\sigma\gsim 1$ measures how large the typical length scale is compared
to the Schwarzschild radii $\sim\mu^{1/(D-3)}$
of the macroscopic objects involved in the scattering.
In that case 
\refb{esti1} \asnote{may be expressed as
\be 
N\sim \mmu^{(D-2)/(D-3)} \, \tau^{D-4} \, \sigma^{-(D-4)}\, \Delta\Omega\, \delta\, .
\ee
 }Even though $\tau$ is a small number and $\sigma\gsim 1$, we see that by taking $\mmu$
to be sufficiently large we can ensure that the number of gravitons emitted in a given bin
is large.

Let us now recall that for our expansion to be valid, we 
also need to ensure that the ratio of the third
term to the second term in \refb{efullgenintro} must be small.
Using \refb{escaling1} we find that for large $M$ two 
subleading terms in \refb{efullgenintro} are of order
\be 
M\, \mmu^M \, \lambda^M \, \tau^{-M+1},
\quad M^2\, \mmu^{M-1}\, \lambda^{M-1}\, \tau^{-M+1}\, ,
\ee
respectively. Here the factors $M$ and $M^2$ arise from the fact that there are $M$ terms in the
sum in
the second term and $M(M-1)/2$ terms in the sum in the third term.
Therefore the ratio of the third term to the second term is given by
\be\label{eratio}
M\, \mmu^{-1}\, \lambda^{-1}\, .
\ee
Since the third term in
\refb{efullgenintro} can receive contribution not just from the pair of gravitons in the same bin, but
also from pair of gravitons in different bins, it is appropriate to take for $M$ in \refb{eratio} the total
number of gravitons emitted during the process. This 
can be estimated by 
setting $\tau, \Delta\Omega, \delta \sim 1$ \asnote{on the right hand side of }\refb{esti1}, 
reflecting that the inverse of the 
typical length
scale $\lambda$ also sets the upper cut-off on $\omega$ for the radiation emitted during the
scattering. This gives
\be \label{eMvalue}
M\sim \mmu^2 \lambda^{-(D-4)}\, .
\ee
Therefore \refb{eratio} reduces to 
\be \label{eratio1}
\mmu \, \lambda^{-(D-3)} \sim \sigma^{-(D-3)}\, .
\ee
This shows that if we can keep $\sigma$ large then the ratio given in \refb{eratio} is small and
our approximation is valid. 
When the macroscopic objects have masses of the same order, then this can be
achieved by taking the typical length scale, {\it e.g.}
the impact parameter, to be large compared to the Schwarzschild radii of each
object.
The other possibility is to take the probe limit that will be discussed shortly. 

We shall now show that this condition is equivalent to keeping small 
the ratio of the total energy carried by the gravitons and the typical
mass of the macroscopic objects involved in the scattering.
The latter is of order $\mmu$ whereas the former
can be estimated by multiplying \refb{esti1} by $\omega\sim \lambda^{-1}\tau$ to get the energy of the
gravitons in a given bin, and then by setting \asnote{$\Delta\Omega\sim1$, $\delta\sim 1$, 
$\tau\sim 1$. }This gives the ratio to be of order 
\be 
\left[\{\mmu^{-1}\} \, \{\lambda^{-1}\, \tau\} \, \{\mmu^2 \, \lambda^{-(D-4)}\, \tau^{\asnote{D-4 }}\}
\right]_{\tau=1}
\sim \sigma^{-(D-3)}\, .
\ee
This is identical to \refb{eratio1} and can be kept small by taking $\sigma$ to be large.

Similar analysis can be done for the ratio of the total angular momentum carried away
by the radiation to the typical angular momenta of the macroscopic objects.
The former is of order $M$ given in \refb{eMvalue} whereas the latter is of order $\mmu\, \lambda$.
Therefore the ratio of the two is given by
\be \label{e321}
\mmu\, \lambda^{-(D-3)}\sim \sigma^{-(D-3)}\, .
\ee
Therefore this remains small for large $\sigma$.

It is not difficult to trace \asnote{why keeping the term proportional to $\MM$ 
in \refb{efullgenintro} small also
ensures }that the angular
momentum and energy carried away by the radiation \asnote{remains }small compared to the angular
momenta and energies of the macroscopic objects.
The contributions to $S^{(0)}_{\rm gr}$ and $S^{(1)}_{\rm gr}$ from a given external state are 
proportional respectively to the momentum and the angular momentum of the state. Now in
writing the expressions for $S^{(0)}_{\rm gr}$ and $S^{(1)}_{\rm gr}$ we have taken the sum over
$a$ to run only over the macroscopic objects and not included
the contributions from the \asnote{radiation, }but
the contact interaction $\MM$ includes these contributions. 
Therefore keeping the effect of $\MM$ small requires that the total energy and
angular momentum carried away by the \asnote{radiation }remain
small compared to the energies and angular momenta of the macroscopic objects
contributing to \refb{esoft0}, \refb{esublead0}.

Let us now describe the other way of keeping $\sigma$ large -- 
namely by taking the probe limit.
For simplicity we shall consider the probe limit 
of the $2\to 2$ scattering although the analysis can be easily generalized
to multi-body scattering. We shall use
a somewhat
more general kinematics 
for finite energy external particles than the one discussed in   
\refb{ep1}, \refb{eonshell}. Instead of assuming that the masses of the probe and
the scatterer remain unchanged during the scattering, we allow them to change.
This will include the situation {\it e.g.} where a part of the probe is stripped off and
absorbed by the scatterer during the scattering process. The new kinematics takes
the form
\be\label{ep1new}
p_{(1)} = (p_{(1)}^0, \vec p_{(1)}), \quad  p_{(2)} = (p_{(2)}^0, \vec p_{(2)}), \quad
p_{(3)}= (M_0,\vec 0), \quad p_{(4)}= (-M_0-p_{(1)}^0-p_{(2)}^0,-\vec p), 
\ee
where
\be \label{eonshellnew}
\vec p\equiv \vec p_{(1)}+\vec p_{(2)}, \qquad p_{(1)}^0 =\sqrt{\vec p_{(1)}^2 + m_{(1)}^2}, 
\quad p_{(2)}^0 =-\sqrt{\vec p_{(2)}^2 + m_{(2)}^2}\, .
\ee
This parametrization manifestly implements the conservation of all components
of momenta.
We now take the $M_0\to\infty$ limit keeping $p_{(a)}$ for $a=1,2$ fixed. 
In this limit \refb{esoft0} reduces to
\be \label{es0}
S_{\rm gr}^{(0)}= \sum_{a=1}^2 \left[{\ve_{\mu\nu} p_{(a)}^\mu p_{(a)}^\nu\over p_{(a)}.k}
+ \ve_{00}{p_{(a)}.k\over (k^0)^2} + 2 \, \ve_{0\nu} {p_{(a)}^\nu\over k^0}\right]\, .
\ee
For analyzing the subleading soft factor $S^{(1)}_{\rm gr}$,
we shall choose the origin of space coordinate to be at the center of mass of the
scatterer  before scattering -- defined such that the $0i$ components of its
angular momentum vanish. In that case we can 
label $\bJ_{(a)}^{\mu\nu}$ of the four external states as follows:
\ben
&& \bJ_{(1)}^{\mu\nu} = \bj_{(1)}^{\mu\nu}, \quad \bJ_{(2)}^{\mu\nu} = 
\bj_{(2)}^{\mu\nu},
\nonumber \\
&& \bJ_{(3)}^{ij}=\bJ^{ij}, \quad \bJ_{(3)}^{0i}=0, \quad \hbox{for} \quad 
1\le i,j\le D-1, \qquad \bJ_{(4)}^{\mu\nu}=
- \bJ_{(1)}^{\mu\nu} - \bJ_{(2)}^{\mu\nu} - \bJ_{(3)}^{\mu\nu}\, .\nonumber \\
\een
Here $\bJ^{ij}$ is a classical angular momentum carried by the 
scatterer before the scattering.  
We now take the limit $M_0\to \infty$ keeping the ratio $\bJ^{ij}/M_0$ fixed. In this limit the
subleading soft factor
reduces to
\be \label{esubleading}
S_{\rm gr}^{(1)} = i\, \sum_{a=1}^2 \left[\left\{
{\ve_{\mu\nu} p_{(a)}^\mu k_\rho \over p_{(a)}. k} 
+ {\ve_{\nu 0} k_{\rho}\over k^0}
\right\}\bj_{(a)}^{\rho\nu}
+\, {\bJ^{ji}\over M_0}\, \left\{ {\ve_{i0} \, k_j \, p_{(a)}. k \over (k^0)^2}
+{\ve_{i\nu} \, p_{(a)}^\nu \, k_j\over k^0} \right\}\right]\, .
\ee
It is easy to see that 
the right hand side of \refb{es0}, \refb{esubleading} are
invariant under the gauge transformation
\be \label{egauge}
\ve_{\mu\nu}\to\ve_{\mu\nu} +\xi_\mu k_\nu + \xi_\nu k_\mu\, ,
\ee
for any vector $\xi$.

We now need to check that in this limit the contribution of the third term in 
\refb{efullgenintro} is
suppressed compared to the second term.
\asnote{We have already seen that when }the impact parameter is large compared
to the Schwarzschild radius of the scatterer then this is true even without the probe limit, 
so let us focus on the case
where the impact parameter is of the same order as the Schwarzschild radius $a$ 
of the scatterer. We also assume that the probe moves with speed of order unity \asnote{and that
$m_{(1)}\sim m_{(2)}\sim m$. }In this case
we have 
\be\label{est1new}
M_0 \sim a^{D-3}, \quad \bJ \sim M_0\, a\sim a^{D-2}, \quad 
p_{(a)}^\mu\sim m, \quad \bj_{(a)}\sim m\, a \quad \hbox{for} \quad a=1,2\, ,
\ee
and the upper cut-off on $\omega$ is of order $a^{-1}$. Therefore we have
\ben\label{eleadsubcomp}
&& S^{(0)}\sim m/\omega, \quad S^{(1)}\sim (m/\omega) \, (\omega \, a),\nonumber \\ &&
\MM(p_{(a)}; \ve_1, k_1, \ve_2, k_2)\sim m^2, \quad 
\{p_{(a)}. (k_r+k_u)\}^{-1}\sim  (m\,\omega)^{-1}\, ,
\qquad \hbox{for $a=1,2$}, \nonumber \\ &&
\sum_{a=3}^4 \{p_{(a)}. (k_r+k_u)\}^{-1}\, 
\MM(p_{(a)}; \ve_1, k_1, \ve_2, k_2)\sim m \, \omega^{-1}\, .
\een
Comparing this with \refb{escaling1} we see that the scalings in \refb{eleadsubcomp} and
\refb{escaling1} are identical if we take $\mmu\sim m$, $\lambda\sim a$ and $\tau\sim \omega a$.
This also gives $\sigma\equiv \lambda\mmu^{-1/(D-3)} \sim a m^{-1/(D-3)}\sim (M_0/m)^{1/(D-3)}$.
Therefore for $M_0>>m$, $\sigma$ is large. This ensures that the contribution to the soft theorem from the
terms proportional to $\MM$ is supressed. Equivalently, the energy and angular momentum carried away by
radiation remains small compared to the energy and angular momentum carried by the macroscopic
objects involved in the scattering. In particular the total amount of energy radiated is of order
\be 
\left[ |S_{\rm gr}^{(0)}(\ve,k) +S_{\rm gr}^{(1)}(\ve,k) |^2  \omega^{D-1}\right]_{\omega=a^{-1}}
\sim m^2 /M_0\, .
\ee
This is small compared to $m$ for $m<<M_0$.

We see from \refb{eleadsubcomp} that the 
subleading contribution is suppressed compared to the leading term by order $\omega a$.
Therefore $\omega a$ is the soft expansion parameter.
This is not
surprising since in this case the Schwarzschild radius 
of the scatterer determines the typical length scale of the problem.

We end with some observations on the final result:
\begin{enumerate}
\item For plane polarized gravitational waves $S_{\rm gr}^{(0)}$ is real 
while $S_{\rm gr}^{(1)}$ is purely imaginary. Therefore the contribution of $S_{\rm gr}^{(1)}$ to the
scattering cross-section vanishes to subleading order. However if we use circularly
polarized gravitational wave then $\ve_{\mu\nu}$ is complex and as a result both 
$S_{\rm gr}^{(0)}$
and $S_{\rm gr}^{(1)}$ are complex. In this case we expect $S_{\rm gr}^{(1)}$ to contribute to the cross
section at subleading order. Related comments can be found in \cite{1502.06120}.
\item As mentioned below \refb{enaem}, even though we have presented our analysis 
as if it relies on the computation of the amplitude with $N$ soft gravitons in a given
momentum bin and no other particles in the final state other than the probe and
the scatterer, our result is valid more generally. Production of other soft gravitons
in other momentum bins will give an additional soft multiplicative factor in the amplitude
\refb{eexpem}, but will not affect $N$ dependence and the analysis leading to 
\refb{enaemgr} remains valid. Inclusion of `hard gravitons' in the final state, carrying momentum of
order $1/a$, may not be represented as a multiplicative factor, and may modify 
$\bM$ appearing in \refb{eexpem} in a non-trivial manner, but it still will not change the
$N$ dependence of \refb{eexpem}. Therefore \refb{enaemgr} 
still remains valid.
\item Under a shift in the origin of space-time by $c^\mu$, 
$\bJ^{\rho\nu}$ gets shifted by $c^\rho p_{(a)}^\nu - c^\nu p_{(a)}^\rho$. This
changes $S_{\rm gr}^{(1)}$ given in \refb{esublead0} 
to $i c. k\, S_{\rm gr}^{(0)}$. Therefore to subleading order
\be
S_{\rm gr}^{(0)}+S_{\rm gr}^{(1)} \to e^{i c. k} \, (S_{\rm gr}^{(0)}+S_{\rm gr}^{(1)})\, .
\ee
Since this is just an overall phase, this does not affect any physical quantities. 
\item
\refb{esoft0} and \refb{esublead0} (and \refb{es0} and \refb{esubleading})
do not depend on the details of the theory or the macroscopic objects involved in the 
scattering except their momenta and angular momenta. However 
the subsubleading contribution will depend on the details
of the macroscopic objects (and the theory)
via the non-universal terms that appear in the subsubleading soft graviton 
theorem. This will be discussed in section \ref{shigh}.
\item If the scattering configuration is such that the angular momenta $\bJ_{(a)}$ 
appearing in
\refb{esublead0} receive dominant contribution from the orbital angular momentum, 
we can express $\bJ_{(a)}$ in terms of the classical trajectories of the incoming and outgoing 
macroscopic objects as follows. Let the trajectory of the \asnote{center of momentum of the }$a$-th 
object be
given by 
\be \label{etraj}
x_{(a)}^\mu = c_{(a)}^\mu + m_{(a)}^{-1} \, p_{(a)}^\mu \, \tau_a\, ,
\ee
where $\tau_a$ are the proper time labelling the trajectories. Then $\bJ_{(a)}^{\mu\nu}$,
dominated by the orbital angular momenta, are given by
\be \label{ebjab}
\bJ_{(a)}^{\mu\nu} = c_{(a)}^\mu p_{(a)}^\nu - c_{(a)}^\nu p_{(a)}^\mu\, .
\ee
\item 
In arriving at the expression \refb{esubleading} for $S_{\rm gr}^{(1)}$ 
we have already fixed the origin
of the spatial coordinate system to be at the initial position of the
scatterer. But we still have the freedom
of shifting the origin of the time coordinate. This shifts 
$c_{(a)}^0$ in \refb{etraj}, \refb{ebjab} for the probe
by some constant $c^0$. It can be checked that under such a shift
$S_{\rm gr}\equiv S_{\rm gr}^{(0)}+S_{\rm gr}^{(1)}$ given in \refb{es0}, \refb{esubleading}
picks up an overall phase $\exp[-ic^0k^0]$
and as a result \refb{enaemgr}, \refb{epowergr}
remains unchanged.

Note that by appropriately choosing the origin of the time coordinate we can set
$c_{(1)}^0$ to 0. But we do not have the freedom of also setting $c_{(2)}^0$ to 0.
Therefore $S_{\rm gr}^{(1)}$ depends non-trivially on the time delay encoded in
$c_{(2)}^0$.
\item The soft factors given in \refb{es0} and \refb{esubleading} 
are gauge invariant without using any
momentum or angular momentum conservation laws. We could in fact replace the
soft factors \refb{esoft0} and \refb{esublead0} before taking 
the probe limit by those in \refb{es0} and
\refb{esubleading} respectively provided the sum over $a$ runs over all macroscopic objects,
since the contribution from the additional terms would
vanish after using momentum and angular momentum conservation. The resulting
soft factor is not manifestly Lorentz covariant, but is manifestly gauge invariant before
using momentum and angular momentum conservation laws. Once
we use momentum and angular momentum conservation laws, then both gauge and
Lorentz invariance are manifest. 
\item The soft factors \refb{es0} and \refb{esubleading} vanish in the $\vec p\to 0$ limit,
i.e.\ in the limit in which the probe is at rest relative to the scatterer. However if we
keep $\vec p/m$ small but fixed as we take the classical limit, our results continue to hold
and give the angular power spectrum of gravitational radiation for non-relativistic scattering.
\end{enumerate}

\sectiono{Sum over polarizations \asnote{for probe scattering }} \label{spol}

Given the angular power spectrum for given polarization, we can sum over polarizations
to get the total angular power spectrum. For this we can use the basis of plane polarized
waves. Since we have argued that in this basis the power spectrum is independent of the
subleading correction to subleading order, we can just focus on the contribution from the
leading \asnote{soft factor. In the probe approximation this is given in \refb{es0}. }

Now the last two terms in \refb{es0} can be thought of as the effect of radiation from
the scatterer since
they come from the $a=3,4$ terms in the sum in \refb{esoft0} -- although there is really
no gauge invariant way of isolating the radiation from the scatterer and the radiation
from the probe. By choosing the transverse gauge in which $\ve_{i0}$ and $\ve_{00}$
vanish, we can make the last two terms in \refb{es0} vanish. Therefore it would seem that
after we sum over polarizations the effect of radiation from the scatterer disappears. 
However we shall now see that this is not so.
For this we express \refb{es0} as
\be 
S^{(0)}(\ve,k) = \ve_{\mu\nu} \, \Sigma^{\mu\nu}(k)\, ,
\ee
where 
\be \label{edefSigma}
\Sigma^{\mu\nu} = \sum_{a=1}^2 \left[
{p_{(a)}^\mu p_{(a)}^\nu\over p_{(a)}.k}
+ \delta^\mu_0 \delta^\nu_0 {p_{(a)}.k\over (k^0)^2} +
\left\{\delta^\mu_0 p_{(a)}^\nu + \delta^\nu_0 p_{(a)}^\mu\right\} {1\over k^0}\right]\, .
\ee
Gauge invariance now translates to
\be\label{egiSigma}
k_\mu \, \Sigma^{\mu\nu}(k)=0, \quad k_\nu \, \Sigma^{\mu\nu}(k)=0\, .
\ee

Let $\ve^{(\alpha)}_{\mu\nu}$ for $1\le\alpha\le D(D-3)/2$ denote the
independent polarization
tensors in some gauge. Then the total angular power spectrum is proportional to
\be\label{ess1}
\sum_{\alpha=1}^{D(D-3)/2}  \left| S^{(0)}(\ve^{(\alpha)},k)\right|^2
= \sum_{\alpha=1}^{D(D-3)/2}  \ve^{(\alpha)*}_{\mu\nu} \ve^{(\alpha)}_{\rho\sigma}
\Sigma^{\mu\nu}(k) \Sigma^{\rho\sigma(k)}\, .
\ee
Now we have
\be\label{ecomplete}
\sum_{\alpha=1}^{D(D-3)/2}  \ve^{(\alpha)*}_{\mu\nu} 
\ve^{(\alpha)}_{\rho\sigma}
= {1\over 2} (\eta_{\mu\rho}\eta_{\nu\sigma} + \eta_{\mu\sigma}\eta_{\nu\rho})
- {1\over D-2} \eta_{\mu\nu} \eta_{\rho\sigma} 
+ \hbox{terms proportional to $k_\mu$, $k_\nu$,
$k_\rho$ or $k_\sigma$}\, .
\ee 
Substituting this into \refb{ess1}, and using \refb{egiSigma} we get
\be \label{efinSig}
\sum_{\alpha=1}^{D(D-3)/2}   \left| S^{(0)}(\ve^{(\alpha)},k)\right|^2
=\Sigma^{\mu\nu}(k) \, \Sigma_{\mu\nu}(k) - {1\over D-2} \Sigma^\mu_\mu(k) \, \Sigma^\nu_\nu(k)\, .
\ee
It is easy to see from this that the last two terms in \refb{edefSigma} does contribute 
to this.

If we had chosen transverse polarizations from the beginning then we could have dropped
the last two terms in \refb{edefSigma}. But the resulting expression will not satisfy 
\refb{egiSigma} and therefore the terms proportional to $k$ in \refb{ecomplete} will
now contribute. The final result of course will remain unchanged.

We can arrive at expression \refb{efinSig} by starting with \refb{esoft0} and performing
the sum over polarizations before taking the $M_0\to \infty$ limit. This will 
give\cite{weinbergbook}
\be \label{etop}
\sum_{\alpha=1}^{D(D-3)/2}  \left| S^{(0)}(\ve^{(\alpha)},k)\right|^2=
\sum_{a,b=1}^4 {1\over p_{(a)}. k \ p_{(b)}. k} \, 
\left[ (p_{(a)}. p_{(b)})^2 - {1\over D-2} p_{(a)}^2 p_{(b)}^2
\right]\, .
\ee
This is a special case of the general formula derived in \cite{0212168}.
It is easy to verify that $M_0\to\infty$ limit on \refb{etop}
reproduces \refb{efinSig}. The contribution
to \refb{efinSig} from the last two terms in \refb{edefSigma} arise from the terms
in \refb{etop} with either $a$ or $b$ or both taking values 3 or 4.

It follows from \refb{ecomplete} that given the radiative part of the soft graviton 
field $h_{\mu\nu}$, its projection to a given polarization tensor $\ve^{\mu\nu}$
can be obtained by contracting $\ve_{\mu\nu}$ to $h_{\rho\sigma}$ by the inverse of
the matrix appearing in \refb{ecomplete}:
\be \label{einverse}
\ve_{\mu\nu} \left\{{1\over 2} (\eta^{\mu\rho}\eta^{\nu\sigma} + \eta^{\mu\sigma}\eta^{\nu\rho})
- {1\over 2} \eta^{\mu\nu} \eta^{\rho\sigma} \right\} h_{\rho\sigma}\, . 
\ee

\sectiono{Plunge} \label{splunge}

We can also consider the case where a pair of macroscopic objects fuse into each other
and form a third object. As in section \ref{sgrav}, we can keep the energy carried by 
gravitational radiation small compared to the initial and the final energies of the macroscopic
objects by making the effective impact parameter large, {\it e.g.} by taking the sizes of the
objects to be large compared to their Schwarzschiid radii so that their centers always remain 
far apart.
In this case the soft factor is given by \refb{esoft0}, \refb{esublead0} 
with $n=3$, representing the two initial objects and one final object.

Alternatively we can  consider
the probe approximation where
the probe fuses to the scatterer and
produces a third object with slightly more mass than the scatterer. This process will have
three external finite energy particles, whose momenta and angular momenta 
can be taken as
\be 
p_{(1)} = (p^0, \vec p), \quad  
p_{(2)}= (M_0,\vec 0), \quad p_{(3)}= (-M_0-p^0, -\vec p)\, ,
\ee
\be 
\bJ_{(1)}^{\mu\nu} =\bj^{\mu\nu}, \quad
\bJ_{(2)}^{ij}=\bJ^{ij}, \quad \bJ_{(2)}^{0i}=0, \quad \hbox{for} \quad 
1\le i,j\le D-1, \qquad \bJ_{(3)}^{\mu\nu}=
- \bJ_{(1)}^{\mu\nu} - \bJ_{(2)}^{\mu\nu} \, .
\ee
We can now compute soft graviton radiation from this process using the soft theorem. The soft
factor up to subleading order is given by
\be 
S_{\rm gr}(k) = \sum_{a=1}^3 {\ve_{\mu\nu} p_{(a)}^\mu p_{(a)}^\nu\over p_{(a)}.k}
+ i\sum_{a=1}^3 {\ve_{\mu\nu} p_{(a)}^\mu k_\rho \over p_{(a)}. k} \bJ_{(a)}^{\rho\nu}\, .
\ee
Taking $M_0\to\infty$ limit with $\bJ/M_0$ fixed gives
\ben\label{eplunge}
S_{\rm gr}(k) &=& \left\{
{\ve_{\mu\nu} p^\mu p^\nu\over p.k}
+ \ve_{00}{p.k\over (k^0)^2} + 2 \, \ve_{0\nu} {p^\nu\over k^0}\right\}\nonumber \\
&+&
 i \left\{\left({\ve_{\mu\nu} p^\mu k_\rho \over p. k} 
+ {\ve_{0\nu} k_\rho\over k^0} \right)\bj^{\rho\nu} +
{J^{ji}\over M_0} \left( {\ve_{0i}k_j\over (k^0)^2} \, p\cdot k + \ve_{\sigma i}\, p^\sigma {k_j\over k^0}
\right)
\right\}
\, .
\een
This formula can also be regarded as a special case of \refb{es00}, 
\refb{esubleading0} in which we take the
limit $p_{(2)}\to 0$, $\bj_{(2)}\to 0$.
The angular power spectrum of soft radiation can now be found from \refb{epowergr}.

Again in the probe limit we can estimate the total radiated energy to be of order $m^2/M_0$. Since this
is small compared to $m$ in the $M_0>>m$ limit, we expect our approximation to be valid.

The reverse of this process is decay, where a macroscopic object splits apart into two objects
due to some internal dynamics. The corresponding formul\ae\ are the same as the ones given 
in this section, but the signs of the $p_{(a)}$'s will be reversed.

\sectiono{Universal parts of higher order non-universal results} \label{shigh}

In this section we shall explore the possibility of extracting universal parts of
higher order soft theorems, both for electromagnetism and for gravity.

\subsection{Subleading soft photon theorem}

Unlike gravitational theories, there is no universal subleading soft photon theorem -- the
$S_{\rm em}^{(1)}$ factor depends on the theory, and can also depend on which external 
particle we consider. 
This can be traced to the non-minimal coupling of the soft photon to the external particles via
the field strength $F_{\mu\nu}$ -- since this has one derivative, it produces a subleading correction.
Nevertheless part of the expression for $S_{\rm em}^{(1)}$ is independent 
of the details of the theory. This takes the form\cite{1404.5551}
\be \label{e62}
S_{\rm em}^{(1)}(\ve, k) = \sum_{a=1}^n q_{(a)} {\ve_\nu k_\rho \JJ^{\rho\nu}\over p_{(a)}. k}
=  i \sum_{a=1}^n q_{(a)} {\ve_\nu k_\rho\over p_{(a)}. k} \,  \bJ_{(a)}^{\rho\nu}\, ,
\ee
in the same convention as was used in \refb{esublead}, \refb{esublead0}. A general derivation of this
formula for any external states in any theory can be given following the procedure described in \cite{1706.00759}.
For large impact parameter $b$ 
the angular momenta of the macroscopic objects are dominated by the orbital part
and have the form given in \refb{ebjab}.  If $m$ denotes the typical mass of the macroscopic
objects involved in the scattering, then $\bJ_{(a)}^{\rho\nu}\sim mb$. On the other hand, the
non-universal corrections to the soft photon theorem are expected to be proportional
to $ma$ where $a$ is the typical size of the objects. Therefore for $b>>a$ the universal part of
$S^{(1)}_{\rm em}$ dominates and we can trust \refb{e62}.

In the probe limit we use
the kinematics given in \refb{ep1} and take the limit $M_0\to\infty$. This gives
\be \label{eph1}
S_{\rm em}^{(1)}(\ve, k) = i \sum_{a=1}^2 q_{(a)} {\ve_\nu k_\rho \over p_{(a)}. k}
\, \bj_{(a)}^{\rho\nu}\, .
\ee
Again, for large impact parameter $b$ 
the angular momenta of the macroscopic objects are dominated by the orbital part
and have the form given in \refb{ebjab}. 
Substituting \refb{ebjab} into \refb{eph1}, using 
$q_{(a)}=(-1)^{a-1} q$ for $a=1,2$, and adding
the result to \refb{ephlead},  we get the total soft factor for photons to subleading order
\be \label{ephcomp}
S_{\rm em}(\ve, k) = q \, \sum_{a=1}^2 (-1)^{a-1}\, {\ve . p_{(a)}\over k. p_{(a)}}
+ i \, q\, \sum_{a=1}^2 (-1)^a \, {1 \over p_{(a)}. k}
\left\{c_{(a)}. k \  p_{(a)}. \ve - c_{(a)}. \ve\  p_{(a)}. k\right\}\, .
\ee
Since $c_{(a)}\sim b$, we see that the subleading contribution is of order $q\, b$.
We expect this to dominate the non-universal corrections to 
$S_{\rm em}^{(1)}$ when the impact parameter $b$ is large compared to the sizes $a$
of the probe and
the scatterer. In this case 
the non-universal terms are expected to be of order $q\, a$, which is small compared to
the subleading contribution of order $q\, b$ given above.

The results described in \refb{eph1} were given for $M_0\to\infty$ limit at fixed $Q$. If we
take the $M_0\to\infty$ limit at fixed $Q/M_0$ and $\bJ^{ij}/M_0$, we have additional
subleading contributions:
\be
\Delta S^{(1)}_{\rm em} = i\, {Q\over M_0} {\bJ^{ij}\over M_0} \ve_j k_i {\vec p.
\vec k\over (k^0)^2} +i\, {Q\over M_0} {1\over k^0} \ve_\nu k_\rho \left(\bj_{(1)}^{\rho\nu}
+\bj_{(2)}^{\rho\nu}\right)\, .
\ee

\subsection{Subsubleading soft graviton theorem}

We can try to carry out a similar analysis for subsubleading soft theorem for gravity.
In our convention the universal part of the soft factor to this order may be 
written as\cite{1103.2981,1312.2229,1401.7026,1404.4091,1706.00759}
\be \label{esoftgr2}
S_{\rm gr}^{(2)} = -{1\over 2} 
\sum_{a=1}^n (p_{(a)}. k)^{-1}   \ve_{\mu\rho} k_\nu k_\sigma \bJ_{(a)}^{\mu\nu} 
\bJ_{(a)}^{\rho\sigma}\,.
\ee
For large impact parameter $b$, the orbital angular momenta are of order 
$mb$ where $m$ is the typical energy of the macroscopic objects, and the spin angular
momenta are of order $ma$ where $a$ is the typical size of these objects. Therefore
the three kinds of terms arising out of the expansion of \refb{esoftgr2}, the square of the 
orbital angular momentum, the square of the spin angular momentum and the cross term,
are respectively of order $m\,\omega\, b^2$, $m\,\omega\,  a^2$ and $m\, \omega\, a\, b$.
The non-universal corrections to \refb{esoftgr2} are expected to be of order $m\, \omega\, a^2$ -- 
of the same order as the square of the spin angular momentum term. Therefore in the
$b>>a$ limit, it is  appropriate to keep the terms involving square of the orbital angular momenta and
the cross term between the orbital and the spin angular momenta, but not the term proportional to
the square of the spin angular momenta.

If we consider $2\to 2$ scattering and take the probe limit
$M_0\to\infty$ with the kinematics described in \refb{ep1new}, \refb{eonshellnew},
then \refb{esoftgr2} becomes
\be\label{es22}
S_{\rm gr}^{(2)} = -{1\over 2} \sum_{a=1}^2 (p_{(a)}. k)^{-1}   
\ve_{\mu\rho} k_\nu k_\sigma \bj_{(a)}^{\mu\nu} 
\bj_{(a)}^{\rho\sigma} - \sum_{a=1}^2 {1\over M_0 k^0} 
 \ve_{i\rho} k_j k_\sigma \bJ^{ij} 
\bj_{(a)}^{\rho\sigma} - {1\over 2}  \sum_{a=1}^2 
{p_{(a)}.k\over M_0^2 (k^0)^2} 
\ve_{i\ell} k_j k_m \bJ^{ij} 
\bJ^{\ell m}\, .
\ee
Now we have $\bJ/M_0\sim a$, where $a$ is the size of the scatterer,
so the last term is suppressed compared to the leading
contribution $S^{(0)}_{\rm gr}$ -- which is of order $m/\omega$ -- 
by a factor of $(\omega a)^2$. We expect the non-universal terms, which
are sensitive to the internal structure of the scatterer, to give contribution of similar order.
Therefore we can trust \refb{es22} only in the regime in which $|\bj_{(a)}^{\rho\nu}|>> |\bJ^{ij}|$ so that
the first two terms on the right hand side of \refb{es22} are large compared to the
last term. This may be achieved {\it e.g.}
by taking the impact parameter $b$ to be large compared
to $a$, since in that case $\bj_{(a)}/m \sim b$ and the contribution from the first two
terms are of order $(m/\omega) (\omega b)^2$ and $(m/\omega) (\omega^2 a b)$
respectively. In this case the soft expansion will be an expansion in powers of 
$\omega b$, and hold when $\omega b$ is small. 
We can now drop the last term in
$S_{\rm gr}^{(2)}$ and express this as
\be
S_{\rm gr}^{(2)} = -{1\over 2} \sum_{a=1}^2 (p_{(a)}. k)^{-1}   
\ve_{\mu\rho} k_\nu k_\sigma \bj_{(a)}^{\mu\nu} 
\bj_{(a)}^{\rho\sigma} - \sum_{a=1}^2 {1\over M_0 k^0} 
 \ve_{i\rho} k_j k_\sigma \bJ^{ij} 
\bj_{(a)}^{\rho\sigma} \, .
\ee
We have to replace $S_{\rm gr}$ 
in \refb{epowergr} by $S_{\rm gr}^{(0)}+S_{\rm gr}^{(1)}+S_{\rm gr}^{(2)}$ to compute the
angular power spectrum of soft gravitons to subsubleading order.

\subsection{Hidden assumptions}

The analysis for subleading soft photon and subsubleading soft graviton theorem 
is based on the assumptions that in the classical limit the soft factor for multiple soft particles
is given by the product of single soft factors. This is the analog of \refb{efinsoft}. These results have
not been established. However given the result for subleading soft gravitons, the fact that
the consecutive soft limit always generates the product formula, and that in the classical limit the
product formula is symmetric under the exchange of any pair of soft particles, we expect 
this result to be true. Extra contact terms would contain the contribution of the soft
radiation to the soft factor, and we expect this contribution to be subdominant as long as the
energy and angular momenta of the radiation remain small compared to those carried by
the macroscopic
objects.
Nevertheless it will be useful to check this by explicit computation of the
`contact terms' similar
to the one given in \cite{1707.06803} and verifying that in the classical limit the 
contact term contribution is small
compared to the product of single soft factors.

\sectiono{Tests of soft theorem from classical radiation analysis} \label{stest}

In this section we shall verify / test the classical limits of soft theorems described above
by explicitly analyzing 
classical electromagnetic / gravitational radiation during various scattering processes.

\subsection{Soft photon theorem} \label{sphoton}

Let us consider a particle of charge $q$ moving along a trajectory $r(\sigma)
=(r^0(\sigma), \vec r(\sigma))$ labelled by the proper time $\sigma$ along the
trajectory. 
Maxwell's equation takes the form
\be \label{emaxwell}
\p^\mu F_{\mu\nu}(x) = -j_\nu(x) = -q\, \int d\sigma \,  \delta^{(D)}(x - r(\sigma))
\, V_\nu(\sigma)\, ,
\ee
where
\be 
V^\alpha(\sigma) = {d r^\alpha\over d\sigma}\, .
\ee
This equation can be solved by setting
\be \label{edefaalpha}
A_\alpha(x) = -\int d\sigma \, G_r(x, r(\sigma)) \, q V_\alpha(\sigma)\, ,
\ee
where
$G_r(x,x')$ is the retarded Green's function, given by
\be\label{eretard}
G_r(x,x') = \int {d^D \ell\over (2\pi)^D} \, e^{i \ell. (x-x')}\, {1\over 
(\ell^0+i\epsilon)^2 - \vec \ell^2}\, ,
\ee
satisfying 
\be 
\eta^{\mu\nu}\p_\mu\p_\nu G_r(x,x')=\delta^{(D)}(x-x'), \quad 
G_r(x,x')=0 \quad \hbox{for $x^0<x^{\prime0}$}\, .
\ee
This allows us to express the time Fourier transform of $A_\alpha(x)$ as
\be
\wt A_\alpha(\omega, \vec x) \equiv
\int dx^0 \, e^{i\omega x^0} A_\alpha(x^0, \vec x) 
=-\int d\sigma \int {d^{D-1} \vec\ell\over (2\pi)^{D-1}}e^{i\omega r^0(\sigma)
+ i\vec \ell. \{\vec x -\vec r(\sigma)\}} {1\over (\omega+i\epsilon)^2 - \vec \ell^2}
\, q V_\alpha(\sigma)\, .
\ee
We shall evaluate this for positive $\omega$ and later recover the result for negative $\omega$ using
the reality of $A_\alpha(x)$.
For fixed $\vec x$ and $\sigma$ we can decompose $\vec \ell$ into its component
$\ell_\parallel$ parallel to $\vec x - \vec r(\sigma)$ and the component $\vec \ell_\perp$
transverse to $\vec x - \vec r(\sigma)$. This gives
\be
\wt A_\alpha(\omega, \vec x) 
=-\int d\sigma \int {d^{D-2} \vec\ell_\perp\over (2\pi)^{D-2}} {d\ell_\parallel\over
2\pi} e^{i\omega r^0(\sigma)
+ i\ell_\parallel |\vec x -\vec r(\sigma)|} {1\over (\omega+i\epsilon)^2 - \ell_\parallel^2-
\vec \ell_\perp^2}\, q V_\alpha(\sigma)\, .
\ee
We can perform the integration over $\ell_\parallel$ by closing its contour at infinity in the
upper half plane, picking up residues at $\ell_\parallel=\sqrt{(\omega+i\epsilon)^2
-\vec\ell_\perp^2}$. This gives
\be 
\wt A_\alpha(\omega, \vec x) 
=i\int d\sigma \int {d^{D-2} \vec\ell_\perp\over (2\pi)^{D-2}}  
e^{i\omega r^0(\sigma)
+ i\sqrt{(\omega+i\epsilon)^2- \vec\ell_\perp^2} |\vec x -\vec r(\sigma)|} {1\over 
2\sqrt{(\omega+i\epsilon)^2 -
\vec \ell_\perp^2}}\, q V_\alpha(\sigma)\,.
\ee
Our goal will be to evaluate this integral in the $|\vec x|\to\infty$ limit since we shall
be interested in computing the radiative part of $\wt A_\alpha$. In this limit the 
integration over $\vec \ell_\perp$ is dominated by the stationary point of the
exponent which is at $\vec \ell_\perp=0$. 
Since the dominant contribution to the integral is expected to come from the region
close to $\vec \ell_\perp=0$, we can expand the exponent in power series in $\vec \ell_\perp$:
\be 
\wt A_\alpha(\omega, \vec x) 
\simeq i\int d\sigma \int {d^{D-2} \vec\ell_\perp\over (2\pi)^{D-2}}  
e^{i\omega r^0(\sigma)
+ i\left\{ \omega+i\epsilon - {\vec\ell_\perp^2\over 2(\omega+i\epsilon)}\right\} 
|\vec x -\vec r(\sigma)| +\cdots} \left\{{1\over 
2(\omega+i\epsilon)} +\cdots \right\} \, q V_\alpha(\sigma)\, ,
\ee
where the $\cdots$ denote terms containing higher powers of $\vec\ell_\perp^2$, which, after
$\vec \ell_\perp$ integration, will turn into higher powers of $1/|\vec x|$. 
The $i\epsilon$ term provides the required damping term $\exp[-\epsilon |\vec x - \vec r(\sigma)| \vec\ell_\perp^2 / (2\omega^2)]$ for the $\vec\ell_\perp$ integral. We can carry out the $\vec\ell_\perp$
integration using the standard formula for gaussian integration. This gives
\ben\label{ereturn}
\wt A_\alpha(\omega, \vec x) 
&\simeq& i\left({\omega\over 2\pi i|\vec x|}\right)^{(D-2)/2} 
{1\over 2\omega} \int d\sigma 
e^{i\omega r^0(\sigma)
+ i \omega |\vec x -\vec r(\sigma)|} \, q V_\alpha(\sigma)\nonumber \\
&\simeq& i\, \NN\, e^{i\omega R}  \int d\sigma 
e^{i\omega \{r^0(\sigma) - \hat n.\vec r(\sigma)\}} \, q V_\alpha(\sigma)\, ,
\een
where $\hat n$ denotes unit vector along $\vec x$, and
\be \label{eNvalue}
R\equiv |\vec x|, \qquad \NN =\left({\omega\over 2\pi i R}\right)^{(D-2)/2} 
{1\over 2\omega}  \, .
\ee
Even though \refb{ereturn} was derived for positive $\omega$, using the reality condition
$(\wt A_\alpha(\omega, \vec x))^*=\wt A_\alpha(-\omega, \vec x)$, we now see that 
\refb{ereturn} holds for negative $\omega$ as well.

We now 
note that for a given particle trajectory, as 
$\sigma\to\pm\infty$, $qV_\alpha(\sigma)$ approaches a constant, and
$\{r^0(\sigma) - \hat n.\vec r(\sigma)\}$ approaches $(V^0 -\hat n.\vec V)\sigma$ plus
a constant. Therefore the integrand is oscillatory. 
This is defined using the standard trick that sets $\int_a^\infty ds \, e^{i\omega s}$
to $- e^{i\omega a}/(i\omega)$, ignoring the boundary terms at infinity. This is 
equivalent to explicitly adding terms that cancel the boundary terms.
Using this insight we make \refb{ereturn} well defined by adding to it appropriate
boundary terms:
\ben\label{eafin}
\wt A_\alpha(\omega, \vec x) 
&\simeq& i\, \NN\, e^{i\omega R}  \int_{\sigma_-}^{\sigma_+} d\sigma 
e^{i\omega \{r^0(\sigma) - \hat n.\vec r(\sigma)\}} \, q V_\alpha(\sigma)
\nonumber \\
&& -\NN \, e^{i\omega R} \, {1\over \omega} \, 
\left[e^{i\omega \{r^0(\sigma) - \hat n.\vec r(\sigma)\}}
 \, \left\{{q V_\alpha(\sigma)\over 
 V^0(\sigma) - \hat n.\vec V(\sigma)}\right\}\right]_{\sigma_-}^{\sigma_+}\, ,
\een
where $\sigma_\pm$ are the limits of integration with the
understanding that we have to take the limit $\sigma_\pm\to\pm\infty$  at the end. 

\asnote{Let }us define
\be \label{e717}
V_\pm = V(\sigma_\pm), \quad r_\pm = r(\sigma_\pm), \quad k=-\omega(1,\hat n), 
\ee
where the minus sign in the expression for $k$ reflects that $k$ represents the
momentum of an outgoing photon. We now expand \asnote{all the terms on 
the right hand side of
\refb{eafin}, except the $e^{i\omega R}$ factor, }in a power series in $\omega$. 
This gives, for any polarization vector
$\ve$,
\be\label{eafin3}
\ve^\alpha \wt A_\alpha =\NN\, e^{i\omega R} \, 
\left[  i \, q\, (\ve.r_+ - \ve. r_-) - {q \, \ve.V_{+} \over k.V_+} (1+ik.r_+)
+ {q \, \ve.V_{-} \over k.V_-} (1+ik.r_-)+\OO(\omega)
\right]\, .
\ee
Taking the trajectories in the far past and far future to be of the form
\be \label{esttraj}
r(\sigma) = c_\pm + V_\pm \, \sigma\, 
\ee
we get 
\be \label{e716}
r_\pm = c_\pm + V_\pm \, \sigma_\pm\, .
\ee
Substituting this into \refb{eafin3} we get
\ben \label{eafin4}
\ve.\wt A &=& \NN\, e^{i\omega R} \, 
\Bigg[ -q \left\{ {\ve.V_{+} \over k.V_+} - {\ve.V_{-} \over k.V_-}
\right\} - {i \,q\over k.V_+}\, \left\{ \ve . V_+ \, k. c_+- \ve . c_+ \, k. V_+\right\}
\nonumber \\ &&+  {i \,q\over k.V_-} \, \left\{ \ve . V_- \, k. c_-- \ve . c_- \, k. V_-\right\}
+\OO(\omega)\Bigg]\, .
\een

Now for this problem, we have
\be 
p_{(1)} = m\, V_-, \quad p_{(2)} = -m \, V_+, \quad 
c_{(1)}=c_-, \quad c_{(2)}= c_+\, ,
\ee
\asnote{where $m$ is the mass of the particle and 
$c_{(1)}$ and $c_{(2)}$ are the constants introduced in \refb{etraj}.
The minus sign in the expression for $p_{(2)}$ accounts for the }fact that it is the momentum
of an outgoing particle. 
Substituting these into \refb{ephcomp}  we get
\be \label{esemcov}
S_{\rm em} = \Bigg[ -q \left\{ {\ve.V_{+} \over k.V_+} - {\ve.V_{-} \over k.V_-}
\right\} - {i \,q\over k.V_+}\, \left\{ \ve . V_+ k. c_+- \ve . c_+ k. V_+\right\}
+  {i \,q\over k.V_-} \, \left\{ \ve . V_- k. c_-- \ve . c_- k. V_-\right\}
\Bigg]\, .
\ee
Comparing this with \refb{eafin4}
we get, up to subleading order in $\omega$,
\be\label{e732}
\ve.\wt A = \NN\, e^{i\omega R} \, S_{\rm em}(\ve,k) \, .
\ee

We shall now use this to compute the energy carried away by
photons of frequency between
$\omega$ and $\omega(1+\delta)$ within a solid angle $\Delta\Omega$ around
$\hat n$. To do this we note that for normalized $\ve$, $\ve.A$ can be regarded as 
a real scalar field $\phi$. 
Therefore $\eps.\wt A$ gives its time Fourier 
transform $\tilde\phi(\omega,\vec x)$. 
Furthermore far away from the 
source the radiation looks like
a plane wave propagating along the direction $\hat n$ with no dependence on
the transverse coordinates. If we denote the radial coordinate along $\hat n$ around the
relevant region
by $z\equiv \hat n.\vec x=|\vec x|=R$, then the relevant part of $\wt\phi(\omega,\vec x)$ 
has the
form
\be \label{ephiz}
\wt\phi(\omega,\vec x) = e^{i \omega z } \,\NN\, 
\, S_{\rm em}(\ve,k)\, .
\ee

Now the energy momentum tensor of $\phi$ is given by
\be
T_{\mu\nu} = \p_\mu\phi \p_\nu\phi - {1\over 2} \eta_{\mu\nu}  \p^\rho\phi \p_\rho\phi
\, .
\ee
Therefore the total energy flow per unit `area' along $z$ direction is
\be \label{eflux}
\int dz \, T^{zt} = \int dz \int{d\omega\over 2\pi} \int 
{d\omega'\over 2\pi} 
(-\omega\omega') e^{-i (\omega+\omega')t} \tilde\phi(\omega, z) \tilde\phi(\omega',z)\, .
\ee
Using \refb{ephiz} we see that the $z$ integral produces a factor of $2\pi \delta(\omega
+\omega')$.
Using the fact that $\tilde\phi(-\omega,z)=\tilde\phi(\omega,z)^*$, we can express
\refb{eflux} as
\be
\int_{-\infty}^\infty {d\omega\over 2\pi} \, \omega^2 |\NN|^2 |S_{\rm em}(\ve,k)|^2
=\int_{0}^\infty {d\omega\over \pi} \, \omega^2 |\NN|^2 |S_{\rm em}(\ve,k)|^2 \, .
\ee
A solid angle $\Delta\Omega$ corresponds to an `area' $R^{D-2} 
\Delta\Omega$.
Therefore the total energy carried by the photons of polarization $\ve$, within solid angle $\Delta\Omega$
of $\hat n$,
with frequency between 
$\omega$ and $\omega(1+\delta)$ is given by
\be\label{ered}
R^{D-2} 
\Delta\Omega \, \omega \, \delta \, {1\over \pi} \, \, 
\omega^2 |\NN|^2 |S_{\rm em}(\ve,k)|^2\, .
\ee
Substituting the value of $\NN$ from \refb{eNvalue} we can reduce 
\refb{ered} to
\be \label{e738}
{1\over 2}\, 
(2\pi)^{-(D-1)}\Delta\Omega \, \omega^{D-1} \, \delta \, |S_{\rm em}(\ve,k)|^2\, .
\ee
This is in perfect agreement with \refb{epower} with $S^{(0)}_{\rm em}$ replaced by
$S_{\rm em}=S^{(0)}_{\rm em}+S^{(1)}_{\rm em}$.
An important point to note is that the normalization constant
$\NN$ contains information about the $\omega$-dependent phase space factor that
was used to get
\refb{epower}.

One could also try to 
give an alternate derivation of \refb{eafin} by using the position space representation
of the retarded Green's function $G_r(x,x')$. In even space-time 
dimensions the retarded Green's function
is proportional to $\theta(x^0 - x^{\prime 0}) \delta ((x-x')^2)$ and one could use the analysis
given in section 14 of \cite{jackson} to derive an expression for the gauge field produced by a
moving charged particle. In odd space-time dimensions the retarded Green's function has a more
complicated form, but presumably the radiative component of the field can be found without too
much hurdle.

\subsection{Soft graviton theorem} \label{sfusion}

Eq.\refb{eafin4} for electromagnetic radiation was derived for an arbitrary trajectory of a
charged particle, moving under some unspecified force. This is possible in
electromagnetism since the conservation of electromagnetic current of a point source
does not require the source to satisfy its equation of motion. However for gravity 
this is not so -- we need the source to satisfy its equation of motion in order that it
generates a conserved energy-momentum tensor. This means that we have to take into
account all forces acting on the source during its motion and take into account the
gravitational radiation generated by the fields responsible for these forces in order to
arrive at a consistent result. We shall now describe some situations where the 
computation can be done and compare the corresponding results with the prediction
of soft theorem. \asnote{Similar calculations of gravitational radiation from classical point sources
have been performed in different context recently in \cite{1611.03493,1705.09263,1712.09250}. }

\subsubsection{Scattering via fusion and decay:} \label{sfusiona}
In \cite{weinbergbook} Weinberg considered the situation where a bunch of particles
moving along straight lines meet at a single point in space-time and come out as
another bunch of particles emerging from the same space-time point and moving along
straight line trajectories. Physically this corresponds to the situation where there is a
repulsive short-range interaction between the particles where the range of the interaction
is large compared to the Schwarzschild radii of the particles. For this reason the 
effect of gravitational interaction between the particles can be ignored and the particles
can be taken to move along straight lines till the repulsive interaction switches on. If we
now consider long wave-length gravitational waves -- with wave-length much larger than
the range of the repulsive interaction -- then the interaction can regarded as contact
interaction. This leads to the picture described above. Weinberg showed that this analysis
produces correctly the leading soft graviton theorem. In fact for this analysis  the probe 
limit was not necessary -- one could reproduce the full result \refb{esoft0}.

\begin{figure}
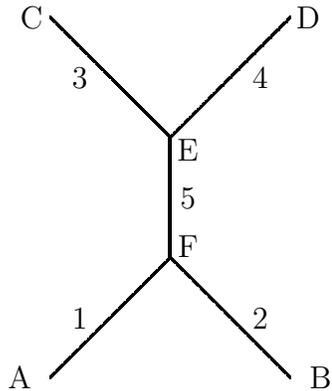

\begin{center}

\figone

\caption{A scattering process in space-time.
\label{fig1}
}

\end{center}

\end{figure}

One could now ask if we can work to higher
order and verify the subleading (or even subsubleading) 
contribution to the soft graviton theorem.  The problem with this is that since the 
external particle trajectories meet at a single point in space-time, all the orbital
angular momenta vanish if we choose the common meeting 
point as the origin of space-time and
there is no subleading contribution left over. We shall now consider a slightly different
type of scattering that  overcomes this problem. 
We shall assume that the interaction is such that once the initial particles collide at
a single space-time point, they fuse and travel as a single entity for certain distance,
and then splits apart into final state particles which travel away in different directions
along straight line trajectories. This has been shown in Fig.~\ref{fig1} for $2\to 2$ 
scattering.
Since the energy momentum tensor is conserved 
locally during this process, this is a consistent source. However 
in this case the orbital angular momentum of the particles
no longer vanish identically and we have a non-zero contribution from the subleading
term in the soft theorem. This can then be compared with the result of direct
computation.

During this analysis we shall not need to take the probe limit. This is a reflection
of the fact that the regime we are considering, in which the interaction length scale
is much larger then the Schwarzschild radius of the black hole, the upper cut-off on the
frequency of gravitational wave remains small compared to the inverse 
Schwarzschild radius 
of the black hole. As a 
result the total energy carried away by the radiation remains
small compared to the incoming and outgoing particle masses even without taking the
probe limit. We also expect the analysis to reproduce the results of soft 
theorem to subsubleading order, since in this case the subsubleading contribution
comes only from the orbital angular momentum part. The possible
non-universal terms would come from 
non-minimal effective coupling of gravity to the external states, and we shall assume that the
external states have only minimal coupling to gravity.

To compute the gravitational radiation from this process, we use the standard formula for the
gravitational field $h_{\alpha\beta}(x)$
of a point source of mass $m$, obtained from the convolution of the retarded Green's 
function $G_r(x,y)$ defined in \refb{eretard}
with $-T_{\alpha\beta} (y)+ (D-2)^{-1} \eta_{\alpha\beta} T^\gamma_\gamma(y)$ 
where $T_{\alpha\beta}$
denotes the energy momentum tensor of the source. This is a straightforward 
generalization of the corresponding formula in four space-time 
dimensions as given in \cite{weinbergbook} . We have
\be \label{erad}
h_{\alpha\beta}(x) = e_{\alpha\beta}(x) - {1\over D-2} \eta_{\alpha\beta} \, 
e^\gamma_\gamma(x), 
\ee
\be \label{edefealbe}
e_{\alpha\beta}(x)\equiv -\int d^D x' \, G_r(x,x') 
T_{\alpha\beta}(x') 
= -\int \, d\sigma \, G_r(x, r(\sigma)) \, P_\alpha(\sigma) \, V_\beta(\sigma)\, ,
\ee
where $r(\sigma)$ describes the trajectory of the particle parametrized by the proper time $\sigma$,
$V=dr/d\sigma$ is the four velocity of the particle and
$P=m V$ is the four momentum of the particle. 
Following the same steps that led from \refb{edefaalpha} to \refb{eafin}, 
we can obtain the long
distance behavior of \refb{edefealbe} to be
\ben\label{epart11}
\tilde e_{\alpha\beta} &=&  i\, \NN \, e^{i\omega R} \, 
\int d\sigma 
e^{i\omega \{r^0(\sigma)  - \hat n.\vec r(\sigma)\}}
\,P_\alpha(\sigma) V_\beta(\sigma)\nonumber \\
&&
- \NN \, e^{i\omega R} \, {1\over \omega} \, 
\left[
e^{i\omega \{r^0(\sigma) - \hat n.\vec r(\sigma)\}}
\left\{{P_\alpha(\sigma) V_\beta(\sigma)
 \over V^0(\sigma) - \hat n.\vec V(\sigma)}\right\}\right]_{\rm boundary}\, , 
\een
where $\NN$ is the same normalization constant as defined in 
\refb{eNvalue}. The integration over $\sigma$ in the first line runs over all the 
trajectory segments,
and the boundary terms given in the second line receives contribution from the outer ends
of all the external lines.
Explicitly evaluating the boundary contributions to \refb{epart11} from the four
end points 
$A$, $B$, $C$ and $D$, we can express \refb{epart11} as 
\be \label{ehabab}
\tilde e_{\alpha\beta} \simeq -\NN \,  e^{i\omega R} \, 
\left[-\sum_{a=1}^2 e^{i k. r_{(a)}} {P_{(a)\alpha} P_{(a)\beta}\over
k. P_{(a)}}  
+ \sum_{a=3}^4 e^{i k. r_{(a)}} {P_{(a)\alpha} P_{(a)\beta}\over
k. P_{(a)}}
- i \sum_{a=1}^5 \int d\sigma_a e^{ik.r(\sigma)}
P_{(a)\alpha} V_{(a)\beta}
\right]
\ee
where $\sigma_a$
is the proper time parametrizing the $a$-th trajectory segment,
$P_{(a)}$ is the momentum flowing along the $a$-th segment along the
direction of increasing $\sigma_a$,  $r_{(a)}$ for $a=1,2,3,4$ are the
locations of the four space-time points $A$, $B$, $C$ and $D$ in Fig.~\ref{fig1}
and $k=-\omega(1,\hat n)$.
Since the external momenta are considered to be positive if they are ingoing,
we have
\be 
P_{(a)}= p_{(a)} \quad \hbox{for $a=1,2$}, \quad P_{(a)}= -p_{(a)} \quad 
\hbox{for $a=3,4$}, \quad P_{(5)}=p_{(1)}+p_{(2)}=-p_{(3)}-p_{(4)}\, ,
\ee
\be
V_{(a)}=v_{(a)} \quad \hbox{for $a=1,2$}, \quad  V_{(a)}=-v_{(a)}\quad \hbox{for $a=3,4$}
\, ,
\ee
where $v_{(a)}=p_{(a)}/m_{(a)}$, $m_{(a)}$ being the mass of the $a$-th external state.
Now let us choose the origin of space-time coordinates to be at $F$, and let
us choose the convention that $\sigma_{1}$, $\sigma_{2}$ and $\sigma_{5}$
vanish at $F$ and $\sigma_{3}$ and $\sigma_{4}$ vanish at $E$. 
Then we have
\be \label{egravtra}
r(\sigma_a) = V_{(a)} \, \sigma_a \quad \hbox{for $a=1,2,5$}, \quad
r(\sigma_a) = r(E)+ V_{(a)} \, \sigma_a \quad \hbox{for $a=3,4$}\, .
\ee
Let us suppose 
further that  we have
\be
\sigma_1 = -s_1 \ \hbox{at} \ A, \quad \sigma_2 = -s_2 \ \hbox{at} \ B, \quad 
\sigma_3 = s_3 \ \hbox{at} \ C, \quad \sigma_4 = s_4 \ \hbox{at} \ D,
\quad \sigma_5 = s_5 \ \hbox{at} \ E\, .
\ee
This gives, by integrating the $dr/d\tau=V$ equation:
\ben
&& r_{(1)} = -s_1 \, v_{(1)} , \quad r_{(2)} = -s_2 \, v_{(2)}, \quad r_{(3)} = -s_3 \, v_{(3)}
+ r(E),
\quad r_{(4)} = -s_4 \, v_{(4)}+r(E), \nonumber \\
&& r(E) = V_{(5)} s_5\, .
\een
Finally, since $v_{(a)}$ is parallel to $p_{(a)}$, we have
\be 
{k.v_{(a)}\over k.p_{(a)}} p_{(a)\alpha} p_{(a)\beta} 
= p_{(a)\alpha} v_{(a)\beta} \, ,
\ee
Using these relations
we can express \refb{ehabab} up to subsubleading order in $k$ as:
\ben\label{eeab}
\tilde e_{\alpha\beta}
&=& \NN \,  e^{i\omega R} \, 
\bigg[\sum_{a=1}^4 {
p_{(a)\alpha} p_{(a)\beta}\over
k. p_{(a)}} + i\sum_{a=3}^4 {1\over k.p_{(a)}}  p_{(a)\alpha} k^\gamma
\left\{p_{(a)\beta} r_\gamma(E) - p_{(a)\gamma} r_\beta(E)
\right\} \nonumber \\
&& - {1\over 2} \sum_{a=3,4} (k.r(E))^2 \, {p_{(a)\alpha}p_{(a)\beta}\over k.p_{(a)}}
+{1\over 2} \sum_{a=3,4}  p_{(a)\alpha} r(E)_\beta \, k.r(E) \bigg]\, .
\een
We can now compute $\tilde h_{\alpha\beta}$ using \refb{erad}:
\be 
\tilde h_{\alpha\beta} = \tilde e_{\alpha\beta} - {1\over D-2} \eta_{\alpha\beta}
\, \tilde e^\gamma_\gamma\, .
\ee
However our interest
will be on the projection of $\tilde h$ along a given polarization $\ve$. This is
obtained by contracting $\ve$ with $\tilde h$ via the metric given on the right hand side
of \refb{einverse}. This gives the projection to be
\be 
\phi(\ve)\equiv \ve^{\alpha\beta} \tilde h_{\alpha\beta} - {1\over 2} \ve^\alpha_\alpha
\tilde h^\beta_\beta=\ve^{\alpha\beta} \tilde e_{\alpha\beta}\, .
\ee
Using \refb{eeab} we now get
\ben \label{epexp}
\phi(\ve) &=& \NN\,  e^{i\omega R} \, 
 \bigg[\sum_{a=1}^4 {\ve^{\alpha\beta} 
p_{(a)\alpha} p_{(a)\beta}\over
k. p_{(a)}} + i\sum_{a=3}^4 {1\over k.p_{(a)}} \ve^{\alpha\beta} p_{(a)\alpha} k^\gamma
\left\{p_{(a)\beta} r_\gamma(E) - p_{(a)\gamma} r_\beta(E)
\right\}
\nonumber \\
&& - {1\over 2} \sum_{a=3,4} (k.r(E))^2 \, {\ve^{\alpha\beta} \, 
p_{(a)\alpha}p_{(a)\beta}\over k.p_{(a)}}
+ {1\over 2} \sum_{a=3,4}  \ve^{\alpha\beta}\, p_{(a)\alpha} r(E)_\beta \, k.r(E) \bigg]\, .
\een
On the other hand here the orbital angular momenta of the external particles,
measured with respect to the origin of space-time situated at $F$, are given by
\be 
\bJ_{(a)\alpha\beta}=0 \quad \hbox{for $a=1,2$}, \quad
\bJ_{(a)\alpha\beta}= r_\alpha(E) \, p_{(a)\beta} - r_\beta(E) \, p_{(a)\alpha} \quad
\hbox{for $a=3,4$}\, .
\ee
Using this, and the fact that $r(E)$ is proportional to $p_{(3)}+p_{(4)}$ so that
$k.r(E) \, (p_{(3)\alpha}+p_{(4)\alpha})=k.(p_{(3)}+p_{(4)})\, r(E)_\alpha$,   
we can express \refb{epexp} as
\ben 
\phi(\ve)
&=&\NN\,  e^{i\omega R} \, 
 \left[\sum_{a=1}^4 {\ve^{\alpha\beta} 
p_{(a)\alpha} p_{(a)\beta}\over
k. p_{(a)}} + i\sum_{a=1}^4 {1\over k.p_{(a)}} \ve^{\alpha\beta} p_{(a)\alpha} k^\gamma
\bJ_{(a)\gamma\beta}
-{1\over 2} \sum_{a=1}^4 {1\over k.p_{(a)}}  \ve_{\alpha\beta}
k_\gamma k_\delta \, \bJ_{(a)}^{\alpha\gamma} \bJ_{(a)}^{\beta\delta}
\right] \nonumber \\ &=& \NN \,  e^{i\omega R} \, 
 \left[S^{(0)}_{\rm gr}(\ve, k) + S^{(1)}_{\rm gr}(\ve,k)+ S^{(2)}_{\rm gr}(\ve, k)\right]\, .
\een
We can now use the same analysis that \asnote{led }from \refb{e732} to \refb{e738} to 
show that the angular power spectrum of soft graviton radiation during this process
is given by eq.\refb{enaemgr}.

\begin{figure}
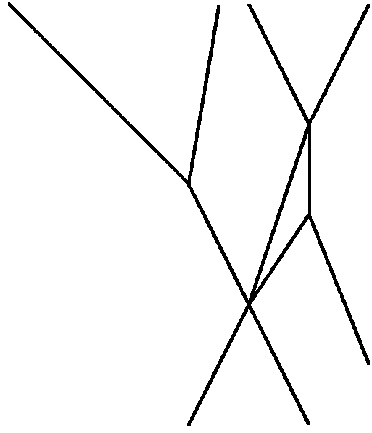

\begin{center}

\figtwo

\caption{A more complicated scattering process in space-time.
\label{fig2}
}

\end{center}

\end{figure}

\subsubsection{Scattering via multiple fusion and decay} \label{smulti}

We shall now consider a more complicated scattering via multiple interactions of the type shown in
Fig.~\ref{fig2}.  The rules are the same as in section \ref{sfusiona}, i.e.\ the energy momentum tensor is
localized along the trajectories and is conserved at each vertex. We shall denote by $p_{(a)}$ for
$a=1,\cdots n$ the external momenta, and all of them will be taken to be ingoing. We shall also denote
by $P_{(s)}$ and $V_{(s)}$ the momenta and velocities  along all the trajectory segments -- both
external and internal, along the direction of increasing proper time. 
Then the analog of \refb{ehabab} takes the form
\be \label{ehababnew}
\tilde e_{\alpha\beta} \simeq -\NN \,  e^{i\omega R} \, 
\left[-\sum_{a=1}^n e^{i k. r_{(a)}} {p_{(a)\alpha} p_{(a)\beta}\over
k. p_{(a)}}  
- i \sum_{s} \int d\sigma_s e^{ik.r(\sigma_s)}
P_{(s)\alpha} V_{(s)\beta}
\right]\, ,
\ee
where $r_{(a)}$ denotes the space-time coordinates of the boundary of the $a$-th external trajectory
and $\sigma_s$ is the proper time along the $s$-th trajectory segment. 
We now expand the integrand in power
series in $k$ up to subsubleading order and carry out integration over $\sigma_s$ using the fact that
along the $s$-th trajectory
\be \label{ebeg2}
r(\sigma_s) = I_s + \sigma_s \, V_{(s)}\, ,
\ee
where we define 
\be \label{ebeg3}
I_s\equiv r(\sigma_{s,i}), \quad  J_s\equiv r(\sigma_{s,f})\, ,
\ee
$\sigma_{s,i}$ and $\sigma_{s,f}$ being the initial and final values of $\sigma_s$ along the $s$-th
trajectory. With this we can express \refb{ehababnew} as
\ben \label{ebeg1}
\tilde e_{\alpha\beta} &\simeq& \NN \,  e^{i\omega R} \, 
\Bigg[\sum_{a=1}^n \left\{1 + i k.r_{(a)}- {1\over 2} (k.r_{(a)})^2\right\} 
{p_{(a)\alpha} p_{(a)\beta}\over
k. p_{(a)}}  \nonumber \\ && \hskip -.5in
+ i \sum_{s} P_{(s)\alpha} V_{(s)\beta} \left\{(\sigma_{s,f}-\sigma_{s,i}) 
+i \, k.I_s \, (\sigma_{s,f}-\sigma_{s,i})
+ {i\over 2} k.V_{(s)} (\sigma_{s,f}-\sigma_{s,i})^2\right\}
\Bigg]\, .
\een
Now it follows from \refb{ebeg2}, \refb{ebeg3} that
\be 
J_s - I_s = V_{(s)} (\sigma_{s,f}-\sigma_{s,i})\, .
\ee
Using this we can express \refb{ebeg1} as
\ben\label{ebeg5}
\tilde e_{\alpha\beta} &\simeq& \NN \,  e^{i\omega R} \, 
\Bigg[\sum_{a=1}^n \left\{1 + i k.r_{(a)}- {1\over 2} (k.r_{(a)})^2\right\} 
{p_{(a)\alpha} p_{(a)\beta}\over
k. p_{(a)}}  \nonumber \\ &&
+ i \sum_{s} P_{(s)\alpha} \left\{(J_s-I_s)_\beta + i\, k.I_s\, (J_s-I_s)_\beta
+ {i\over 2} k.(J_s-I_s) (J_s-I_s)_\beta \right\}
\Bigg]\, .
\een

Consider now the sum
$\sum_{s} P_{(s)\alpha} (J_s-I_s)_\beta$.
We can rearrange this by fixing a given vertex and summing over all the lines connected to the
vertex. This will give a sum over all the momenta entering a vertex which will vanish by momentum 
conservation. The only exceptions are the outer ends of the external trajectories. This contribution
gives
\be\label{e749}
\sum_{s} P_{(s)\alpha} (J_s-I_s)_\beta= -\sum_{a=1}^n r_{(a)\beta} p_{(a)\alpha}\, ,
\ee
where the minus sign reflects the fact that the momenta $p_{(a)}$ 
are directed away from these ends.

Next consider the sum
\ben\label{e750}
&& \sum_{s} P_{(s)\alpha}  \, \left\{k.I_s\, (J_s-I_s)_\beta
+{1\over 2} \, k.(J_s-I_s) \, (J_s-I_s)_\beta
\right\}\nonumber \\
&=& {1\over 2} \sum_{s} P_{(s)\alpha}  \, \left\{k.J_s\, (J_s)_\beta - k.I_s\, (I_s)_\beta
-k.J_s\, (I_s)_\beta + k.I_s\, (J_s)_\beta
\right\}
\, .
\een
Substituting \refb{e749} and \refb{e750} into \refb{ebeg5} we get
\be \label{ebeg6}
\ve^{\alpha\beta} \, \tilde e_{\alpha\beta} \simeq \NN \,  e^{i\omega R} \, \bar S_{\rm gr}\, ,
\ee
where
\ben \label{ebeg67}
\bar S_{\rm gr} &=& \ve^{\alpha\beta}\, 
\Bigg[\sum_{a=1}^n \left\{1 + i k.r_{(a)}\right\} 
{p_{(a)\alpha} p_{(a)\beta}\over
k. p_{(a)}}  -i\sum_{a=1}^n r_{(a)\beta} p_{(a)\alpha}
\nonumber \\
&& \hskip -1in - {1\over 2} \sum_{a=1}^n (k.r_{(a)})^2\, 
{p_{(a)\alpha} p_{(a)\beta}\over
k. p_{(a)}} - {1\over 2} 
\sum_{s} P_{(s)\alpha}  \, \left\{k.J_s\, (J_s)_\beta - k.I_s\, (I_s)_\beta
-k.J_s\, (I_s)_\beta + k.I_s\, (J_s)_\beta
\right\}
\Bigg] \nonumber \\
&\equiv& \bar S^{(0)}_{\rm gr} + \bar S^{(1)}_{\rm gr} + \bar S^{(2)}_{\rm gr}\, .
\een
\asnote{$\bar S^{(0)}_{\rm gr}$, $\bar S^{(1)}_{\rm gr}$ and  $\bar S^{(2)}_{\rm gr}$ represent
respectively the leading, subleading and subsubleading contributions to \refb{ebeg67}. }

Our goal will be to show that $\bar S_{\rm gr}$ given in \refb{ebeg67} 
agrees with the soft factor $S^{(0)}_{\rm gr}+S^{(1)}_{\rm gr}+S^{(2)}_{\rm gr}$.
Now for this problem we have
\be \label{ejamn}
\bJ_{(a)}^{\mu\nu} = r_{(a)}^\mu p_{(a)}^\nu - r_{(a)}^\nu p_{(a)}^\mu\, .
\ee
Using this and \refb{esoft0}, \refb{esublead0}, \refb{esoftgr2} we get
\ben \label{ebeg68}
S_{\rm gr} &\equiv& S^{(0)}_{\rm gr} + S^{(1)}_{\rm gr} + S^{(2)}_{\rm gr}\nonumber \\
&=& \ve^{\alpha\beta}\, 
\Bigg[\sum_{a=1}^n 
{p_{(a)\alpha} p_{(a)\beta}\over
k. p_{(a)}}  + i\sum_{a=1}^n \left\{ {p_{(a)\alpha} p_{(a)\beta}\over
k. p_{(a)}} k. r_{(a)} -
r_{(a)\beta} p_{(a)\alpha}\right\}
\nonumber \\
&& - {1\over 2} \sum_{a=1}^n \, 
{1\over
k. p_{(a)}}  \bigg\{ p_{(a)\alpha}\,  p_{(a)\beta}  (k.r_{(a)})^2 +
(k.p_{(a)})^2 r_{(a)\alpha} \, r_{(a)\beta} \nonumber \\ && - k.p_{(a)} \, k.r_{(a)}\,  p_{(a)\alpha} r_{(a)\beta} 
- k.p_{(a)} \, k.r_{(a)}\,  r_{(a)\alpha} p_{(a)\beta}
\bigg\}
\Bigg]\, . 
\een
Comparing the first line of \refb{ebeg67} and the second line of
\refb{ebeg68} we see that $\bar S_{\rm gr}$ and $S_{\rm gr}$
agree up to subleading order. We shall now show that this equality also holds at the subsubleading order.

One property of \refb{ebeg67} that we shall use to simplify our analysis is that this expression
is independent of the length of the external legs, i.e.\ it remains invariant
under the transformation $r_{(a)}\to r_{(a)}
+ \asnote{b_{(a)} }p_{(a)}$ for arbitrary constants \asnote{$b_{(a)}$. }This 
can be verified explicitly, but also follows from
the original expression \refb{ehababnew} which by construction is independent of the outer limits
of $\sigma$ integration for the external lines. Using this property we can even take the external lines
to have zero length. \refb{ebeg68} shares a similar property that follows from the fact that 
$\bJ_{(a)}^{\mu\nu}$ defined in \refb{ejamn} is invariant under $r_{(a)}\to r_{(a)}
+ \asnote{b_{(a)} }p_{(a)}$.

Our proof of the equality of $S^{(2)}_{\rm gr}$ and $\bar S^{(2)}_{gr}$ will proceed by
induction. First we shall assume that it holds for all tree diagrams with $n$ external legs and prove that
it also holds for all tree diagrams with $n+1$ external legs. Then we shall assume that it holds for all
diagrams with at most $\ell$ loops and then prove that it also holds for all diagrams with at most $\ell+1$
loops.

Let us begin with the first step. Suppose we have an $n+1$ point amplitude in which the external lines
carrying momentum $p_{(u)}$ and $p_{(v)}$ join at a space-time point $r$ to produce an internal line
of momentum $p_{(u)}+p_{(v)}$. By taking the external legs carrying momenta $p_{(u)}$ and 
$p_{(v)}$ to have zero length, we take $r_{(u)}=r_{(v)}=r$. 
We compare this with another diagram with $n$ external legs 
where the pair of external lines carrying momenta $p_{(u)}$ and $p_{(v)}$ are removed and the
internal line carrying momentum $p_{(u)}+p_{(v)}$ is regarded as an external line. 
From \refb{ebeg68} we see that the difference in
$S^{(2)}_{\rm gr}$  between the first and the second diagram is given by
\ben \label{esdiff}
\Delta S^{(2)}_{\rm gr} 
&=& -{1\over 2} (k.r)^2 \ve_{\alpha\beta} \left\{ {p_{(u)}^\alpha p_{(u)}^\beta\over k.p_{(u)}}
+ {p_{(v)}^\alpha p_{(v)}^\beta\over k.p_{(v)}}\right\}
-{1\over 2} \ve^{\alpha\beta} r_\alpha r_\beta \, k.\{p_{(u)}+p_{(v)}\}
\nonumber \\
&& + \ve^{\alpha\beta} r_\alpha \{p_{(u)\beta} + p_{(v)\beta}\} \, k.r
+{1\over 2}  (k.r)^2 \ve_{\alpha\beta} {\{p_{(u)}^\alpha +p_{(v)}^\alpha\} \{p_{(u)}^\beta
+ p_{(v)}^\beta\} \over k.\left\{p_{(u)}+ p_{(v)}\right\}}
\nonumber \\ && +{1\over 2} \ve^{\alpha\beta} r_\alpha r_\beta \, k.\{p_{(u)}+p_{(v)}\}
- \ve^{\alpha\beta} r_\alpha \{p_{(u)\beta} + p_{(v)\beta}\} \, k.r\nonumber \\
&=& -{1\over 2} (k.r)^2 \ve_{\alpha\beta} \left\{ {p_{(u)}^\alpha p_{(u)}^\beta\over k.p_{(u)}}
+ {p_{(v)}^\alpha p_{(v)}^\beta\over k.p_{(v)}}\right\}
+{1\over 2}  (k.r)^2 \ve_{\alpha\beta} {\{p_{(u)}^\alpha +p_{(v)}^\alpha\} \{p_{(u)}^\beta
+ p_{(v)}^\beta\} \over k.\left\{p_{(u)}+ p_{(v)}\right\}}\, . \nonumber \\
\een
We shall now compare this with the difference in $\bar S^{(2)}_{\rm gr}$, given in the second line
of \refb{ebeg67}, between the first and the second configurations. Since for $r_{(u)}=r_{(v)}=r$ the
two diagrams have the same set of lines, the difference comes solely from the terms proportional to
$(k.r_{(a)})^2$.This gives
\be \label{ebsdiff}
\Delta \bar S^{(2)}_{\rm gr} = -{1\over 2} (k.r)^2 \ve_{\alpha\beta} 
\left\{ {p_{(u)}^\alpha p_{(u)}^\beta\over k.p_{(u)}}
+ {p_{(v)}^\alpha p_{(v)}^\beta\over k.p_{(v)}}\right\}
+{1\over 2}  (k.r)^2 \ve_{\alpha\beta} {\{p_{(u)}^\alpha +p_{(v)}^\alpha\} \{p_{(u)}^\beta
+ p_{(v)}^\beta\} \over k.\left\{p_{(u)}+ p_{(v)}\right\}}\, .
\ee
Comparing \refb{esdiff} and \refb{ebsdiff} we see that $\Delta S^{(2)}_{\rm gr} =\Delta 
\bar  S^{(2)}_{\rm gr}$. Since
we have assumed that for tree diagrams with $n$ external states $S^{(2)}_{\rm gr}=\bar S^{(2)}_{\rm gr}$, this
establishes the equality of $S^{(2)}_{\rm gr}$ and $\bar S^{(2)}_{\rm gr}$ for tree diagrams with $n+1$ external
states.\footnote{In making this inference we have assumed that the tree diagram with $(n+1)$ external
states has at least two external lines that join to an internal line via a three point vertex. One may worry
that this proof does not hold for diagrams in which all external states connect to vertices with more than
three legs. This can be addressed by noting that a vertex with more than three legs can be 
realized by starting with a diagram containing only three point vertices, and then taking the lengths of
some of the internal lines to zero.
}
Repeating this analysis we can establish the equality of $S^{(2)}_{\rm gr}$ and $\bar S^{(2)}_{\rm gr}$ for
all tree diagrams.

We now turn to the second step -- the loop diagrams. Our starting assumption is that the equality
of $\bar S^{(2)}_{\rm gr}$ and $S^{(2)}_{\rm gr}$ holds for all diagrams with $\ell$ or less loops. Let us 
consider a diagram with $\ell+1$ loops. We can get from this a diagrams with $\ell$ loops 
by picking an internal line that forms part of a loop and regarding this as a pair of external
lines carrying opposite momenta. If the internal line extends from $r_1$ to $r_2$ and carries
momentum $p$, then the difference in $S^{(2)}_{\rm gr}$ 
between the first diagram with $\ell+1$ loops and the second diagram with $\ell$ loops 
comes from the two extra external lines that the second diagram carries.  As before in the
second diagram we shall take the external lines with momenta $\pm p$ to have zero length.
This gives
\ben\label{essgg}
\Delta S^{(2)}_{\rm gr} &=& 
{1\over 2} 
{1\over
k. p} \, \ve^{\alpha\beta} \bigg[ p_{\alpha}\,  p_{\beta}  \{(k.r_{2})^2 - (k.r_{1})^2 \}+
(k.p_{})^2 \{r_{2\alpha} \, r_{2\beta} -r_{1\alpha} \, r_{1\beta}
\} \nonumber \\ && - 2\, k.p \, p_{\alpha} \, \{k.r_{2}\,   r_{2\beta} - k.r_{1}\,   r_{1\beta} 
\}
\bigg]\, .
\een
On the other hand the difference in the contribution to $\bar S^{(2)}_{\rm gr}$ from the first and the second
diagrams comes from two sources. 
The first diagram will have an extra line carrying
momentum $p$ and extending from $r_1$ to $r_2$ and
will therefore receive a contribution from that line. 
On the other hand
the second diagram will have two additional external lines carrying
momenta $\pm p$ and will therefore receive additional contribution from the terms proportional to
$(k.r_{(a)})^2$ in \refb{ebeg67}. This gives
\ben \label{ewcn}
\Delta  \bar S^{(2)}_{\rm gr} &=& 
-{1\over 2} \ve^{\alpha\beta}\, p_\alpha \, \{k.r_2 \, r_{2\beta}- k.r_1 \, r_{1\beta}
- k.r_2 \, r_{1\beta} + k.r_1 \, r_{2\beta} \nonumber \\ &&
+
{1\over 2} \ve^{\alpha\beta} {p_\alpha p_\beta\over k.p} \{ (k.r_2)^2 
- (k.r_1)^2\}\,.
\een
Now using that fact that the vector $r_2-r_1$ is proportional to $p$, we can write
\ben
&& p_\alpha \, k.r_2\, r_{1\beta} = p_\alpha \, k.(r_2-r_1)\, r_{1\beta} + p_\alpha \, k.r_1\, r_{1\beta}
= k.p \, (r_2-r_1)_\alpha \, r_{1\beta} + p_\alpha \, k.r_1\, r_{1\beta}\, , \nonumber \\
&& p_\alpha \, k.r_1\, r_{2\beta} = p_\alpha \, k.(r_1-r_2)\, r_{2\beta} + p_\alpha \, k.r_2\, r_{2\beta}
= k.p \, (r_1-r_2)_\alpha \, r_{2\beta} + p_\alpha \, k.r_2\, r_{2\beta}\, . 
\een
Using this we can express \refb{ewcn} as
\ben\label{ebsggrr}
\Delta  \bar S^{(2)}_{\rm gr} &=& 
-{1\over 2} \ve^{\alpha\beta}\, p_\alpha \, \{2 k.r_2 \, r_{2\beta}- 2 k.r_1 \, r_{1\beta}\}
+{1\over 2} \, \ve^{\alpha\beta} \, k.p \{ (r_2-r_1)_\alpha r_{1\beta} + (r_2-r_1)_\alpha r_{2\beta}
\}
\, 
 \nonumber \\ &&
+
{1\over 2} \ve^{\alpha\beta} {p_\alpha p_\beta\over k.p} \{ (k.r_2)^2 
- (k.r_1)^2\}\,.
\een
Since $\ve^{\alpha\beta}$ is symmetric, \refb{ebsggrr} and \refb{essgg} are identical. Using the
fact that for $\ell$-loop diagrams $\bar S^{(2)}_{\rm gr}$ and $S^{(2)}_{\rm gr}$ are identical, this establishes that
for the $(\ell+1)$-loop diagrams also the two are  
identical. By induction, this establishes the equality of 
$S^{(2)}_{\rm gr}$ and $\bar S^{(2)}_{\rm gr}$ for all diagrams. 

\subsubsection{Inclusion of \asnote{spin }} \label{sspin}

In the analysis of the previous two subsections we have not included any contribution from the 
intrinsic spin of the objects. In this subsection we shall rectify this situation. For this we use the fact that
if $\Sigma^{\alpha\beta}(\sigma)$ denotes the spin angular momentum of an object moving along the
trajectory $r(\sigma)$, then its contribution to the stress tensor is given by 
(see {\it e.g.} \cite{1712.09250}
for recent \asnote{review): }
\be
\Delta T^{\alpha\beta}(x) = 
\int d\sigma\, {dr^{(\alpha}(\sigma)\over d\sigma} \, \Sigma^{\beta)\gamma}(\sigma) \, \p_\gamma
\delta^{(D)}(x - r(\sigma))\, .
\ee
Its contribution to the  \asnote{metric }is given by
\be 
\Delta h_{\alpha\beta}(x) = \Delta e_{\alpha\beta}(x) - {1\over D-2} \eta_{\alpha\beta} \, 
\Delta e^\gamma_\gamma(x)\, ,
\ee
\be 
\Delta e_{\alpha\beta}(x)\equiv -\int d^D x' \, G_r(x,x') \,
\Delta T_{\alpha\beta}(x')\, .
\ee
This can be easily evaluated following the
procedure described in sections \ref{sphoton}, \ref{sfusiona}, \ref{smulti}. Let us
consider the general process described in section \ref{smulti}. Then the 
\asnote{radiative component of the }time Fourier
transform of $\Delta e_{\alpha\beta}$, denoted by $\Delta \tilde e_{\alpha\beta}$,
is obtained by replacing the
$P_\alpha(\sigma) V_\beta(\sigma)$ factor in \refb{epart11} by $-i \, k^\gamma \,
V_{(\alpha}(\sigma)\, \Sigma_{\beta)\gamma}(\sigma)$.
This gives
\ben\label{epart11A}
\Delta \tilde e_{\alpha\beta} &=&  \NN \, e^{i\omega R} \, \sum_s
\int d\sigma_s\,
e^{i\omega \{r^0(\sigma_s)  - \hat n.\vec r(\sigma_s)\}}
\, k^\gamma \,
V_{(\alpha}(\sigma_s)\, \Sigma_{\beta)\gamma}(\sigma_s) \nonumber \\
&&
+ \NN \, e^{i\omega R} \, {1\over \omega} \, 
\left[
e^{i\omega \{r^0(\sigma) - \hat n.\vec r(\sigma)\}}
\left\{{i k^\gamma \,
V_{(\alpha}(\sigma)\, \Sigma_{\beta)\gamma}(\sigma)
 \over V^0(\sigma) - \hat n.\vec V(\sigma)}\right\}\right]_{\rm boundary}\, .
\een
Here the sum over $s$ in the first term runs over all the trajectory segments, whereas the boundary
terms receive contribution from the outer ends of all the external lines. 
Due to the explicit $k^\gamma$ factor, the first term begins contributing at the subsubleading
order, whereas the second term begins contributing at the subleading order. Therefore, in evaluating
this expression to the subsubleading order we can replace the 
$e^{i\omega \{r^0(\sigma_s)  - \hat n.\vec r(\sigma_s)\}}$ factor in the first term by 1. Since 
$\Sigma_{\beta\gamma}(\sigma_s)$ remains constant along a given trajectory segment and
$V^\alpha(\sigma_s) = dr^\alpha(\sigma_s)/d\sigma_s$, we get 
\be \label{evertkx}
\sum_s \int d\sigma_s V_{\alpha}(\sigma_s)\, \Sigma_{\beta\gamma}(\sigma_s) 
= \sum_s (J_s-I_s)_\alpha \, \Sigma_{\beta\gamma}(\sigma_s) \, ,
\ee
where, as in section \ref{smulti}, we denote by $I_s$ and $J_s$ the beginning and the end
points of the $s$-th trajectory segment. We can now rearrange the sum by first summing over all
lines that enter a vertex and then summing over vertices. Since by conservation of angular
momentum the sum over all $\Sigma_{\alpha\beta}$ entering a vertex vanishes, \refb{evertkx}
receives contribution only from the outer ends of the external lines situated \asnote{at
 }$r_{(a)}$ for $1\le a\le n$. This gives the contribution from the first term on the right hand
side of \refb{epart11A} to be 
\be \label{eadad1}
-{1\over 2} \NN \, e^{i\omega R} \, k^\gamma\, \sum_{a=1}^n 
\left\{ r_{(a)\alpha} \Sigma_{(a)\beta\gamma}+r_{(a)\beta} \Sigma_{(a)\alpha\gamma}\right\}\, .
 \ee
On the other hand, the contribution from the second term on the right hand side of
\refb{epart11A} to this order is given by
\be \label{eadad2}
-{1\over 2}\, i \, \NN\, e^{i\omega R} \, \sum_{a=1}^n (1 + i k.r_{(a)}) \, {1\over k.p_{(a)}} \, k^\gamma
\left\{ p_{(a)\alpha} \Sigma_{(a)\beta\gamma} + p_{(a)\beta} \Sigma_{(a)\alpha\gamma}\right\}\, .
\ee
Adding \refb{eadad1} and \refb{eadad2} and contracting the result with the polarization tensor
$\ve^{\alpha\beta}$, we get
\be
\ve^{\alpha\beta}\Delta \tilde e_{\alpha\beta}
= \NN\, e^{i\omega R} \, \Delta \bar S_{\rm gr}\, ,
\ee
where
\be \label{edefdbs}
\Delta \bar S_{\rm gr}=-i \sum_{a=1}^n {1\over k.p_{(a)}} \ve^{\alpha\beta}\, 
k^\gamma \, p_{(a)\alpha}\, \Sigma_{(a)\beta\gamma}
+ \sum_{a=1}^n {1\over k.p_{(a)}}  \ve^{\alpha\beta} \, k^\gamma \, \left\{
k.r_{(a)}\, p_{(a)\alpha} - k.p_{(a)}\, r_{(a)\alpha}\right\}
\, \Sigma_{(a)\beta\gamma}\, .
\ee

We shall now compare this to the change in the soft factor when we replace $\bJ^{\mu\nu}_{(a)}$
in \refb{ejamn} by 
\be \label{ejamnchange}
\bJ_{(a)}^{\mu\nu} = r_{(a)}^\mu p_{(a)}^\nu - r_{(a)}^\nu p_{(a)}^\mu + \Sigma^{\asnote{\mu\nu }}_{(a)}\, ,
\ee
and collect terms that are linear in $\Sigma^{\mu\nu}$. Using \refb{esublead0} and \refb{esoftgr2}, 
we get
\be \label{edda}
\Delta S_{\rm gr} = i\sum_{a=1}^n {\ve_{\mu\nu} p_{(a)}^\mu k_\rho \over p_{(a)}. k} 
\Sigma_{(a)}^{\rho\nu} - \sum_{a=1}^n (p_{(a)}. k)^{-1}   
\ve_{\mu\rho} k_\nu k_\sigma \{ r_{(a)}^\mu p_{(a)}^\nu - r_{(a)}^\nu p_{(a)}^\mu\}
\Sigma_{(a)}^{\rho\sigma}\, .
\ee
Using the antisymmetry of $\Sigma_{(a)\beta\gamma}$ one can easily verify that \refb{edefdbs}
and \refb{edda} are equal. This establishes the validity of the soft theorem to subsubleading order for
terms linear in $\Sigma_{(a)\beta\gamma}$. As already argued before, we do not expect this relation
to hold for terms quadratic in the spin, since their contribution is of the same order as the 
non-universal terms.

\subsubsection{Probe scattering from a charged scatterer} \label{sem}

The examples in sections \ref{sfusiona} and \ref{smulti}  
involve scattering in which the stress tensor -- the source
of gravity wave -- is localized on a one dimensional subspace of space-time. 
We shall now consider the case of scattering
of a probe of charge $q$ by a  large mass scatterer of charge $Q$. We shall assume that the probe
is sufficiently light so that we can ignore the gravitational force compared to the electromagnetic force.
We shall furthermore assume that there are two $U(1)$ gauge fields, one of them with wrong sign
kinetic term, and that the charges $q$ and $Q$ are both null vectors in this two dimensional charge 
space so that $q^2$ and $Q^2$ vanish but $q\cdot Q\ne 0$.
Such a theory of course is not fully consistent, but it can be regarded as a toy model for checking the
soft graviton theorem. In this case the stress tensor for the fields produced by either the probe or
the scatterer vanish, but the cross term survives. 
We emphasize that while this choice of charges simplifies our analysis, we expect soft theorem to work
even without these assumptions.

We shall now compute the radiative component of the gravitational field during the scattering
to \asnote{subsubleading} order in the soft momentum. It follows
from the same analysis that led to  \refb{erad}, \refb{edefealbe} that we have
\be\label{econser}
e_{\alpha\beta}\asnote{(x) } = -\int d^D x' \, G_r(x,x') \, T_{\alpha\beta}(x')\, ,
\ee
where $T_{\alpha\beta}$ is the total stress tensor. 
We shall first focus on the computation of the spatial components $e_{ij}$ for $1\le i,j
\le (D-1)$ for which the source is $-T_{ij}$. The contribution to $T_{ij}$ due to the  scatterer vanishes
in the infinite mass limit so that
$T_{ij}$ is
given by a sum of two terms -- one due to the probe and the other due
to the stress energy tensor of the electromagnetic field. We shall denote the corresponding contributions
to $e_{\ia\ja}$ by $e^1_{\ia\ja}$ and $e^2_{\ia\ja}$ respectively. 
The time Fourier transform of $e^1_{\ia\ja}$, denoted by $\tilde e^1_{\ia\ja}$, 
is given by the same formula as \refb{ehababnew} where the sum over $a$ now runs 
over only the initial and the final state of the probe:
\be
\tilde e^1_{\ia\ja} = \NN\, e^{i\omega R} \, \sum_{a=1}^2 \, e^{ik.r_{(a)}} {p_{(a)\ia} p_{(a)\ja}\over k.p_{(a)}}
+ i \, \NN\, e^{i\omega R} \, \int d\sigma \, e^{ik.r(\sigma)} \, P_\ia\, V_\ja\, .
\ee
\asnote{Here the integral over $\sigma$ }runs over the whole trajectory from $r(\sigma)=r_{(1)}$ to
$r(\sigma)=r_{(2)}$. We shall expand $e^{ik.r(\sigma)}$ in power series in $k$ and work up to 
subsubleading order. 
After writing $V=dr/d\sigma$,  integrating the second term by parts 
and taking into account the fact that for the outgoing particle $P_{(2)}=-p_{(2)}$,
we get, up to subsubleading order in the expansion in powers of $k$,
\ben\label{epartpartpre}
\tilde e^1_{\ia\ja} &=& \NN\, e^{i\omega R} \, \sum_{a=1}^2 \, \Bigg[\left\{ 1 + i k. r_{(a)}
-{1\over 2} (k.r_{(a)})^2\right\}
{p_{(a)\ia} p_{(a)\ja}
\over k.p_{(a)}} \nonumber \\ &&
- {i\over 2}  \sum_{a=1}^2 \, \{ p_{(a)\ia} r_{(a)\ja}  + p_{(a)\ja} r_{(a)\ia}\} (1 + ik.r_{(a)})
\nonumber \\
&& - {i\over 2} \, 
\int d\sigma \,  \left\{ {d P_\ia\over d\sigma} \, r_\ja(\sigma)
+ {d P_\ja\over d\sigma} \, r_\ia(\sigma)\right\} \{1+ik.r(\sigma)\} 
\nonumber \\ && 
+ {1\over 2} \, 
\int d\sigma \, k.V(\sigma) \, \left\{ P_\ia \, r_\ja + P_\ja \, r_\ia\right\} \Bigg]\, .
\een
Now using the fact that $P$ and $V$ are proportional to each other, the last term may be
manipulated as
\ben
&&{1\over 2} \int d\sigma \, k.V(\sigma) \, \left\{ P_\ia \, r_\ja + P_\ja r_\ia\right\}
= {1\over 2} \int d\sigma \, k.P(\sigma) \, \left\{ V_\ia \, r_\ja + V_\ja r_\ia\right\}
\nonumber \\ &=& 
{1\over 2} \int d\sigma \, k.P(\sigma) \, {d\over d\sigma} \left\{ r_\ia \, r_\ja \right\}
=-{1\over 2} \sum_{a=1}^2 \, k.p_{(a)} \, r_{(a)\ia} \, r_{(a)\ja}
- {1\over 2} \int d\sigma \, k.{dP(\sigma)\over d\sigma} \, \left\{ r_{\ia} \, r_{\ja} \right\}\, ,
\nonumber \\
\een
where in the last step we have carried out an integration by parts. With this \refb{epartpartpre}
can be written as
\ben\label{epartpart}
\tilde e^1_{\ia\ja} &=& \NN\, e^{i\omega R} \, \sum_{a=1}^2 \, \Bigg[\left\{ 1 + i k. r_{(a)}
-{1\over 2} (k.r_{(a)})^2\right\}
{p_{(a)\ia} p_{(a)\ja}
\over k.p_{(a)}} \nonumber \\ &&
- {i\over 2}  \sum_{a=1}^2 \, \{ p_{(a)\ia} r_{(a)\ja}  + p_{(a)\ja} r_{(a)\ia}\} (1 + ik.r_{(a)})
-{1\over 2} \sum_{a=1}^2 \, k.p_{(a)} \, r_{(a)\ia} \, r_{(a)\ja}
\nonumber \\
&& \hskip -.2in
- {i\over 2} \, 
\int d\sigma \,  \left\{ {d P_\ia\over d\sigma} \, r_\ja(\sigma)
+ {d P_\ja\over d\sigma} \, r_\ia(\sigma)\right\} \{1+ik.r(\sigma)\} 
- {1\over 2} \int d\sigma \, k.{dP(\sigma)\over d\sigma} \, \left\{ r_{\ia}(\sigma) \, 
r_{\ja}(\sigma) \right\}
\Bigg]\, .\nonumber \\ 
\een

On the other hand,
$e^2_{\ia\ja}$ is given by
\be\label{e766}
e^2_{\ia\ja} =  -\int d^D x' \, 
G_r(x,x')\, \left[F_{\ia \rho}(x') \cdot F_{\ja}^{~\rho}(x') -
{1\over 4} \, \delta_{\ia\ja} \, F_{\rho\sigma}(x') \cdot F^{\rho\sigma}(x')\right]\, ,
\ee
where $\cdot$ denotes the inner product in the two dimensional space labelling the two $U(1)$ gauge
fields. Manipulating this expression in the same way that led from \refb{edefaalpha} to 
\refb{ereturn} we see that the
radiative part of \refb{e766} is given by
\be\label{epart3}
\tilde e^2_{\ia\ja}
= i\, \NN\, e^{i\omega R}  \int d^D x'  
\, e^{ik.x'} \,  \left[F_{\ia \rho}(x') \cdot F_{\ja}^{~\rho}(x') -
{1\over 4} \, \delta_{\ia\ja} \, F_{\rho\sigma}(x') \cdot F^{\rho\sigma}(x')\right]\, .
\ee
Now $F_{\mu\nu}$ can be written as the sum of the gauge field strength 
$F^S_{\mu\nu}$ produced by the
scatterer and the gauge field strength $F^P_{\mu\nu}$ produced by the probe.
Using the fact that the charges carried by the probe and the scatterers are null, we can
express \refb{epart3} as
\be\label{epart2}
\tilde e^2_{\ia\ja}
= i\, \NN\, e^{i\omega R}  \int d^D x'  
\,  e^{ik.x'} \, \left[F^P_{\ia \rho}(x') \cdot {F^S}_{\ja}^{~\rho}(x') 
+  F^S_{\ia \rho}(x') \cdot {F^P}_{\ja}^{~\rho}(x') 
-
{1\over 2} \, \delta_{\ia\ja} \, F^S_{\rho\sigma}(x') 
\cdot F^{P\rho\sigma}(x')\right]\, .
\ee

Now the prediction of soft theorem for $\tilde e_{ij}$ according to \refb{es00}-\refb{ehabexp}
(with $\bJ_{ij}=0$)
is
\ben \label{e767}
&& \NN \,  e^{i\omega R}\,\Bigg[
\sum_{a=1}^2 {p_{(a)i} p_{(a)j}\over p_{(a)}.k} + {i\over 2} 
\sum_{a=1}^2 {1\over p_{(a)}.k} \left\{p_{(a)i} k_\rho (\bj_{(a)})^\rho_{~j} 
+ p_{(a)j} k_\rho (\bj_{(a)})^\rho_{~i}\right\} \nonumber \\ &&
\hskip 1in -{1\over 2} \sum_{a=1}^2 \, (p_{(a)}.k)^{-1} \, \ve_{ij}\, k_\nu k_\sigma \, 
\bj_{(a)}^{i\nu} \bj_{(a)}^{j\sigma}
\Bigg]
\, .
\een
Using the result $\bj_{(a)}^{\mu\nu} = r_{(a)}^\mu p_{(a)}^\nu -  r_{(a)}^\nu p_{(a)}^\mu$ 
it is easy to check that the
terms in the first two lines on the 
right hand side of \refb{epartpart} produce the desired contribution \refb{e767}. Therefore
we need to show that the sum 
of \refb{epart2} and the term in the third line of the right hand side of
\refb{epartpart} vanishes. 

\asnote{Let }us denote by $I_{ij}$ the contribution to $\tilde e^1_{ij}$ from the 
third line of the right hand side of
\refb{epartpart}.
Using the equations of motion of the probe\footnote{The sign in \refb{efs1} is convention
dependent. It has been chosen so as to be consistent with \refb{emaxwell}.}
\be \label{efs1}
{dP_\alpha\over d\sigma} = q \cdot F^S_{\alpha\rho}(r(\sigma)) \, {d r^\rho\over d\sigma}\, ,
\ee
and the fact that the only non-vanishing components of the Coulomb field of the scatterer are
$F^S_{i0}=- F^S_{0i}$, 
we can express $I_{ij}$ as
\ben \label{eAij}
I_{ij} &=&  -{i\over 2}  \, \NN\, e^{i\omega R} \, 
\int d\sigma \, \left\{
q \cdot F^S_{i0}(r(\sigma)) \, {d r^0\over d\sigma} \, r_j(\sigma) 
+ q \cdot F^S_{j0}(r(\sigma)) \, {d r^0\over d\sigma} \, r_i(\sigma) \right\} \{1+ik.r(\sigma)\}
\nonumber \\ &&
- {1\over 2} \, \NN\, e^{i\omega R}\,
\int d\sigma \, \left\{ k^0 \, q \cdot F^S_{0m}(r(\sigma)) 
\, {d r^m\over d\sigma} 
+ k^m \, q \cdot F^S_{m0}(r(\sigma)) 
\, {d r^0\over d\sigma} \right\}\, \left\{ r_i(\sigma) \, r_j(\sigma) \right\}\, . \nonumber \\
\een
Let us denote by $\wt F_{\mu\nu}(\ell)$ the Fourier transform of $F_{\mu\nu}$ in all variables:
\be \label{effour}
\wt F_{\mu\nu}(\ell) = \int d^D x \, e^{-i\ell.x} \, F_{\mu\nu}(x)\, .
\ee
Then \refb{eAij} may be expressed as
\ben \label{e770}
&& \hskip -.3in I_{ij} =  -{i\over 2}  \, \NN\, e^{i\omega R} \, 
\int d\sigma \, \int {d^D \ell \over (2\pi)^D} \, 
e^{i\ell . r(\sigma)} \, {d r^0\over d\sigma} \, 
\left\{
q \cdot \wt F^S_{i0}(\ell) \, r_j(\sigma) 
+ q \cdot \wt F^S_{j0}(\ell)  \, r_i(\sigma) \right\} \{1+ik.r(\sigma)\}
\nonumber \\ &&
- {1\over 2} \, \NN\, e^{i\omega R}\,
\int d\sigma \, \int {d^D \ell \over (2\pi)^D} \, 
e^{i\ell . r(\sigma)} \, \left\{ k^0 \, q \cdot \wt F^S_{0m}(\ell) 
\, {d r^m\over d\sigma} 
+ k^m \, q \cdot \wt F^S_{m0}(\ell) 
\, {d r^0\over d\sigma} \right\}\, \left\{ r_i(\sigma) \, r_j(\sigma) \right\}\, . \nonumber \\
\een
Now using the Euclidean Green's function in $(D-1)$ dimension we see that the Fourier 
transform of the Coulomb field $\wt A^0$ produced by a static source of charge $Q$ at the
origin is given by
\be 
\wt A^0 = {Q\over \vec \ell^2} \, 2\pi \, \delta(\ell^0)\, ,
\ee
leading to 
\be \label{e775a}
\wt F^S_{i0}(\ell) = -i\, \ell_i\, Q \, {1\over \vec \ell^2} \, 2\pi \, \delta(\ell^0)\, .
\ee
\asnote{Substituting \refb{e775a} into \refb{e770} we get
\ben \label{e773}
I_{ij} &=&  {i\over 2}  \, \NN\, e^{i\omega R} \, 
\int d\sigma \, \int {d^{D-1} \ell \over (2\pi)^{D-1}} \, 
e^{i\vec\ell . \vec r(\sigma)} \, {1\over \vec \ell^2}\, q \cdot Q \, {d r^0\over d\sigma} \, 
\left\{
i\, \ell_i \, r_j(\sigma) 
+  i \, \ell_j \, r_i(\sigma) \right\} \{1+ik.r(\sigma)\}
\nonumber \\  && \hskip -.5in
+ {i\over 2} \, \NN\, e^{i\omega R}\,
\int d\sigma \, \int {d^{D-1}\ell\over (2\pi)^{D-1}} \, e^{i\vec\ell. \vec r(\sigma)} \, {1\over \vec\ell^2}\, q \cdot Q \, 
\left\{ -k^0  \, \ell_m 
\, {d r^m\over d\sigma} 
+ \vec k.\vec\ell  \,  {d r^0\over d\sigma} \right\}\, \left\{ r_i(\sigma) \, r_j(\sigma) \right\}\, . \nonumber \\
\een }

\asnote{We now turn to the analysis of $\tilde e^2_{ij}$ given in \refb{epart2}. We have
\ben \label{ee2def}
\tilde e^2_{ij} &=&  i \, \NN\, e^{i\omega R}\int d^D x'  \, e^{ik.x'}
\,  \left[- F^P_{i 0}(x') \cdot F^S_{j0}(x') 
-  F^P_{j 0}(x') \cdot F^S_{i0}(x') 
+
 \delta_{ij} \, F^S_{k0}(x') 
\cdot F^{P}_{k0}(x')\right]\, . \nonumber \\
\een
It follows from }\refb{edefaalpha}, \refb{eretard} 
that the Fourier transform $\wt F^P_{i0}$ of the
field generated by the probe is given by
\be \label{e776}
\wt F^P_{i0}(-\ell-k) =-q\, 
{1\over (\ell^0+k^0-i\eps)^2 - (\vec \ell+\vec k)^2} \, \int d\sigma \, e^{i (\ell+k). r(\sigma)}
\left\{ -i \, (\ell_i+k_i) \, {dr_0\over d\sigma} + i \, (\ell_0+k_0) \, {dr_i\over d\sigma} \right\}\, .
\ee
Using \refb{effour}, \refb{e775a}, \refb{e776}, \asnote{$\tilde e^2_{ij}$ given in \refb{ee2def}  }may 
be expressed as (to subsubleading order)
\ben \label{exx77}
&& \hskip -.3in 
\tilde e^2_{ij} = i \, \NN\, e^{i\omega R}\int {d^D\ell\over (2\pi)^D} 
\,  \Bigg[- \wt F^P_{i 0}(-\ell-k) \cdot \wt F^S_{j0}(\ell) 
-  \wt F^P_{j 0}(-\ell-k) \cdot \wt F^S_{i0}(\ell) 
+
 \delta_{ij} \, \wt F^P_{k0}(-\ell-k) 
\cdot \wt F^{S}_{k0}(\ell)\Bigg]
\nonumber \\
&& \hskip -.3in 
= i \, \NN\, e^{i\omega R} \int d\sigma \int {d^{D-1}\ell\over (2\pi)^{D-1}} \, e^{i \vec \ell.
\vec r(\sigma) + i k . r(\sigma)}
\,  q\cdot Q\, {1\over (\vec\ell^2)(\vec\ell^2 + 2\vec\ell.\vec k)} 
\nonumber \\ && 
\left[\left\{2\, \ell_i \ell_j +\ell_i k_j + \ell_j k_i 
-(\vec\ell^2 +\vec \ell . \vec k) \, \delta_{ij}\right\} \, 
{d r_0\over d\sigma} + \left\{ - k_0 \ell_j {d r_i\over d\sigma} - k_0 \, \ell_i \, {d r_j\over d\sigma} 
+ k_0 \, \ell_m \, {d r_m\over d\sigma} \, \delta_{ij}
\right\}
\right]\nonumber \\
&& \hskip -.3in 
= i \, \NN\, e^{i\omega R} \int d\sigma \int {d^{D-1}\ell\over (2\pi)^{D-1}} \, e^{i \vec \ell.
\vec r(\sigma)}
\,  q\cdot Q\, {1\over (\vec\ell^2)^2 }
\{2\, \ell_i \ell_j 
-\vec\ell^2  \, \delta_{ij}\} \, 
{d r_0\over d\sigma} \left\{ 1 + i  k.  r(\sigma) \right\} \nonumber \\
&& \hskip -.3in + i \, \NN\, e^{i\omega R} \int d\sigma \int {d^{D-1}\ell\over (2\pi)^{D-1}} \, e^{i\vec \ell.
\vec r(\sigma)}
\,  q\cdot Q\, {1\over (\vec\ell^2)^3}
\left\{-4\, \ell_i \ell_j \, \vec\ell . \vec k + \vec\ell^2 (\ell_i k_j + \ell_j k_i) 
+ \vec \ell. \vec k\, \vec\ell^2  \, \delta_{ij}\right\} \, 
{d r_0\over d\sigma}
\nonumber \\ 
&& \hskip -.3in + i \, \NN\, e^{i\omega R} \int d\sigma \int {d^{D-1}\ell\over (2\pi)^{D-1}} \, e^{i\vec \ell.
\vec r(\sigma)}
\,  q\cdot Q\, {1\over (\vec\ell^2)^2}
\, \left\{ - k_0 \ell_j {d r_i\over d\sigma} - k_0 \, \ell_i \, {d r_j\over d\sigma} 
+ k_0 \, \ell_m \, {d r_m\over d\sigma} \, \delta_{ij}
\right\}
\, . 
\een
The integrand of the last term proportional to $\delta_{ij}$ is proportional to
$
{d\over d\sigma} e^{i\vec\ell.\vec r(\sigma)}
$
and vanishes by integration by parts. In this case there is no boundary terms since the 
$e^{i\vec\ell.\vec r(\sigma)}$ is oscillatory and as a result the boundary terms vanish after integration over
$\vec\ell$.  We now use the relations
\ben
&&{1\over (\vec\ell^2)^2} \{2\, \ell_i \ell_j -\vec\ell^2 \, \delta_{ij}\}
= -{1\over 2} \left[{\p \over \p \ell_i} \left({\ell_j\over \vec\ell^2}\right) 
+ {\p \over \p \ell_j} \left({\ell_i\over \vec\ell^2}\right)
\right], 
\quad {\ell_m\over (\vec\ell^2)^2} = -{1\over 2} \, {\p\over \p\ell_m} \, \left({1\over \vec \ell^2}\right)\, , 
\nonumber \\
&& {1\over 2} {\p^2\over \p \ell_i\p\ell_j} \left( {\vec\ell.\vec k \over \vec\ell^2}\right)
=
{1\over (\vec\ell^2)^3} \left\{ 4\ell_i \ell_j \, \vec\ell.\vec k 
- (k_i\, \ell_j+k_j\, \ell_i)\, \vec\ell^2- \delta_{ij} \, \vec\ell^2 \, \vec\ell.\vec k 
\right\}\, ,
\een
and integration by parts, to express \refb{exx77} as
\ben\label{elast1}
&& \hskip -.3in 
\tilde e^2_{ij} ={i\over 2} \, \NN\, e^{i\omega R} \int d\sigma \int {d^{D-1}\ell\over (2\pi)^{D-1}} \, e^{i \vec \ell.
\vec r(\sigma)}
\,  q\cdot Q\, {1\over \vec\ell^2 }
\{i\ell_i r_j + i\ell_j r_i\} \, 
{d r_0\over d\sigma} \left\{ 1 + i k. r(\sigma)\right\} \nonumber \\
&& + {i\over 2} \, \NN\, e^{i\omega R} \int d\sigma \int {d^{D-1}\ell\over (2\pi)^{D-1}} \, e^{i\vec \ell.
\vec r(\sigma)}
\,  q\cdot Q\, {1\over \vec\ell^2} \,
r_i(\sigma) \, r_j(\sigma) \, 
{d r_0\over d\sigma} \, \vec\ell.\vec k \nonumber \\
&& + {i\over 2} \, \NN\, e^{i\omega R} \int d\sigma \int {d^{D-1}\ell\over (2\pi)^{D-1}} \, e^{i\vec \ell.
\vec r(\sigma)}
\,  q\cdot Q\, {1\over \vec\ell^2} \left\{- i \, k_0 \, r_j {d r_i\over d\sigma}- i \, k_0 \, r_i {d r_j\over d\sigma}
\right\} \, .
\een
The terns in the last line can be combined into a single derivative with respect to $\sigma$. Integrating
by parts over the $\sigma$ variable, we can bring \refb{elast1} to the form
\ben\label{elast2}
&& \hskip -.3in 
\tilde e^2_{ij} ={i\over 2} \, \NN\, e^{i\omega R} \int d\sigma \int {d^{D-1}\ell\over (2\pi)^{D-1}} \, e^{i \vec \ell.
\vec r(\sigma)}
\,  q\cdot Q\, {1\over \vec\ell^2 }
\{i\ell_i r_j + i\ell_j r_i\} \, 
{d r_0\over d\sigma} \left\{ 1 + i k. r(\sigma)\right\} \nonumber \\
&& + {i\over 2} \, \NN\, e^{i\omega R} \int d\sigma \int {d^{D-1}\ell\over (2\pi)^{D-1}} \, e^{i\vec \ell.
\vec r(\sigma)}
\,  q\cdot Q\, {1\over \vec\ell^2} \, 
r_i(\sigma) \, r_j(\sigma) \, 
{d r_0\over d\sigma} \, \vec k.\vec\ell  \nonumber \\
&& + {i\over 2} \, \NN\, e^{i\omega R} \int d\sigma \int {d^{D-1}\ell\over (2\pi)^{D-1}} \, e^{i\vec \ell.
\vec r(\sigma)}
\,  q\cdot Q\, {1\over \vec\ell^2} \left\{-  k_0 \, \ell_m {d r_m\over d\sigma} \,  r_i(\sigma) \, r_j(\sigma)
\right\} \, .
\een

Eq.\refb{e773}, \refb{elast2} and the relations $r_0=-r^0$, $k_0=-k^0$ now give
\be 
I_{ij} + \tilde e^2_{ij}=0\, .
\ee
This shows that all the extra terms cancel and 
that $\tilde e_{ij}$ has the correct form as given in \refb{e767}.

Now it follows from \refb{econser} and the conservation of the total stress tensor that 
$\p^\mu e_{\mu\nu}=0$. 
Using this condition we get
\be \label{e778}
\tilde{e}_{00}(\omega,\vec{k})\ =\ \frac{k^{i}k^{j}}{k_{0}^{2}}\tilde e_{ij}, \qquad
\tilde{e}_{0i}(\omega,\vec{k})\ =\ -\frac{k^{j}}{k^{0}}\tilde e_{ij}\, .
\ee
Using the form \refb{e767} for $\tilde e_{ij}$ we now get
\be \label{e779}
\ve^{\mu\nu} \tilde e_{\mu\nu} 
= \NN \, e^{i\omega R} \, \left\{ S^{(0)}_{\rm gr}(\ve, k) + S^{(1)}_{\rm gr}(\ve, k)+ S^{(2)}_{\rm gr}
(\ve, k)\right\}\, ,
\ee
with $S^{(0)}_{\rm gr}$, $S^{(1)}_{\rm gr}$ and $S^{(2)}_{\rm gr}$ 
given by \refb{es00}, \refb{esubleading0} and \refb{esubsubleading0} (with
$\bJ_{ij}=0$) respectively.
It now follows from the same analysis that led from \refb{e732} to 
\refb{e738} that the angular power
spectrum of soft graviton radiation emitted during this process agrees with the prediction of
soft graviton theorem to subsubleading order.

\asnote{The next two examples will test soft theorem only to leading order, but have been included
since they take into account the effect of the gravitational field, produced by the macroscopic objects
involved in the scattering, on each other. }

\subsubsection{Probe scattering from Schwarzschild black hole}

Ref.\cite{peter} considered the 
scattering of a 
probe particle of mass $m$ in the background geometry produced by 
a Schwarzschild blackhole of mass $M_0>>m$ in the limit where 
the impact parameter $b$ is much larger then the Schwarzschild radius $a$ of the blackhole. 
In particular 
\cite{peter} gives the expression for the time Fourier transform of 
the radiative  component of the metric fluctuation
$h_{\alpha\beta}$. The computation was more sophisticated than the one given in 
\cite{weinbergbook} (as reported at the beginning of section \ref{sfusiona}) 
in that it took into account the effect of Schwarzschild geometry on the probe particle. 
Therefore we can in principle compare the small $\omega$ expansion of the results of 
\cite{peter} with our results. Unfortunately since the computation was done in four space-time 
dimensions, the soft expansion suffers from infrared divergences as discussed in section
\ref{sir}. For this reason we have to limit our comparison to the 
leading order term in the expansion
which is not affected by infrared divergences.\footnote{Although there are some results in
higher dimensions (see {\it e.g.} \cite{0212168,0309203}), 
we do not have enough analytic results to compare them with the 
predictions of soft theorem.}

Starting point of our analysis is eqn.(2.11) of \cite{peter} which gives the expression for 
the spatial component of $\tilde e_{\alpha\beta}\equiv \tilde h_{\alpha\beta}-{1\over 2}
\eta_{\alpha\beta} \tilde h^\rho_{~\rho}$ (called $\tilde h_{\alpha\beta}$ in \cite{peter}).
There is a difference in the normalization factor of 2 between the our definition of
$h_{\alpha\beta}$ (which takes the metric to be $\eta_{\alpha\beta}+2\, h_{\alpha\beta}$) 
and the definition used in \cite{peter} (which takes the metric to be 
$\eta_{\alpha\beta}+h_{\alpha\beta}$). There is also an overall sign difference since 
\cite{peter} uses metric with mostly minus signature whereas we are using metric with mostly plus
signature. Taking these into account and setting $8\pi G=1$,
the result of eq.(2.11), (2.12)
of \cite{peter} \asnote{for large $|\vec x|$ }may
be expressed as 
\be \label{epet0}
\tilde e_{ij}(\vec x, \omega) =\tilde e^{(1)}_{ij}(\vec x, \omega) +\tilde e^{(2)}_{ij}(\vec x, \omega) 
+ \hbox{terms subleading in $\omega\to 0$ limit}\, ,
\ee
where
\be \label{epet1}
\tilde e^{(1)}_{ij}(\vec x, \omega) 
= {m\over 4\pi (1+2\vp)} \int {dt'\over 1+2\vp'} \, {dt'\over ds}\, v_i v_j \, {e^{i\omega (t'+R')}
\over R'} \, ,
\ee
and 
\ben \label{epet2}
\tilde e^{(2)}_{ij}(\vec x, \omega) &=& i\, {M_0 m\over 32\, \pi^2\omega} {e^{i\omega R}\over R}
\int dt' \, {dt'\over ds}\, (1+\vec v^2) \, \left(\p'_i\p'_j -{1\over 2} \delta_{ij} \, \p'_k\p'_k\right)\, 
\bigg\{ \ln (r'+\hat n.\vec r^{\, \prime}) \, e^{i\omega (t' - \hat n.\vec r^{\, \prime})} \nonumber \\ &&
\hskip 1in+ \int_{r'+\hat n.\vec r^{\, \prime}}^\infty {du\over u}
e^{i\omega (t' - \hat n.\vec r^{\, \prime}+u)} 
\bigg\}\bigg|_{\vec r^{\, \prime}=\vec x\, '(t')}\, .
\een
Here  $\vec x\, '(t')$ denotes the trajectory of the
particle, $\vp=-h_{00}(\vec x)$, $\vp'=-h_{00}(\vec x\,'\asnote{(t') })$, $s$ is the proper time along the
trajectory, $v_i = dx'_i/dt'$ is the velocity along the trajectory, $R'=|\vec x-\vec x\, '(t')|$,
$R=|\vec x|$, $r'=|\vec r^{\, \prime}|$ 
and $\p'_i \equiv \p/\p r_i^{\, \prime}$.
For large $|\vec x|=R$, we can approximate $R'$ by $R - \hat n.\vec x\, '$ where $\hat n$
denotes unit vector along $\vec x$. Also we can set $\vp(\vec x)$ to 0.

Let us first evaluate $\tilde e^{(1)}_{ij}$.
Since as $t'\to\pm\infty$, $v_i$ approaches a constant
value, $dt'/ds=(1-\vec v^2)^{-1/2}$ also approaches a constant value, and $\vp'=-h_{00}(\vec x\, '(t'))$
approaches zero, we see that the integrand in \refb{epet1} is oscillatory. Therefore we need to define
this by adding boundary terms as in going from \refb{ereturn} to \refb{eafin}. 
Using the fact that leading term in
$\vec x\, '(t')$ \asnote{for large $t'$ }is $\vec v \, t'$ we see that the required boundary term is
\be \label{epetsing}
- {m\over 4\pi} \, \, {e^{i\omega R} \over R}  \, 
{v_i v_j \over \sqrt{1-\vec v^2}} \, {1\over  i\omega (1-\hat n.\vec v)}
\, e^{i\omega t'(1 - \hat n.\vec v)}\bigg|_{t'=t_-}^{t'=t_+}  \, .
\ee
If $p_\pm$ denote the momenta of the particle at $t_\pm$, both counted as positive for ingoing
momenta, then we have 
\be 
p_\pm = \mp\, m (1, \vec v_\pm)  / \sqrt{1-\vec v_\pm^2}\, .
\ee
Also denoting by $k = -\omega(1,\hat n)$ the momentum of the outgoing gravitons, we get
\be
k. p_\pm = \omega\, p^0_\pm (1-\hat n.\vec v_\pm)\, .
\ee
Using these results we
can now extract the $1/\omega$ term in \refb{epetsing} which is the only source of 
$1/\omega$ contribution
to $\tilde e^{(1)}_{ij}$:
\be 
\tilde e^{(1)}_{ij} = {1\over 4\pi i R}
\, e^{i\omega R} \, \sum_{a=\pm} {p_{ai} p_{aj}\over k.p_a} + \hbox{subleading}= \NN\, e^{i\omega R} \, \sum_{a=\pm} {p_{ai} p_{aj}\over k.p_a} + \hbox{subleading}\, ,
\ee
where in the second step we have used the fact that in $D=4$ the normalization factor $\NN$
given in \refb{eNvalue} is $1/ (4\pi i R)$.  

Next we turn to the computation of $\tilde e^{(2)}_{ij}$ \asnote{given in \refb{epet2}. }Since 
the derivatives $\p'_k$ acting on the 
exponential factors will bring down factors of $\omega$ making the contribution subleading, we 
consider those terms in which the derivatives act on the other $\vec r\, '$ dependent factors. Now we
have
\ben
&& \p_k' \bigg\{ \ln (r'+\hat n.\vec r^{\, \prime}) \, e^{i\omega (t' - \hat n.\vec r^{\, \prime})} 
+ \int_{r'+\hat n.\vec r^{\, \prime}}^\infty {du\over u}
e^{i\omega (t' - \hat n.\vec r^{\, \prime}+u)} 
\bigg\}\nonumber \\
&=& \{\p_k' (r'+\hat n.\vec r^{\, \prime})\} {1\over (r'+\hat n.\vec r^{\, \prime})}
\left\{e^{i\omega (t' - \hat n.\vec r^{\, \prime})}  - e^{i\omega (t'+ r^{\prime})} 
\right\} = \OO(\omega)\, .
\een
This shows that the contribution from $\tilde e^{(2)}_{ij}$ is subleading and $\tilde e^{(1)}_{ij}$ 
represents the full contribution to $\tilde e_{ij}$.

This determines the spatial components of the metric. The $e_{0i}$ and $e_{00}$ components
are determined by recalling that the constraint equations on the metric to linear order are given by
$\p^\mu e_{\mu\nu}=0$. In momentum space this translates to
\be 
k^\mu \tilde e_{\mu\nu}=0\, .
\ee
We can now use the arguments given in \refb{e778}, \refb{e779} to conclude that
\be \label{epetfin}
\phi(\ve)\equiv \ve^{\mu\nu} \tilde e_{\mu\nu} = \NN\, e^{i\omega R}\, S^{(0)}_{\rm gr}(\ve, k)\, ,
\ee
with $S^{(0)}_{\rm gr}$ given in \refb{es00}. The same analysis that led to the conclusion that 
\refb{e732} is consistent with the power spectrum produced according to the soft photon theorem
now tells us that \refb{epetfin} is consistent with the power spectrum produced by the 
leading soft graviton theorem.

\subsubsection{Scattering on the brane-world}

In \cite{1210.6976} the authors analyzed gravitational bremsstrahlung during ultra-Planckian 
collision of two massive particles localized on a $d$ dimensional brane, which in turn is
embedded in $D$ dimensional space-time. For definiteness we shall take the extra $(D-d)$ 
dimensions to be non-compact. The impact parameter $b$ is assumed to be larger than the 
Schrwarzchild radius associated to the 
center of mass energy of the particles. 
In this limit the deflection in the particle trajectories are small. The authors 
consider an iterative scheme where beginning with free inertial trajectories 
of the two particles localized on the brane, they compute the first order perturbation $h^{(1)}$ to the 
$D$-dimensional (flat) space-time metric. This is  then used to compute the corrected trajectories of 
the particles and subsequently the correction to the corresponding stress tensors. They then compute 
the second order perturbations to the metric by considering the corrected stress tensor of the two particles 
and the stress tensor of the gravitational field $h^{(1)}$. These formulae lead them to energy spectrum 
of gravitational radiation in any (as opposed to low frequency) bin. To this order in perturbation theory, 
the authors show that it is consistent to restrict the particle trajectories to lie on the brane. 

We shall show that the results of \cite{1210.6976} to leading order in the frequency of 
emitted radiation are in agreement with
the leading soft graviton theorem. 
For this we need to determine $\tilde{e}_{\alpha\beta}(\omega,\vec{x})$ to leading order 
in $\omega$. 
Now  the stress tensor for the two gravitating particles is given in eq.~(2.25) of
\cite{1210.6976} as
\ben
T_{\alpha\beta}(x)\ =\ T^{(0)}_{\alpha\beta}(x)\ +\ T^{(1)}_{\alpha\beta}(x)
\een
where
\ben\label{galstov1}
T^{(0)}_{\alpha\beta}(x)\ &=&\ \sum_{i=1}^{2}\ \int\ d\sigma P_{(i) \alpha} V_{(i) \beta} \,
\delta^{(D)}(x\ -\ r^{(0)}(\sigma))\nonumber\\
T^{(1)}_{\alpha\beta}(x)\ &=&\ \sum_{i=1}^{2} \int\ d\sigma\ \bigg[\ 2(\delta V)_{(i) (\alpha}
(\sigma)P_{(i) \beta)}
\ +\ m_{(i)}\, {\cal F}^{\mu\nu}_{(i) \alpha\beta}\ h_{(i)\mu\nu}(r^{(0)}(\sigma))\ 
\nonumber \\
&& \hskip 1in -\ P_{(i) \alpha}
V_{(i) \beta} V^{(i) \mu}\partial_{\mu}\bigg]\delta^{(D)}(x\ -\ r^{(0)}(\sigma)) + S_{\alpha\beta}\, .
\een
Here $V_{(i)}$ and $P_{(i)}$ are intial velocities and momenta of the two particles, 
$m_{(i)}$ are their masses 
and $\delta V_{(i)\alpha}(\sigma)$ are the corrections to the velocities as a result of 
gravitational back-reaction. ${\cal F}_{(i)}$ are tensors which depend on the initial velocities:
\ben
{\cal F}^{\mu\nu}_{(i) \alpha\beta}\ =\ V_{(i)}^{\mu}\delta^{\nu}_{(\alpha}V_{(i) \beta)}\ -\ \frac{1}{2}\eta^{\mu\nu}V_{(i) \alpha}V_{(i) \beta}\, ,
\een
and $h_{(i)}$ gives the metric fluctuation produced at the location of the $i$-th particle due
to the other particle. $S_{\alpha\beta}$ denotes the contribution to the stress tensor from the gravitational
field, and the relevant part of $S_{\alpha\beta}$ that produces radiation involves the product of the gravitational
fields from the two particle trajectories. 
As before, $\tilde e_{\alpha\beta}$ is computed from the stress tensor using the formula:
\ben
\tilde{e}_{\alpha\beta}(\omega,\vec{x})\ =\ \int dx^{0}e^{i\omega  x^{0}}\int d^{D}x^{\prime}\ 
G_{r}(x,x^{\prime}) 
T_{\alpha\beta}(x^{\prime})\, ,
\een
where $G_r$ denotes $D$-dimensional retarded Green's function. 

Now using the fact that $V_{(i)}(\sigma) = V_{(i)}+\delta V_{(i)}(\sigma)$ and that 
$V_{(i)}(\sigma) \propto P_{(i)}(\sigma)$, the sum of $T^{(0)}_{\alpha\beta}(x)$ and the first term
in the expression for $T^{(1)}_{\alpha\beta}(x)$ may be written as
\be
\sum_{i=1}^{2}\ \int\ d\sigma P_{(i) \alpha}(\sigma) V_{(i) \beta} (\sigma)
\, \delta^{(D)}(x\ -\ r^{(0)}(\sigma))\, ,
\ee
up to terms of order $(\delta V_{(i)}(\sigma))^2$. 
By standard analysis reviewed in the previous sections it
follows that its contribution to the radiative part of the metric reproduces the result of the leading 
soft theorem up to subleading corrections.

We shall now show that the rest of the terms in $T_{\mu\nu}$ 
do not contribute to $\tilde{e}_{\alpha\beta}(\omega,\vec{x})$ at leading order in $\omega$. 
As we know from our previous analysis these terms will only contribute 
to leading order if the source approaches constant values as $t\to\pm\infty$. 
In the second term in the expression for $T^{(1)}_{\mu\nu}$, the 
$ h^{(i)}_{\alpha\beta}(r(\sigma))$ factor, representing the gravitational field at the location of the
$i$-th particle due to the other particle, vanishes as $\sigma\rightarrow \pm\infty$ since in this limit the
two particles get widely separated. The same remark holds for the $S_{\mu\nu}$ term -- since
this involves product of the gravitational fields of the two particles, in the $t\to\pm\infty$ limit
this contribution vanishes. Finally  the 
contribution to the radiative component of $\tilde{e}_{\alpha\beta}(\omega,\vec{x})$ from the third term in 
$T^{(1)}_{\alpha\beta}$ 
may be obtained by replacing the
$V_{(\alpha}\Sigma_{\beta)\gamma}$ term in \refb{epart11A} by 
$P_{(i)\alpha}V_{(i)\beta}V_{(i)\gamma}$. We now see that due to the presence of the term proportional
to $k^\gamma$ in \refb{epart11A}, its contribution begins at the subleading order. Therefore it also does 
not contribute at the leading order.

\sectiono{Modification of soft expansion in four dimensions} \label{sir}

In four space-time dimensions the usual S-matrix is ill defined due to 
infrared divergences which arise due to long range nature of the interactions, 
as is the case for electromagnetism and gravity. 
Therefore soft theorems, that give the relation between S-matrix elements, also become
ill-defined except perhaps as relation between the S-matrix elements in an infra-red regulated
theory\cite{1405.1015}.
The long range nature of interaction implies that scattering states -- or in other words asymptotic particles -- are never free and their trajectories display log-divergences at late times. 
In this section
we shall examine how infrared divergence affects classical scattering process, and soft theorem that
gives the expression for soft radiation during such scattering.

We begin by examining the analysis of soft electromagnetic radiation 
given in section \ref{sphoton}. As already mentioned, soft photon theorem holds irrespective
of the nature of the scattering, {\it e.g.} even when the scatterer has no charge
and the scattering takes place via non-electromagnetic interaction. However let us 
consider the case where the scatterer carries a charge. In this case there will be a long
range Coulomb force acting \asnote{on }the 
probe even when it is far away from the scatterer. 
To see the effect of this we first express the 
late and early time straight line 
 trajectories of the probe in the absence of long range Coulomb force, given in
\refb{esttraj}, as 
\be\label{eintr}
\vec r(t) = \vec b_\pm  + \vec \beta_\pm\, (t -a_\pm) 
\,  , \quad \hbox{as $t\to\pm\infty$}\, ,
\ee
with
\be
c_-= (a_-, \vec b_-), \quad  c_+=(a_+, \vec b_+), \quad 
V_\pm = V^0_\pm (1, \vec\beta_\pm)\, .
\ee
It
is easy to see that in the presence of long range Coulomb force on the probe,
the late and early time straight line 
 trajectories of the probe given in
\refb{eintr} will be modified to the form
\be\label{emodtraj}
\vec r(t) = \vec b_\pm  + \vec \beta_\pm\, (t -a_\pm) 
- C_\pm \, \vec\beta_\pm\, \ln |t|\,  , \quad \hbox{as $t\to\pm\infty$}\, ,
\ee
where $C_\pm$ are constants that depend on the strength of the Coulomb interaction
between the probe and the scatterer. 
In order to see the effect of these logarithmic terms on the soft theorem, we 
label $r_\pm$ defined in \refb{e717}
as $(t_\pm, \vec r_\pm)$ and rewrite \refb{eafin3} as
\ben\label{eafin3aa}
\ve^\alpha \wt A_\alpha &=& \NN\, e^{i\omega R} \, 
\Bigg[ -{q \, \ve.V_{+} \over k.V_+} (1- ik^0 t_++ i\vec k.\vec r_+)
+ {q \, \ve.V_{-} \over k.V_-} (1-ik^0t_-+ i\vec k.\vec r_-) 
\nonumber \\ &&
+ i \, q\, (-\ve^0 t_+ 
+ \vec \ve.\vec r_+ + \ve^0 t_- - \vec \ve .\vec r_-)
\Bigg]\, .
\een
Substituting \refb{emodtraj} into \refb{eafin3aa} 
one can easily see that 
the soft factor $\bar S_{\rm em}$ computed from classical radiation
formula, defined as the term inside the square bracket
in \refb{eafin3aa},  acquires an extra term given by
\be\label{edelsb}
\Delta \bar S_{\rm em} = -iq\left\{
C_+\, \vec\ve.\vec \beta_+ \, \ln |t_+| \, {1\over 1-\hat n.\vec\beta_+}
- C_-\, \vec\ve.\vec \beta_- \, \ln |t_-| \, {1\over 1-\hat n.\vec\beta_-}
\right\}
\, ,
\ee
where we have chosen $\ve^0=0$ gauge.
On the other hand the change in the form of $\vec r_\pm(t)$ also modifies the
form of $\vec j_{(a)}^{\mu\nu}$ given in \refb{ebjab}. It is easy to see that the terms
proportional to $C_\pm$ in \refb{emodtraj} leaves $\bj_{(a)}^{ij}$ unchanged for
$1\le i,j\le 3$ but changes $\bj_{(a)}^{i0}$ by
\be
\Delta \bj_\pm^{i0} = \Delta r^i_\pm \, p_\pm^0
= - C_\pm \, \beta^i_\pm \, p^0_\pm \, \ln |t|\, .
\ee
Under this change the soft factor \refb{eph1} computed from soft theorem 
changes by
\be \label{edels}
\Delta S_{\rm em} = -iq\left\{
C_+\, \vec\ve.\vec \beta_+ \, \ln |t_+| \, {1\over 1-\hat n.\vec\beta_+}
- C_-\, \vec\ve.\vec \beta_- \, \ln |t_-| \, {1\over 1-\hat n.\vec\beta_-}
\right\}
\, .
\ee
Since this is equal to \refb{edelsb} the soft theorem still holds formally. 
This agreement is not surprising since in both computations we effectively replace 
$\vec b_\pm$ by $\vec b_{\pm} - C_\pm \, \vec \beta_\pm \, \ln|t_\pm|$.
However the
presence of $\ln |t_\pm|$ terms in the soft factor indicates that the soft expansion
itself breaks down. Instead of expanding the right hand side of \refb{eafin} 
in a power series expansion in $\omega$ before integrating over $\sigma$,
one should \asnote{express this as
\be
-{1\over \omega} \, \NN\, e^{i\omega R} \int_{-\infty}^\infty d\sigma \, e^{i\omega (r^0(\sigma)
-\hat n.\vec r(\sigma))}\, {\p\over \p\sigma} \left( {q V_\alpha(\sigma)\over V^0(\sigma) -
\hat n. \vec V(\sigma)}\right) \, ,
\ee
 }carry out the integration keeping the 
$\omega(r^0(\sigma)-\hat n.\vec r(\sigma))$ in the
exponent, and then study the small $\omega$ behavior of the amplitude. The presence
of the $\ln |t_\pm|$ term \refb{edels} is an indication that this expansion is non-analytic
at $\omega=0$, involving $\ln\omega$ terms in the expansion. In the quantum theory this is reflected
in the fact that derivatives of the amplitude with respect to the external momenta, hidden in the
definition of $\JJ^{\mu\nu}$, diverge in the soft limit.

A similar analysis can be carried out for soft graviton theorem. Since everything 
gravitates, in this case it is impossible to switch off the long range interactions in the
far past and far future. If for simplicity 
we work in the probe limit in which the particles 2, 4 and 5 of section 
\ref{sfusiona} are
infinitely heavy, then in \refb{egravtra} we shall get terms proportional to $\ln |t|$
in the expression for $\vec r(\sigma_1)$ and $\vec r(\sigma_3)$. 
This will generate logarithmic
corrections to the soft factor indicating that the expansion in power series in $\omega$
breaks down at the subleading order. In this case there will also be additional effects
due to long range gravitational force acting on the soft gravitons.

\sectiono{Discussion} \label{sdis}

In this paper we have analyzed the consequences of soft theorem in the classical limit
and have shown that they can predict the spectrum of long wavelength radiation 
during a classical scattering. We shall end \asnote{by first recalling some salient features 
of our analysis and then making a few comments on some open issues. }
\begin{enumerate}

\item \asnote{{\bf The probe limit and beyond:}
 }Much of our discussion on the soft theorem has been carried out in the probe limit
where one of the objects is much heavier than the other. This ensures that the total
amount of radiation emitted during the scattering is small compared to the mass of the
lighter object. However there are other ways to ensure this, e.g. by keeping the impact
parameter $b$ to be larger than the Schwarzschild radii of the objects, as
in the example of sections \ref{sfusiona} and \ref{smulti}. Soft theorem can be
used to compute classical gravity wave radiation during such cases as well -- the 
only change \asnote{is }that we now have to use the soft factors given in \refb{esoft0} and
\refb{esublead0} instead of \refb{es0} and \refb{esubleading}. 
Alternatively we can use \refb{es0}, \refb{esubleading} but allow the
sum over $a$ to run over all external states so that the additional terms vanish by momentum 
conservation.
The soft expansion will now be an
expansion in powers of $\omega b$. 

If we take the external objects to have masses of the same order and allow them to come
within a distance of the order of their Schwarzschild radii, then the total 
radiated energy does not remain small compared to the energies of the incident objects and we can
no longer ignore the contribution of the `hard gravitons' carrying this energy to the
soft factors given in \refb{esoft0}, \refb{esublead0}, 
\refb{esoftgr2}. Inclusion of these factors will require
detailed knowledge of the spectrum of these \asnote{gravitons }in the final state.
\item {\bf Nature of interactions:}
Our result does not depend on the nature of the interaction that 
takes place during the scattering process -- it need not be only electromagnetic or
gravitational.
\item {\bf Multi-body scattering:}
Our analysis can also be applied to multi-body scattering. For example if \asnote{some object(s)
 }breaks apart into several pieces during the scattering process due to some internal force,
we can still compute the soft radiation by knowing the initial states before the scattering
and the final states after the scattering, without any detailed knowledge of the process
that broke the \asnote{object }apart during the scattering process. 
\item {\bf Hidden assumptions:}
The derivation of soft theorem in generic theories of gravity, given 
in \cite{1703.00024,1706.00759}, implicitly
assumed that the mass spectrum of particles in the theory is discrete. 
In the case of continuous spectrum the emission or absorption
of soft radiation could take place via a series of transitions between real 
intermediate states instead of via the virtual intermediate states as considered
in \cite{1703.00024,1706.00759}.
For classical
objects this assumption of discrete spectrum 
breaks down since the spectrum becomes almost continuous.
However we do not expect emission of \asnote{soft }gravity 
waves to cause transition between the
internal states of the classical object, and therefore the fact that the spectrum is not
discrete should not affect the derivation of the soft theorem. A possible exception arises
when the scatterer is a black hole.
In this case some of the gravity waves emitted during the scattering 
could be absorbed by the black hole, changing the internal state of the black hole. 
Can this affect our analysis? 
Soft radiations of wave-length $\omega^{-1}$ are expected to be produced
at a distance of order $\omega^{-1}$ from the scatterer. For $D=4$
the solid angle subtended by the
scatterer of linear size $a$ to a point at a distance $\omega^{-1}$ is of order 
$a^2 \omega^2$. This is a subsubleading effect and is of the same order as the
non-universal terms in the soft theorem. 
In higher dimensions the effect is even smaller.
Therefore we do not expect our results 
up to subleading order (and the universal part at the subsubleading order) to
be affected due to the absorption of the soft radiation by the black hole.

\item {\bf Observation:} 
Given the observation of gravitational wave at LIGO\cite{1602.03837}, 
one natural 
question is whether we can actually test soft graviton
theorem from the observation of gravitational waves. \asnote{Unfortunately the direct application
of soft theorem requires us to consider a scattering event instead of the merger of a bound pair of
objects that is the common source of gravitational waves. The 
most likely event }is the fusion of two massive objects as discussed 
in section \ref{splunge}
or their scattering at close encounter discussed in section \ref{sgrav}. However
since the soft factor is small when the relative velocity between the incoming objects is small, 
an appreciable soft radiation will require reasonably large initial 
relative velocity between the two
objects when they are far away from each other, and at the same time their trajectories must
come close to each other so that there is appreciable amount of gravitational radiation.  
Such events are rare.
If we want to go beyond the leading order, we also
need to understand properly the infrared modification of the soft theorem in four dimensions. Despite
these difficulties it is not inconceivable that soft graviton theorem may be tested in 
actual  gravitational wave
experiments in the distant future.

\end{enumerate}

\bigskip

{\bf Acknowledgement:}
We wish to thank K.G. Arun, Sayantani Bhattacharyya, Miguel Campiglia, Gary Horowitz, Sachin Jain,
Shilpa Kastha, Prahar Mitra, Siddharth Prabhu,
Madhusudan Raman,  K.P. Yogendran  and Yogesh Srivastava for 
useful discussions and K.G. Arun for pointing out very useful references. 
A.S. would like to acknowledge the  hospitality of IMSc,
Chennai where this work was initiated, ICTS Bangalore where this 
work was nearing completion and KITP, Santa Barbara for
a visit during the course of this work during which the
research was supported in part by the National Science Foundation under Grant No. NSF PHY-1125915. 
A.S. would also
like to thank the theory group at NISER, Bhubaneswar for hospitality during a visit
when this work was presented.  A.L. would like to thank IIT Gandhinagar for their hospitality during the course of this work and  ICTS Bangalore during the final stages of this work. 
Work of A.L is supported  in part by Ramanujan Fellowship. 
The work of A.S. was
supported in part by the 
J. C. Bose fellowship of 
the Department of Science and Technology, India.

\appendix

\sectiono{Alternative analysis of the test of soft photon theorem in four dimensions} 
\label{sphotonfour}

In this appendix we shall recall the result of \cite{jackson} for classical 
electromagnetic radiation from particle
scattering in four space-time dimensions
and compare the result with the prediction of the soft photon
theorem. From eq.(15.1) of \cite{jackson} we get
the formula for the total energy radiated in the frequency range 
between $\omega$ and $\omega(1+\delta)$, carrying polarization $\ve$,
within a solid angle $\Delta\Omega$
around the direction labelled by $\hat n$:
\be \label{eja1}
{q^2\over 16\pi^3}\, \omega \, \delta\, \Delta\Omega\, 
\left|\int{d\over dt}\left[ {\vec\ve. \left\{
\hat n \times (\hat n\times \vec\beta)\right\} \over 1 - \hat n. \vec\beta}\right]
e^{i\omega (t - \hat n. \vec r(t))} dt
\right|^2\, .
\ee
Here  $\vec r(t)$ is the 
location of the probe at time $t$ and $\vec\beta=d\vec r/ dt$ is the velocity. 
The polarization four vector has been taken to be of the form $(0,\vec\ve)$ with
$k^\mu \ve_\mu = -\omega\hat n. \vec \ve =0$.
In writing
\refb{eja1} we have taken into account the factor of $4\pi$ mentioned in footnote
\ref{fo1}. Comparing this with \refb{epower} for $D=4$
we see that the soft factor should be
given by
\be 
\bar S_{\rm em}(\ve, k) = e^{i\alpha}\, {q\over \omega} \int{d\over dt}\left[ {\vec\ve. \left\{
\hat n \times (\hat n\times \vec\beta)\right\} \over 1 - \hat n. \vec\beta}\right]
e^{i\omega (t - \hat n. \vec r(t))} dt\, ,
\ee
where $e^{i\alpha}$ is a phase that will be fixed by comparing this with the leading
soft factor. The bar on $S_{\rm em}$ indicates that this is the soft factor computed
from the classical radiation formula.
Using the gauge condition $\vec \ve. \hat n=0$, we can express this as
\be \label{eppex}
\bar S_{\rm em}(\ve, k) =- e^{i\alpha}\, {q\over \omega} \int{d\over dt}\left[ {\vec\ve. \vec\beta 
\over 1 - \hat n. \vec\beta}\right]
e^{i\omega (t - \hat n. \vec r(t))} dt\, .
\ee
We now expand $e^{i\omega (t - \hat n. \vec r(t))}$ to first order in $\omega$ and
integrate by parts the integral over $t$. This gives
\be \label{epp1}
\bar S_{\rm em}(\ve, k) =- e^{i\alpha}\, {q\over \omega} \left[ {\vec\ve. \vec\beta 
\over 1 - \hat n. \vec\beta} 
\left\{1 + i\omega (t - \hat n. \vec r(t))\right\}
\right]_{t_-}^{t_+}
+ e^{i\alpha}\, {q\over \omega} \int_{t_-}^{t_+} dt \, {\vec\ve. \vec\beta 
\over 1 - \hat n. \vec\beta} \,  i \omega\, \left(1 - \hat n. \vec \beta\right)
\ee
where we have to take $t_\pm \to\pm \infty$ limit at the end of the calculation. 
We now note
that using the relation $\vec\beta = d\vec r/dt$ 
the second term can be integrated explicitly, and we can express
\refb{epp1} as
\ben \label{epp2}
\bar S_{\rm em}(\ve, k) &=& - e^{i\alpha}\, {q\over \omega} \left[ {\vec\ve. \vec\beta 
\over 1 - \hat n. \vec\beta} 
\left\{1 + i\omega (t - \hat n. \vec r(t))\right\}
\right]_{t_-}^{t_+}
+ e^{i\alpha}\, {q\over \omega}  \, i \omega\,  \ \Big[ \vec\ve. \vec r\Big]_{t_-}^{t_+}
\nonumber \\ &=&
- e^{i\alpha} \, q\, \omega^{-1} \left[{\vec\ve. \vec\beta_+
\over 1 - \hat n. \vec\beta_+} - {\vec\ve. \vec\beta_-
\over 1 - \hat n. \vec\beta_-}
\right] \nonumber \\ &&
+ i\, q\, e^{i\alpha}\, \left[{1\over 1-\hat n. \vec \beta_+}
\left\{ -\vec\ve. \vec\beta_+\,  t_+ +  \vec\ve. \vec\beta_+ \,
\hat n. \vec r(t_+) 
+ \vec\ve . \vec r(t_+) - \hat n. \vec \beta_+ \  \vec\ve . \vec r(t_+)
\right\}
\right] \nonumber \\ &&
- i\, q\, e^{i\alpha}\, \left[{1\over 1-\hat n. \vec \beta_-}
\left\{- \vec\ve. \vec\beta_- \, t_- +  \vec\ve. \vec\beta_- \hat n. \vec r(t_-) 
+ \vec\ve . \vec r(t_-) - \hat n. \vec \beta_- \ \vec\ve . \vec r(t_-)
\right\}
\right]
 \, .\nonumber \\
\een
Using covariant notation,
\be
V_\pm = V^0_\pm (1, \vec \beta_\pm), \quad V^0_\pm = (1-\vec\beta_\pm^2)^{-1/2},
\quad r_{\pm} = (t_\pm, \vec r_\pm), \quad k = -\omega(1,\hat n)\, ,
\ee
\refb{epp2} may be written as
\be \label{esemcova}
\bar S_{\rm em} = e^{i\alpha}
\Bigg[ -q \left\{ {\ve.V_{+} \over k.V_+} - {\ve.V_{-} \over k.V_-}
\right\} - {i \,q\over k.V_+}\, \left\{ \ve . V_+ \, k. r_+- \ve . r_+ \, k. V_+\right\}
+  {i \,q\over k.V_-} \, \left\{ \ve . V_- \, k. r_- - \ve . r_- \, k. V_-\right\}
\Bigg]\, .
\ee
Upon using \refb{e716} this becomes 
identical to the right hand side of \refb{esemcov} for $\alpha=0$. Therefore we see that
$\bar S_{\rm em}=S_{\rm em}$.


\begin{thebibliography}{99}



\small

\baselineskip=14pt

\parskip=0pt

\bibitem{Gell-Mann}
M.~Gell-Mann and M.~L.~Goldberger, Phys.\ Rev.\ {\bf 96}, 1433 (1954).

\bibitem{low}
F.~E.~Low, Phys.\ Rev.\ {\bf 110}, 974 (1958).

\bibitem{saito}
S.~Saito, Phys.\ Rev.\ {\bf 184}, 1894 (1969).

\bibitem{burnett}
T.~H.~Burnett and N.~M.~Kroll, Phys.\ Rev.\ Lett.\ {\bf 20}, 86 (1968).

\bibitem{bell}
J.~S.~Bell and R. Van Royen, Nuovo Cim.\ {\bf A60}, 62 (1969).

\bibitem{duca}
V.~Del Duca, Nucl. Phys. {\bf B345}, 369 (1990).


\bibitem{weinberg1} 
  S.~Weinberg,
  ``Photons and Gravitons in s Matrix Theory: 
  Derivation of Charge Conservation and Equality of Gravitational and Inertial Mass,''
  Phys.\ Rev.\  {\bf 135}, B1049 (1964).
  doi:10.1103/PhysRev.135.B1049

\bibitem{weinberg2} 
  S.~Weinberg,
  ``Infrared photons and gravitons,''
  Phys.\ Rev.\  {\bf 140}, B516 (1965).
  doi:10.1103/PhysRev.140.B516

\bibitem{jackiw1} 
  D.~J.~Gross and R.~Jackiw,
  ``Low-Energy Theorem for Graviton Scattering,''
  Phys.\ Rev.\  {\bf 166}, 1287 (1968).
  doi:10.1103/PhysRev.166.1287
  
\bibitem{jackiw2} 
  R.~Jackiw,
  ``Low-Energy Theorems for Massless Bosons: Photons and Gravitons,''
  Phys.\ Rev.\  {\bf 168}, 1623 (1968).
  doi:10.1103/PhysRev.168.1623
  
  
\bibitem{ademollo} 
  M.~Ademollo, A.~D'Adda, R.~D'Auria, F.~Gliozzi, E.~Napolitano, S.~Sciuto and P.~Di Vecchia,
  ``Soft Dilations and Scale Renormalization in Dual Theories,''
  Nucl.\ Phys.\ B {\bf 94}, 221 (1975).
  doi:10.1016/0550-3213(75)90491-5

\bibitem{shapiro} 
  J.~A.~Shapiro,
  ``On the Renormalization of Dual Models,''
  Phys.\ Rev.\ D {\bf 11}, 2937 (1975).
  doi:10.1103/PhysRevD.11.2937


\bibitem{1312.2229} 
  A.~Strominger,
  ``On BMS Invariance of Gravitational Scattering,''
  JHEP {\bf 1407}, 152 (2014)
  doi:10.1007/JHEP07(2014)152
  [arXiv:1312.2229 [hep-th]].

\bibitem{1401.7026} 
  T.~He, V.~Lysov, P.~Mitra and A.~Strominger,
  ``BMS supertranslations and WeinbergÕs soft graviton theorem,''
  JHEP {\bf 1505}, 151 (2015)
  doi:10.1007/JHEP05(2015)151
  [arXiv:1401.7026 [hep-th]].

\bibitem{1408.2228} 
  M.~Campiglia and A.~Laddha,
  ``Asymptotic symmetries and subleading soft graviton theorem,''
  Phys.\ Rev.\ D {\bf 90}, no. 12, 124028 (2014)
  doi:10.1103/PhysRevD.90.124028
  [arXiv:1408.2228 [hep-th]].

\bibitem{1411.5745} 
  A.~Strominger and A.~Zhiboedov,
  ``Gravitational Memory, BMS Supertranslations and Soft Theorems,''
  JHEP {\bf 1601}, 086 (2016)
  doi:10.1007/JHEP01(2016)086
  [arXiv:1411.5745 [hep-th]].

\bibitem{1502.02318} 
  M.~Campiglia and A.~Laddha,
  ``New symmetries for the Gravitational S-matrix,''
  JHEP {\bf 1504}, 076 (2015)
  doi:10.1007/JHEP04(2015)076
  [arXiv:1502.02318 [hep-th]].
 
 \bibitem{1505.05346} 
  M.~Campiglia and A.~Laddha,
  ``Asymptotic symmetries of QED and Weinberg?s soft photon theorem,''
  JHEP {\bf 1507}, 115 (2015)
  doi:10.1007/JHEP07(2015)115
  [arXiv:1505.05346 [hep-th]].
  
\bibitem{1506.05789} 
  S.~G.~Avery and B.~U.~W.~Schwab,
  ``Burg-Metzner-Sachs symmetry, string theory, and soft theorems,''
  Phys.\ Rev.\ D {\bf 93}, 026003 (2016)
  doi:10.1103/PhysRevD.93.026003
  [arXiv:1506.05789 [hep-th]].

\bibitem{1509.01406} 
  M.~Campiglia and A.~Laddha,
  ``Asymptotic symmetries of gravity and soft theorems for massive particles,''
  JHEP {\bf 1512}, 094 (2015)
  doi:10.1007/JHEP12(2015)094
  [arXiv:1509.01406 [hep-th]].

\bibitem{1605.09094} 
  M.~Campiglia and A.~Laddha,
  ``Sub-subleading soft gravitons: New symmetries of quantum gravity?,''
  Phys.\ Lett.\ B {\bf 764}, 218 (2017)
  doi:10.1016/j.physletb.2016.11.046
  [arXiv:1605.09094 [gr-qc]].

\bibitem{1605.09677} 
  M.~Campiglia and A.~Laddha,
  ``Subleading soft photons and large gauge transformations,''
  JHEP {\bf 1611}, 012 (2016)
  doi:10.1007/JHEP11(2016)012
  [arXiv:1605.09677 [hep-th]].

\bibitem{1608.00685} 
  M.~Campiglia and A.~Laddha,
  ``Sub-subleading soft gravitons and large diffeomorphisms,''
  JHEP {\bf 1701}, 036 (2017)
  doi:10.1007/JHEP01(2017)036
  [arXiv:1608.00685 [gr-qc]].

\bibitem{1612.08294} 
  E.~Conde and P.~Mao,
  ``BMS Supertranslations and Not So Soft Gravitons,''
  arXiv:1612.08294 [hep-th].

\bibitem{1701.00496} 
  T.~He, D.~Kapec, A.~M.~Raclariu and A.~Strominger,
  ``Loop-Corrected Virasoro Symmetry of 4D Quantum Gravity,''
  arXiv:1701.00496 [hep-th].


\bibitem{1612.05886} 
  M.~Asorey, A.~P.~Balachandran, F.~Lizzi and G.~Marmo,
  ``Equations of Motion as Constraints: Superselection Rules, Ward Identities,''
  arXiv:1612.05886 [hep-th].

\bibitem{1703.05448} 
  A.~Strominger,
  ``Lectures on the Infrared Structure of Gravity and Gauge Theory,''
  arXiv:1703.05448 [hep-th].

\bibitem{1709.03850} 
  A.~Laddha and P.~Mitra,
  ``Asymptotic Symmetries and Subleading Soft Photon Theorem in Effective Field Theories,''
  arXiv:1709.03850 [hep-th].
  
\bibitem{1711.04371} 
  D.~Kapec and P.~Mitra,
  ``A $d$-Dimensional Stress Tensor for Mink$_{d+2}$ Gravity,''
  arXiv:1711.04371 [hep-th].

\bibitem{progress}
Anupam H, Arpan Kundu, Alok Laddha, Work in progress.



\bibitem{1103.2981} 
  C.~D.~White,
  ``Factorization Properties of Soft Graviton Amplitudes,''
  JHEP {\bf 1105}, 060 (2011)
  doi:10.1007/JHEP05(2011)060
  [arXiv:1103.2981 [hep-th]].

\bibitem{1404.4091} 
  F.~Cachazo and A.~Strominger,
  ``Evidence for a New Soft Graviton Theorem,''
  arXiv:1404.4091 [hep-th].

\bibitem{1404.5551} 
  E.~Casali,
  ``Soft sub-leading divergences in Yang-Mills amplitudes,''
  JHEP {\bf 1408}, 077 (2014)
  doi:10.1007/JHEP08(2014)077
  [arXiv:1404.5551 [hep-th]].

\bibitem{1404.7749} 
  B.~U.~W.~Schwab and A.~Volovich,
  ``Subleading Soft Theorem in Arbitrary Dimensions from Scattering Equations,''
  Phys.\ Rev.\ Lett.\  {\bf 113}, no. 10, 101601 (2014)
  doi:10.1103/PhysRevLett.113.101601
  [arXiv:1404.7749 [hep-th]].

\bibitem{1405.1015} 
  Z.~Bern, S.~Davies and J.~Nohle,
  ``On Loop Corrections to Subleading Soft Behavior of Gluons and Gravitons,''
  Phys.\ Rev.\ D {\bf 90}, no. 8, 085015 (2014)
  doi:10.1103/PhysRevD.90.085015
  [arXiv:1405.1015 [hep-th]].

\bibitem{1405.1410} 
  S.~He, Y.~t.~Huang and C.~Wen,
  ``Loop Corrections to Soft Theorems in Gauge Theories and Gravity,''
  JHEP {\bf 1412}, 115 (2014)
  doi:10.1007/JHEP12(2014)115
  [arXiv:1405.1410 [hep-th]].

\bibitem{1405.2346} 
  A.~J.~Larkoski,
  ``Conformal Invariance of the Subleading Soft Theorem in Gauge Theory,''
  Phys.\ Rev.\ D {\bf 90}, no. 8, 087701 (2014)
  doi:10.1103/PhysRevD.90.087701
  [arXiv:1405.2346 [hep-th]].


\bibitem{1405.3413} 
  F.~Cachazo and E.~Y.~Yuan,
  ``Are Soft Theorems Renormalized?,''
  arXiv:1405.3413 [hep-th].

\bibitem{1405.3533} 
  N.~Afkhami-Jeddi,
  ``Soft Graviton Theorem in Arbitrary Dimensions,''
  arXiv:1405.3533 [hep-th].

\bibitem{1406.6574} 
  J.~Broedel, M.~de Leeuw, J.~Plefka and M.~Rosso,
  ``Constraining subleading soft gluon and graviton theorems,''
  Phys.\ Rev.\ D {\bf 90}, no. 6, 065024 (2014)
  doi:10.1103/PhysRevD.90.065024
  [arXiv:1406.6574 [hep-th]].

\bibitem{1406.6987} 
  Z.~Bern, S.~Davies, P.~Di Vecchia and J.~Nohle,
  ``Low-Energy Behavior of Gluons and Gravitons from Gauge Invariance,''
  Phys.\ Rev.\ D {\bf 90}, no. 8, 084035 (2014)
  doi:10.1103/PhysRevD.90.084035
  [arXiv:1406.6987 [hep-th]].
  
\bibitem{1406.7184} 
  C.~D.~White,
  ``Diagrammatic insights into next-to-soft corrections,''
  Phys.\ Lett.\ B {\bf 737}, 216 (2014)
  doi:10.1016/j.physletb.2014.08.041
  [arXiv:1406.7184 [hep-th]].
  
\bibitem{1407.5936} 
  M.~Zlotnikov,
  ``Sub-sub-leading soft-graviton theorem in arbitrary dimension,''
  JHEP {\bf 1410}, 148 (2014)
  doi:10.1007/JHEP10(2014)148
  [arXiv:1407.5936 [hep-th]].

\bibitem{1407.5982} 
  C.~Kalousios and F.~Rojas,
  ``Next to subleading soft-graviton theorem in arbitrary dimensions,''
  JHEP {\bf 1501}, 107 (2015)
  doi:10.1007/JHEP01(2015)107
  [arXiv:1407.5982 [hep-th]].

\bibitem{1408.4179} 
  Y.~J.~Du, B.~Feng, C.~H.~Fu and Y.~Wang,
  ``Note on Soft Graviton theorem by KLT Relation,''
  JHEP {\bf 1411}, 090 (2014)
  doi:10.1007/JHEP11(2014)090
  [arXiv:1408.4179 [hep-th]].

\bibitem{1410.6406} 
  D.~Bonocore, E.~Laenen, L.~Magnea, L.~Vernazza and C.~D.~White,
  ``The method of regions and next-to-soft corrections in DrellÐYan production,''
  Phys.\ Lett.\ B {\bf 742}, 375 (2015)
  doi:10.1016/j.physletb.2015.02.008
  [arXiv:1410.6406 [hep-ph]].

\bibitem{1412.3699} 
  A.~Sabio Vera and M.~A.~Vazquez-Mozo,
  ``The Double Copy Structure of Soft Gravitons,''
  JHEP {\bf 1503}, 070 (2015)
  doi:10.1007/JHEP03(2015)070
  [arXiv:1412.3699 [hep-th]].




\bibitem{1503.04816} 
  F.~Cachazo, S.~He and E.~Y.~Yuan,
  ``New Double Soft Emission Theorems,''
  Phys.\ Rev.\ D {\bf 92}, no. 6, 065030 (2015)
  doi:10.1103/PhysRevD.92.065030
  [arXiv:1503.04816 [hep-th]].

\bibitem{1504.01364} 
  A.~E.~Lipstein,
  ``Soft Theorems from Conformal Field Theory,''
  JHEP {\bf 1506}, 166 (2015)
  doi:10.1007/JHEP06(2015)166
  [arXiv:1504.01364 [hep-th]].

\bibitem{1507.08882} 
  S.~D.~Alston, D.~C.~Dunbar and W.~B.~Perkins,
  ``$n$-point amplitudes with a single negative-helicity graviton,''
  Phys.\ Rev.\ D {\bf 92}, no. 6, 065024 (2015)
  doi:10.1103/PhysRevD.92.065024
  [arXiv:1507.08882 [hep-th]].

\bibitem{1509.07840} 
  Y.~t.~Huang and C.~Wen,
  ``Soft theorems from anomalous symmetries,''
  JHEP {\bf 1512}, 143 (2015)
  doi:10.1007/JHEP12(2015)143
  [arXiv:1509.07840 [hep-th]].


\bibitem{1604.00650} 
  J.~Rao and B.~Feng,
  ``Note on Identities Inspired by New Soft Theorems,''
  JHEP {\bf 1604}, 173 (2016)
  doi:10.1007/JHEP04(2016)173
  [arXiv:1604.00650 [hep-th]].
  
 
  \bibitem{1604.02834} 
  S.~He, Z.~Liu and J.~B.~Wu,
  ``Scattering Equations, Twistor-string Formulas and Double-soft Limits 
  in Four Dimensions,''
  JHEP {\bf 1607}, 060 (2016)
  doi:10.1007/JHEP07(2016)060
  [arXiv:1604.02834 [hep-th]].

\bibitem{1604.03893} 
  F.~Cachazo, P.~Cha and S.~Mizera,
  ``Extensions of Theories from Soft Limits,''
  JHEP {\bf 1606}, 170 (2016)
  doi:10.1007/JHEP06(2016)170
  [arXiv:1604.03893 [hep-th]].

\bibitem{1607.02700} 
  A.~P.~Saha,
  ``Double Soft Theorem for Perturbative Gravity,''
  JHEP {\bf 1609}, 165 (2016)
  doi:10.1007/JHEP09(2016)165
  [arXiv:1607.02700 [hep-th]].

\bibitem{1611.02172} 
  A.~Luna, S.~Melville, S.~G.~Naculich and C.~D.~White,
  ``Next-to-soft corrections to high energy scattering in QCD and gravity,''
  JHEP {\bf 1701}, 052 (2017)
  doi:10.1007/JHEP01(2017)052
  [arXiv:1611.02172 [hep-th]].

\bibitem{1611.07534} 
  H.~Elvang, C.~R.~T.~Jones and S.~G.~Naculich,
  ``Soft Photon and Graviton Theorems in Effective Field Theory,''
  arXiv:1611.07534 [hep-th].

\bibitem{1611.03137} 
  C.~Cheung, K.~Kampf, J.~Novotny, C.~H.~Shen and J.~Trnka,
  ``A Periodic Table of Effective Field Theories,''
  arXiv:1611.03137 [hep-th].



\bibitem{1702.02350} 
  A.~P.~Saha,
  ``Double Soft Theorem for Perturbative Gravity II: Some Details on CHY Soft Limits,''
  arXiv:1702.02350 [hep-th].

\bibitem{1406.4172} 
  B.~U.~W.~Schwab,
  ``Subleading Soft Factor for String Disk Amplitudes,''
  JHEP {\bf 1408}, 062 (2014)
  doi:10.1007/JHEP08(2014)062
  [arXiv:1406.4172 [hep-th]].

\bibitem{1406.5155} 
  M.~Bianchi, S.~He, Y.~t.~Huang and C.~Wen,
  ``More on Soft Theorems: Trees, Loops and Strings,''
  Phys.\ Rev.\ D {\bf 92}, no. 6, 065022 (2015)
  doi:10.1103/PhysRevD.92.065022
  [arXiv:1406.5155 [hep-th]].

\bibitem{1411.6661} 
  B.~U.~W.~Schwab,
  ``A Note on Soft Factors for Closed String Scattering,''
  JHEP {\bf 1503}, 140 (2015)
  doi:10.1007/JHEP03(2015)140
  [arXiv:1411.6661 [hep-th]].

\bibitem{1502.05258}
 P.~Di Vecchia, R.~Marotta and M.~Mojaza,
  ``Soft theorem for the graviton, dilaton and the Kalb-Ramond field in the bosonic string,''
  JHEP {\bf 1505}, 137 (2015)
  doi:10.1007/JHEP05(2015)137
  [arXiv:1502.05258 [hep-th]].
  
  
\bibitem{1504.05558} 
  T.~Klose, T.~McLoughlin, D.~Nandan, J.~Plefka and G.~Travaglini,
  ``Double-Soft Limits of Gluons and Gravitons,''
  JHEP {\bf 1507}, 135 (2015)
  doi:10.1007/JHEP07(2015)135
  [arXiv:1504.05558 [hep-th]].

\bibitem{1504.05559} 
  A.~Volovich, C.~Wen and M.~Zlotnikov,
  ``Double Soft Theorems in Gauge and String Theories,''
  JHEP {\bf 1507}, 095 (2015)
  doi:10.1007/JHEP07(2015)095
  [arXiv:1504.05559 [hep-th]].

\bibitem{1505.05854} 
  M.~Bianchi and A.~L.~Guerrieri,
  ``On the soft limit of open string disk amplitudes with massive states,''
  JHEP {\bf 1509}, 164 (2015)
  doi:10.1007/JHEP09(2015)164
  [arXiv:1505.05854 [hep-th]].

\bibitem{1507.00938} 
  P.~Di Vecchia, R.~Marotta and M.~Mojaza,
  ``Double-soft behavior for scalars and gluons from string theory,''
  JHEP {\bf 1512}, 150 (2015)
  doi:10.1007/JHEP12(2015)150
  [arXiv:1507.00938 [hep-th]].

\bibitem{1507.08829} 
  A.~L.~Guerrieri,
  ``Soft behavior of string amplitudes with external massive states,''
  Nuovo Cim.\ C {\bf 39}, no. 1, 221 (2016)
  doi:10.1393/ncc/i2016-16221-2
  [arXiv:1507.08829 [hep-th]].

\bibitem{1511.04921} 
  P.~Di Vecchia, R.~Marotta and M.~Mojaza,
  ``Soft Theorems from String Theory,''
  Fortsch.\ Phys.\  {\bf 64}, 389 (2016)
  doi:10.1002/prop.201500068
  [arXiv:1511.04921 [hep-th]].
  
  \bibitem{1512.00803} 
  M.~Bianchi and A.~L.~Guerrieri,
  ``On the soft limit of closed string amplitudes with massive states,''
  Nucl.\ Phys.\ B {\bf 905}, 188 (2016)
  doi:10.1016/j.nuclphysb.2016.02.005
  [arXiv:1512.00803 [hep-th]].

\bibitem{1601.03457} 
  M.~Bianchi and A.~L.~Guerrieri,
  ``On the soft limit of tree-level string amplitudes,''
  arXiv:1601.03457 [hep-th].

\bibitem{1604.03355} 
  P.~Di Vecchia, R.~Marotta and M.~Mojaza,
  ``Subsubleading soft theorems of gravitons and dilatons in the bosonic string,''
  JHEP {\bf 1606}, 054 (2016)
  doi:10.1007/JHEP06(2016)054
  [arXiv:1604.03355 [hep-th]].

\bibitem{1610.03481} 
  P.~Di Vecchia, R.~Marotta and M.~Mojaza,
  ``Soft behavior of a closed massless state in superstring and universality in the soft behavior of the dilaton,''
  JHEP {\bf 1612}, 020 (2016)
  doi:10.1007/JHEP12(2016)020
  [arXiv:1610.03481 [hep-th]].

\bibitem{1702.03934} 
  A.~Sen,
  ``Soft Theorems in Superstring Theory,''
  arXiv:1702.03934 [hep-th].


\bibitem{1705.06175} 
  P.~Di Vecchia, R.~Marotta and M.~Mojaza,
  ``Double-soft behavior of the dilaton of spontaneously broken conformal invariance,''
  arXiv:1705.06175 [hep-th].

\bibitem{1703.00024} 
  A.~Sen,
  ``Subleading Soft Graviton Theorem for Loop Amplitudes,''
  arXiv:1703.00024 [hep-th].

\bibitem{1706.00759} 
  A.~Laddha and A.~Sen,
  ``Sub-subleading Soft Graviton Theorem in Generic Theories of Quantum Gravity,''
  arXiv:1706.00759 [hep-th].
 
 \bibitem{1707.06803} 
  S.~Chakrabarti, S.~P.~Kashyap, B.~Sahoo, A.~Sen and M.~Verma,
  ``Subleading Soft Theorem for Multiple Soft Gravitons,''
  arXiv:1707.06803 [hep-th].

\bibitem{jackson}
J.D.~Jackson, ``Classical Electrodynamics,''
  
\bibitem{1308.6285} 
  J.~Ware, R.~Saotome and R.~Akhoury,
  ``Construction of an asymptotic S matrix for perturbative quantum gravity,''
  JHEP {\bf 1310}, 159 (2013)
  doi:10.1007/JHEP10(2013)159
  [arXiv:1308.6285 [hep-th]].

\bibitem{1612.03290} 
  S.~Hollands, A.~Ishibashi and R.~M.~Wald,
  ``BMS Supertranslations and Memory in Four and Higher Dimensions,''
  Class.\ Quant.\ Grav.\  {\bf 34}, no. 15, 155005 (2017)
  doi:10.1088/1361-6382/aa777a
  [arXiv:1612.03290 [gr-qc]].
  
\bibitem{1702.00095} 
  D.~Garfinkle, S.~Hollands, A.~Ishibashi, A.~Tolish and R.~M.~Wald,
  ``The Memory Effect for Particle Scattering in Even Spacetime Dimensions,''
  Class.\ Quant.\ Grav.\  {\bf 34}, no. 14, 145015 (2017)
  doi:10.1088/1361-6382/aa777b
  [arXiv:1702.00095 [gr-qc]].

\bibitem{1712.00873} 
  G.~Satishchandran and R.~M.~Wald,
  ``The Memory Effect For Particle Scattering in Odd Spacetime Dimensions,''
  arXiv:1712.00873 [gr-qc].

\bibitem{1712.01204} 
  M.~Pate, A.~M.~Raclariu and A.~Strominger,
  ``Gravitational Memory in Higher Dimensions,''
  arXiv:1712.01204 [hep-th].


\bibitem{weinbergbook}
S.~Weinberg,
 ``Gravitation and Cosmology : Principles and Applications 
 of the General Theory of Relativity,''

\bibitem{1502.06120} 
  S.~Pasterski, A.~Strominger and A.~Zhiboedov,
  ``New Gravitational Memories,''
  JHEP {\bf 1612}, 053 (2016)
  doi:10.1007/JHEP12(2016)053
  [arXiv:1502.06120 [hep-th]].

\bibitem{0409156} 
  W.~D.~Goldberger and I.~Z.~Rothstein,
  ``An Effective field theory of gravity for extended objects,''
  Phys.\ Rev.\ D {\bf 73}, 104029 (2006)
  doi:10.1103/PhysRevD.73.104029
  [hep-th/0409156].
  
\bibitem{0605238} 
  W.~D.~Goldberger and I.~Z.~Rothstein,
  ``Towers of Gravitational Theories,''
  Gen.\ Rel.\ Grav.\  {\bf 38}, 1537 (2006)
  [Int.\ J.\ Mod.\ Phys.\ D {\bf 15}, 2293 (2006)]
  doi:10.1007/s10714-006-0345-7, 10.1142/S0218271806009698
  [hep-th/0605238].

\bibitem{0511061} 
  R.~A.~Porto,
  ``Post-Newtonian corrections to the motion of spinning bodies in NRGR,''
  Phys.\ Rev.\ D {\bf 73}, 104031 (2006)
  doi:10.1103/PhysRevD.73.104031
  [gr-qc/0511061].

\bibitem{1601.04914} 
  R.~A.~Porto,
  ``The effective field theorist?s approach to gravitational dynamics,''
  Phys.\ Rept.\  {\bf 633}, 1 (2016)
  doi:10.1016/j.physrep.2016.04.003
  [arXiv:1601.04914 [hep-th]].

\bibitem{1502.07644} 
  D.~Kapec, V.~Lysov, S.~Pasterski and A.~Strominger,
  ``Higher-Dimensional Supertranslations and Weinberg's Soft Graviton Theorem,''
  Annals of Mathematical Sciences and Applications, Volume 2 (2017),
  pp 69-94
  doi:10.4310/AMSA.2017.v2.n1.a2
  [arXiv:1502.07644 [gr-qc]].

\bibitem{1707.08016} 
  M.~Pate, A.~M.~Raclariu and A.~Strominger,
  ``Color Memory,''
  arXiv:1707.08016 [hep-th].


\bibitem{mem1}
Ya. B. Zeldovich and A. G. Polnarev, Sov. Astron. {\bf 18}, 17 (1974)

\bibitem{mem2}
V. B. Braginsky and L. P. Grishchuk, ``Kinematic Resonance and Memory Effect in Free Mass Gravitational Antennas,'' Sov. Phys. {\bf JETP 62}, 427 (1985), [Zh. Eksp. Teor.
Fiz. 89, 744 (1985)].

\bibitem{mem3}
V. B. Braginsky, K. S. Thorne, ``Gravitational-wave bursts with memory and experi-
mental prospects,'' Nature {\bf 327.6118}, 123-125  (1987).

\bibitem{mem4}
M. Ludvigsen, ``Geodesic Deviation At Null Infinity And The Physical Effects Of Very
Long Wave Gravitational Radiation,'' Gen. Rel. Grav. {\bf 21}, 1205 (1989).

\bibitem{0212168} 
  V.~Cardoso, O.~J.~C.~Dias and J.~P.~S.~Lemos,
  ``Gravitational radiation in D-dimensional space-times,''
  Phys.\ Rev.\ D {\bf 67}, 064026 (2003)
  doi:10.1103/PhysRevD.67.064026
  [hep-th/0212168].

\bibitem{1611.03493} 
  W.~D.~Goldberger and A.~K.~Ridgway,
  ``Radiation and the classical double copy for color charges,''
  Phys.\ Rev.\ D {\bf 95}, no. 12, 125010 (2017)
  doi:10.1103/PhysRevD.95.125010
  [arXiv:1611.03493 [hep-th]].
  
  \bibitem{1705.09263} 
  W.~D.~Goldberger, S.~G.~Prabhu and J.~O.~Thompson,
  ``Classical gluon and graviton radiation from the bi-adjoint scalar double copy,''
  Phys.\ Rev.\ D {\bf 96}, no. 6, 065009 (2017)
  doi:10.1103/PhysRevD.96.065009
  [arXiv:1705.09263 [hep-th]].

\bibitem{1712.09250} 
  W.~D.~Goldberger, J.~Li and S.~G.~Prabhu,
  ``Spinning particles, axion radiation, and the classical double copy,''
  arXiv:1712.09250 [hep-th].

\bibitem{peter}
P.~C.~Peters, ``Relativistic Gravitational Bremsstrahlung.'', Phys.\ Rev.\ {\bf D1}, 1559 (1970).

\bibitem{0309203} 
  E.~Berti, M.~Cavaglia and L.~Gualtieri,
  ``Gravitational energy loss in high-energy particle collisions: Ultrarelativistic plunge into a multidimensional black hole,''
  Phys.\ Rev.\ D {\bf 69}, 124011 (2004)
  doi:10.1103/PhysRevD.69.124011
  [hep-th/0309203].

\bibitem{1210.6976} 
  D.~Gal'tsov, P.~Spirin and T.~N.~Tomaras,
  ``Gravitational bremsstrahlung in ultra-planckian collisions,''
  JHEP {\bf 1301}, 087 (2013)
  doi:10.1007/JHEP01(2013)087
  [arXiv:1210.6976 [hep-th]].


\bibitem{1602.03837} 
  B.~P.~Abbott {\it et al.} [LIGO Scientific and Virgo Collaborations],
  ``Observation of Gravitational Waves from a Binary Black Hole Merger,''
  Phys.\ Rev.\ Lett.\  {\bf 116}, no. 6, 061102 (2016)
  doi:10.1103/PhysRevLett.116.061102
  [arXiv:1602.03837 [gr-qc]].

\end{thebibliography}
\end{document}